\documentclass[12pt]{article}
\usepackage{amsmath}
\usepackage{bm}
\usepackage{amsfonts}
\usepackage{graphicx,psfrag,epsf}
\usepackage{enumerate}
\usepackage[round]{natbib}
\usepackage{url} 
\usepackage{subcaption}
\usepackage{multirow}
\usepackage[]{algorithm2e}
\usepackage[dvipsnames]{xcolor}
\usepackage{units}
\usepackage{hyperref}
\usepackage{cleveref}
\usepackage{diagbox}
\usepackage{authblk}
\usepackage{booktabs}
\usepackage{xcolor}

\newcommand{\blind}{1}

\title{Misspecification-robust likelihood-free inference in high dimensions}

\author[1]{Owen Thomas}

\author[2]{Raquel S\'{a}-Le\~{a}o}

\author[3, 4]{Herm\'{i}nia de Lencastre}

\author[5, 6]{Samuel Kaski}

\author[1, 7, 8]{Jukka Corander}

\author[9]{Henri Pesonen}

\affil[1]{Oslo Centre for Biostatistics and Epidemiology, University of Oslo, Oslo, Norway}

\affil[2]{Laboratory of Molecular Microbiology of Human Pathogens, Instituto de Tecnologia Qu\'{i}mica e Biol\'{o}gica Ant\'{o}nio Xavier, Universidade Nova de Lisboa, Oeiras, Portugal}

\affil[3]{Laboratory of Molecular Genetics, Instituto de Tecnologia Qu\'{i}mica e Biol\'{o}gica Ant\'{o}nio Xavier, Universidade Nova de Lisboa, Oeiras, Portugal}

\affil[4]{Laboratory of Microbiology and Infectious Diseases, The Rockefeller University, New York, USA}

\affil[5]{Helsinki Institute of Information Technology, Department of Computer Science, Aalto University, Finland}

\affil[6]{Department of Computer Science, University of Manchester, UK}

\affil[7]{Helsinki Institute of Information Technology, Department of Mathematics and Statistics, University of Helsinki, Finland}

\affil[8]{Parasites and Microbes, Wellcome Sanger Institute, UK}

\affil[9]{Oslo Centre for Biostatistics and Epidemiology, Oslo University Hospital, Oslo, Norway}

\begin{document}
\maketitle

\def\spacingset#1{\renewcommand{\baselinestretch}%
{#1}\small\normalsize} \spacingset{1}

\if0\blind
{
  \title{\bf Misspecification-robust likelihood-free inference in high dimensions}
  \author{Owen Thomas \\
    Department of Biostatistics, University of Oslo, Norway \hspace{.5cm}\\
    Raquel S\'{a}-Le\~{a}o \\
    Laboratory of Molecular Microbiology of Human Pathogens, \\
    Instituto de Tecnologia Qu\'{i}mica e Biol\'{o}gica Ant\'{o}nio Xavier,\\ 
    Universidade Nova de Lisboa,\\
    Oeiras, Portugal\hspace{.5cm}\\
    Herm\'{i}nia de Lencastre \\
    Laboratory of Molecular Genetics,\\
    Instituto de Tecnologia Qu\'{i}mica e Biol\'{o}gica Ant\'{o}nio Xavier,\\
    Universidade Nova de Lisboa,\\
    Oeiras, Portugal\hspace{.5cm}\\
    Laboratory of Microbiology and Infectious Diseases,\\
    The Rockefeller University, New York, USA \hspace{.5cm}\\
    Samuel Kaski \\
    Department of Computer Science, Aalto University, Finland \hspace{.5cm}\\
    and  \hspace{.5cm}\\
    Jukka Corander\\
    Department of Biostatistics, University of Oslo, Norway \hspace{.5cm} \\
    Henri Pesonen \\
    Oslo Centre for Biostatistics and Epidemiology, Oslo University Hospital, Norway \hspace{.5cm}\\
    }
  \maketitle
} \fi

\if0\blind
{
  \bigskip
  \bigskip
  \bigskip
  \begin{center}
    {\LARGE\bf Split-BOLFI for scalable, misspecification-robust likelihood free inference}
\end{center}
  \medskip
} \fi

\clearpage

\begin{abstract}
Likelihood-free inference for simulator-based statistical models has developed rapidly from its infancy to a useful tool for practitioners. However, models with more than a handful of parameters still generally remain a challenge for the Approximate Bayesian Computation (ABC) based inference. To advance the possibilities for performing likelihood-free inference in higher dimensional parameter spaces, we introduce an extension of the popular Bayesian optimisation based approach to approximate discrepancy functions in a probabilistic manner which lends itself to an efficient exploration of the parameter space. Our approach achieves computational scalability for higher dimensional parameter spaces by using separate acquisition functions,  discrepancies, and associated summary statistics for distinct subsets of the parameters. The efficient additive acquisition structure is combined with exponentiated loss-likelihood to provide a misspecification-robust characterisation of posterior distributions for subsets of model parameters. The method successfully performs computationally efficient inference in a moderately sized parameter space and compares favourably to existing modularised ABC methods. We further illustrate the potential of this approach by fitting a bacterial transmission dynamics model to a real data set, which provides biologically coherent results on strain competition in a 30-dimensional parameter space.
\end{abstract}

\noindent%
{\it Keywords:}  Loss Likelihoods, Approximate Bayesian Computation, Bacterial Transmission Dynamics, High-Dimensional Inference
\vfill

\newpage
\spacingset{1.45} 
\section{Introduction}
\label{sec:intro}

Likelihood free inference (LFI) framework for simulator-based models with an intractable likelihood has witnessed a remarkable development over the past couple of decades and a growing interest from a number of diverse application fields \citep{secrier2009abc,itan2009origins,drovandi2011estimation,corander2017frequency,cameron2012approximate,jarvenpaa2019bayesian,shen2019pneumococcal,simola2019machine}. Methodologically, the field has its roots in Approximate Bayesian Computation (ABC), followed by and subsequent work synthesising ABC with established Monte Carlo inference algorithms \citep{beaumont2009adaptive,del2012adaptive}. For general overviews of the computational and statistical aspects of ABC, see \citep{marin2012approximate,lintusaari2017fundamentals,sisson2018handbook}. 

Here we build a new method for likelihood-free inference (LFI) motivated by the concept of generalised Bayesian updates, defined by a loss function derived from the subjective judgement of the statistician regarding the properties of the statistical model and true data generating process. Such a method is robust against model misspecification, in the sense that the updates are naturally tempered when the observed data and best-case simulated data do not resemble each other. 
Such an approach naturally scales to a higher-dimensional context by asserting an additive structure in the loss function, bringing about gains in computational efficiency for both simulation acquisition and belief distribution characterisation. We build upon the Bayesian optimization approach to accelerate LFI \citep{gutmann2016bayesian} to perform efficient acquisitions in high-dimensional parameter spaces. We call this method Split-BOLFI.

The remainder of the paper is structured as follows. Section \ref{sec:meth} explains the background of the methods developed in this paper, while Section \ref{sec:splitBOLFI} introduces the new algorithm. Section \ref{sec:verify} presents the performance of the method on both synthetic and real data and Section \ref{sec:conc} concludes the study with a discussion and proposals of future work. 

\section{Methodology}
\label{sec:meth}

In this section we present some related methods relevant to this article, the methodology of Split-BOLFI, described through its relevance to misspecification, scalable inference and Bayesian Optimisation, in addition to some discussion of the expected computational costs of the methods considered.

\subsection{Scalable likelihood-free methods}

Scaling LFI to higher dimensional parameter spaces remains a challenging and open question. Methods developed for low-dimensional models cannot be assumed to scale effectively to higher-dimensional parameters, in part due to the challenge defining of defining informative summary statistics and exploring large parameters spaces, suggesting that specialized methods are more appropriate for such problems. The state of the field has been reviewed in \citep{chapter8sisson2018handbook}. One such method is Gaussian Copula ABC, which defines distinct discrepancy functions for each parameter value, using a copula approach to characterize pairwise joint distributions of the variables and performing rejection sampling from the prior \citep{li2017extending}.

A separate but related scalability question is that of efficient use of draws from the simulator model. A key insight to developing accelerated LFI was demonstrated with Bayesian Optimization for Likelihood-Free Inference (BOLFI) by replacing standard random draws from an importance distribution by efficient acquisitions delivered by probabilistic Bayesian optimization \citep{gutmann2016bayesian}. 

\subsection{Model misspecification}
\label{sec:misspec}

Misspecification remains an important problem for successful Bayesian inference: a likelihood-based belief update according to Bayes' theorem implicitly selects parameter values from the prior that are closest in KL divergence to the process generating the observed data, which may exhibit undesirable behaviour as an inference strategy \citep{walker2013bayesian}.

A key contribution from previous research concerned with misspecification is a method \citep{bissiri2016general} for belief updates based on Bayesian principles, without assuming a likelihood function, but only defining a loss function $l(\bm{\theta},X)$ associated with a parameter vector $\bm{\theta}$ and the observed data $X$. Such reasoning concludes that, in the absence of a true likelihood, a prior distribution $\pi(\bm{\theta})$ can be updated to a data-conditional belief distribution $\tilde{\pi}(\bm{\theta}|X)$ via an exponentiation of the negative loss divided by a tempering constant $\delta > 1$:
\begin{equation}\label{eqn:gbupdate}
 \tilde{\pi}(\bm{\theta}|X) \propto \exp(-l(\bm{\theta},X) / \delta)\pi(\bm{\theta}).
\end{equation}

An exponentiated loss update distribution becomes a traditional Bayes update when the loss function $l(\bm{\theta},X)$ is chosen to be the negative log-likelihood and the tempering constant $\delta=1$. Such an update maintains the principle of coherency, i.e.~the update can perform with the data in any order through the implicit use of the KL divergence to impose the influence of the prior on the posterior \citep{bissiri2016general}.

Of additional concern for this work is the relationship between misspecification and modularity: it is relatively common for different data features described by one model to exist at differing degrees of misspecification, and as such it is advantageous to pursue modularised inference procedures that prevent one particularly poorly-specified component of the model from sabotaging the entire inference procedure \citep{plummer2015cuts,jacob2017better,carmona2020semi,chakraborty2023modularized}. Examples of situations in which models may exhibit heterogeneous levels of misspecification between modules are inverse probability weighting for non-randomised data \citep{hogan2004instrumental}, regression models with distinctly specified processes for the mean and noise \citep{korendijk2008influence}, and the specification of random and fixed effects in a mixed effects model \citep{bayarri2009modularization}.


Model misspecification is a concern when no models contained in the prior distribution are capable of generating data matching the properties of the observed data. It should be assumed that this is possible whenever a statistician is working with real data, given that they do not know the true process from which the observed data is generated. The loss between the simulated data and observed data used by a traditional update using Bayes' theorem is the KL divergence, implicit through the use of the likelihood function. The field of LFI has developed methods for going beyond the likelihood function, initially developed for simulator models without tractable likelihood functions, but now becoming useful for avoiding the assumptions implicit in the use of Bayes' theorem with an explicit likelihood function.

The choice of summary statistics $\bm{\phi}(X)$ from which to form a discrepancy has previously been regarded as an attempt to capture the features of the data necessary to best approximate the structure of the unknown likelihood function. Instead, we suggest a  definition of summary statistics and discrepancies which reflect the subjective judgement of the statistician, conditional on their modelling priorities and suspicions of misspecification. A related approach has been explored recently in Focused Bayesian Prediction (FBP) \citep{loaiza2021focused}, in which a loss function is constructed using the predictive target of interest to the statistician.

In contrast to FBP, we propose an inference-motivated approach, in which the end target of the algorithm is to provide belief distributions, conditional on summaries that the statistician considers relevant to the statistical question. Such an approach has been explored with a general choice of divergences in \citep{jewson2018principles,knoblauch2019generalized,lewis2021bayesian}, in which the statistician can choose a divergence that gives less influence than the KL to the potentially misspecified tails of the distribution. Here we extend the reasoning to the usage of summaries representing more specific features of the data that the statistician can choose to include or exclude from consideration in the belief update.

We define a discrepancy $d(\bm{\theta})$ as a function of the parameter vector $\bm{\theta}$, as an Euclidean distance between simulated summary statistic vector $\bm{\phi}(X_{\bm{\theta}})$ and observed summary statistics $\bm{\phi}(X)$:
\begin{equation}\label{eqn:discdef}
d(\bm{\theta})=||\bm{\phi}(X_{\bm{\theta}})-\bm{\phi}(X)||_2.
\end{equation}

The use of discrepancies within a likelihood-free context is reminiscent of a loss function $l(\bm{\theta},X)$, a concept central to the derivation of Bayes' theorem from decision-theoretic principles \citep{bernardo2009bayesian}. In order to define a belief update, it is necessary to define a scalar factor $\delta$ defining both the level of tempering and the  difference in scale between the loss $l(\bm{\theta})$ and discrepancy $d(\bm{\theta})$. 
Analogously to the original BOLFI method, which used the discrepancy minimum to define a KDE bandwidth, we suggest using the minimum value $d_{min}$ of the discrepancy from the simulator to set this parameter, resulting in the following belief update:
\begin{equation}\label{eqn:dminupdate}
 \tilde{\pi}(\bm{\theta}|X) \propto  \exp (-d(\bm{\theta})/\delta) p(\bm{\theta}) = \exp (-d(\bm{\theta})/d_{min}) p(\bm{\theta})
\end{equation}

The use of the minimum discrepancy for the tempering constant $\delta$ is desirable because it helps to avoid a confident belief update when the statistical model is misspecified under a given loss: in this sense Split-BOLFI is misspecification-robust. The minimum observed discrepancy captures the natural lengthscale of the discrepancy function relative to the degree of misspecification of the model under the asserted loss. If the variation in the discrepancy is small relative to the minimum value, this is a sign that the model is poorly specified and we should perform a tempered update in order to avoid an overly confident belief distribution. Alternatively, if the minimum value of the discrepancy is small relative to its variation, this suggests that the model is well-specified under the given loss and can encourage a confident belief update.

As an example, consider performing a belief update for a univariate Gaussian model with unknown mean but variance assumed equal to one. The sufficient statistic for the statistical model is simply the mean of the observed data: with this we can perform a traditional Bayesian update, but the use of additional summaries in a loss function allows for alternative belief updates. In the situation where the observed data is from a Gaussian distribution with a standard deviation not equal to one, the observed variance will be notably different between the observed data and simulated data drawn from the statistical model, even when the posterior over the mean is concentrated and apparently converged.

\begin{figure}
    \centering
    \begin{subfigure}[b]{0.45\textwidth}
        \includegraphics[width=\textwidth]{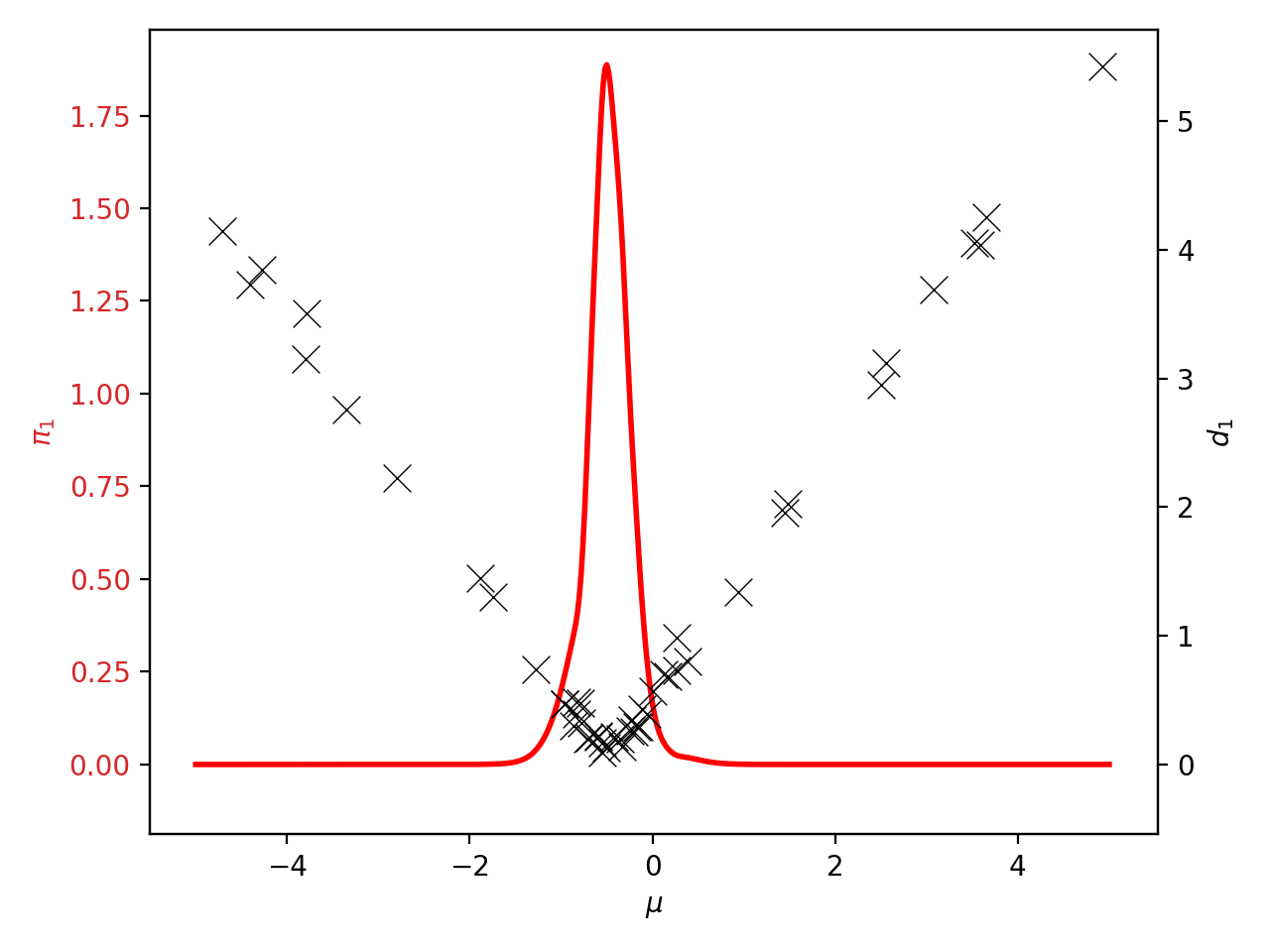}
        \caption{$\sigma=1, \sigma_{g}=1, \mu_{g}=0$}
        \label{fig:well_specified_mean_gauss}
    \end{subfigure}
    ~ 
    \begin{subfigure}[b]{0.45\textwidth}
        \includegraphics[width=\textwidth]{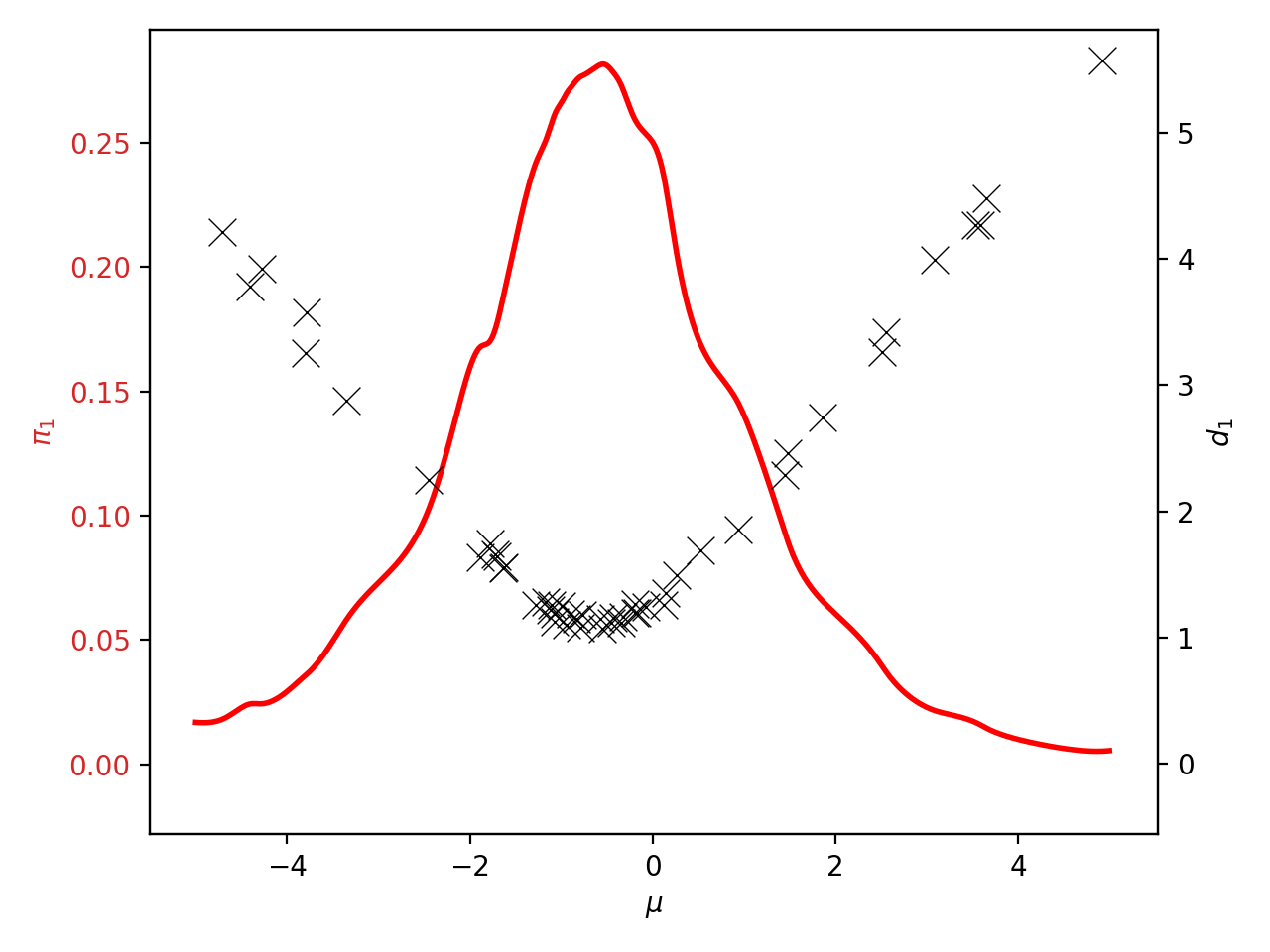}
        \caption{$\sigma=1, \sigma_{g}=2, \mu_{g}=0$}
        \label{fig:misspecified_mean_gauss}
    \end{subfigure}
    
    \caption{Acquired discrepancies and proxy likelihood for inference for the mean $\mu$ of a univariate Gaussian with an assumed standard deviation $\sigma=1$. The generative model has mean $\mu_{g}=0$ and a standard deviation $\sigma=1$ in the well-specified case and $\sigma=2$ in the misspecified case. The discrepancy was defined using the sample mean and standard deviation. The misspecified model demonstrates substantial posterior tempering compared to the well-specified model, indicating that the inference method has identified the mode misspecification and reduced the confidence of the update.}\label{fig:d_misspecification}
\end{figure}

A statistician assessing a statistical model consisting of a Gaussian distribution with an assumed fixed variance would reasonably suspect that it is not necessarily reflective of the real world. As such, the question of whether to include the variance as a summary statistic is now left to the statistician's judgement. They may prefer a belief update with a dispersion that reflects the degree of confidence in the model, in which case including the variance as a summary will help prevent the convergence of the belief distribution to a single model parameter value when none of the values correspond to well-specified models. This phenomenon is demonstrated in Figure \ref{fig:d_misspecification}, in which the update under misspecification exhibits more tempering relative to the well-specified situation, given the information present in the sample variance. Given that the Gaussian model with an assumed variance value has no asymptotic ability to predict different variance values, our approach distinguishes itself from the goals of FBP \citep{loaiza2021focused}.

\subsection{Scalable Modularity}\label{sec:highdimmeth}

Loss-based inference methods naturally extend to higher-dimensional parameter spaces through the imposition of additive structure on the loss function. Such structure associates distinct losses with different subsets of parameters, recasting the high-dimensional loss function as a combination of low-dimensional functions defined over different regions of parameter space.

Previous work on high-dimensional LFI has used the association of specific summary statistics with specific parameters in order to perform effective inference \citep{li2017extending}. We consider a mapping from parameters to a collection of summary statistics. In the simplest case, each parameter maps to disjoint groups of summaries, and the likelihood proxy factorises as a product across all dimensions.

We further pursue the modularised inference framework described in Section \ref{sec:misspec}, for two main reasons. Firstly, it is possible that misspecification may be present heterogeneously within the simulator model, and as such we would like to modularise the inference procedure to allow components at differing levels of misspecification to see the relevant data separately. Secondly, the modularisation of inference enables the algorithmic scalability of the inference procedure to higher-dimensional spaces, by effectively imposing structure to reduce the joint problem to a tractable collection of lower-dimensional problems.

Formally, we can divide the $p$-dimensional parameter vector $\bm{\theta}$ into $q\leq p$ disjoint subsets $[\theta]_j$ with each subset sharing a corresponding subset $[\phi(X)]_j$ of the summary statistic vector $\bm{\phi}(X)$. Each subset is then assigned its own discrepancy $d_j([\theta]_j)$, which contributes to the full discrepancy function as an additive combination, weighted according to scalars $\delta_j$
\begin{equation}\label{eqn:additiveloss}
  d(\bm{\theta}) \propto  \sum_{j=1}^{q} d_j([\theta]_j)/\delta_j
\end{equation}

\begin{figure}
    \centering
    \begin{subfigure}[b]{0.48\textwidth}
        \includegraphics[width=\textwidth]{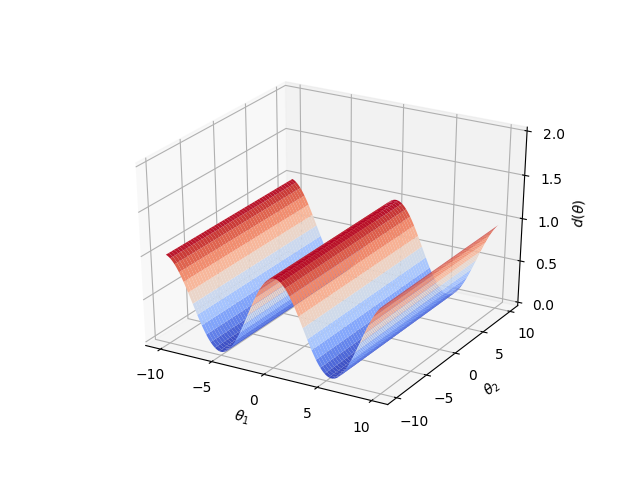}
        \caption{$d(\bm{\theta})=d_1([\theta]_1)=\sin^{2}(0.3\bm{\theta}_1 + 5)$}
        \label{fig:d_1}
    \end{subfigure}
    \begin{subfigure}[b]{0.48\textwidth}
        \includegraphics[width=\textwidth]{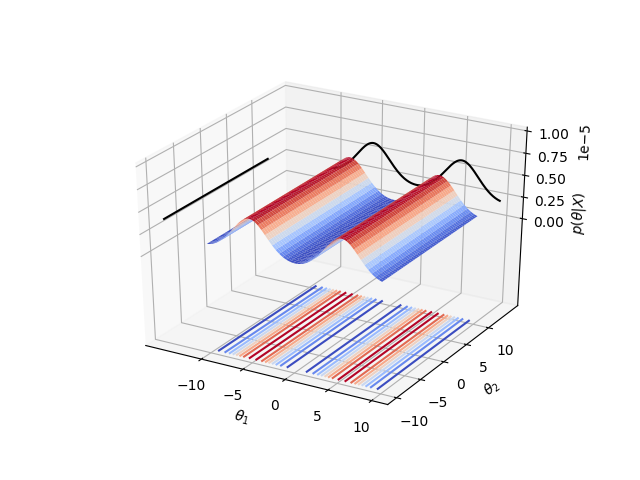}
        \caption{$p(\bm{\theta}|X)\propto \exp(-d_1([\theta]_1))$}
        \label{fig:L_1}
    \end{subfigure}
    \begin{subfigure}[b]{0.48\textwidth}
        \includegraphics[width=\textwidth]{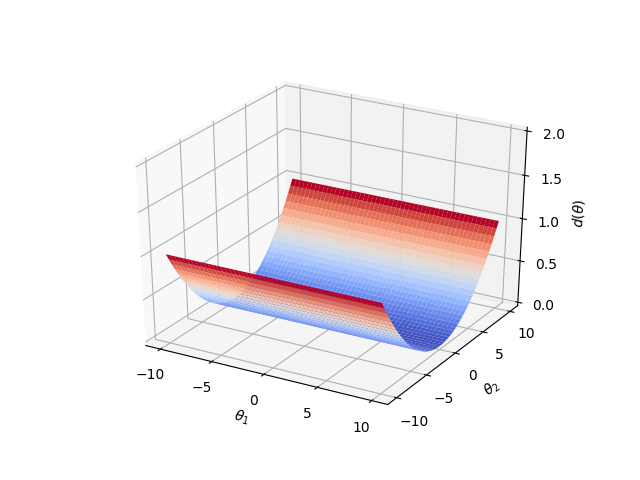}
        \caption{$d(\bm{\theta})=d_2([\theta]_2)=0.01\bm{\theta}_2^2$}
        \label{fig:d_2}
    \end{subfigure}
    \begin{subfigure}[b]{0.48\textwidth}
        \includegraphics[width=\textwidth]{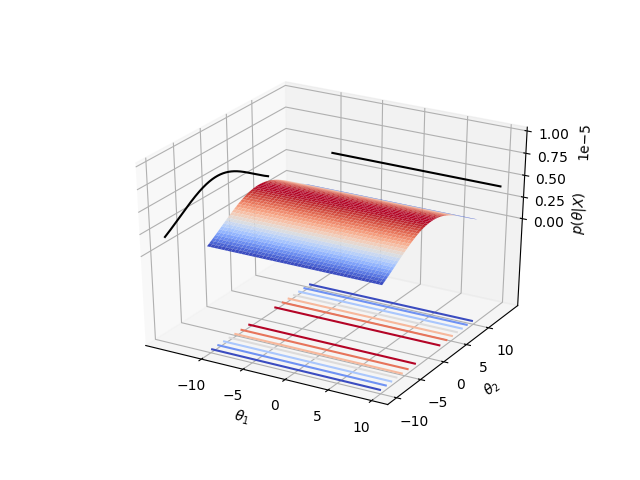}
        \caption{$p(\bm{\theta}|X)\propto \exp(-d_2([\theta]_2))$}
        \label{fig:L_2}
    \end{subfigure}
    \begin{subfigure}[b]{0.48\textwidth}
        \includegraphics[width=\textwidth]{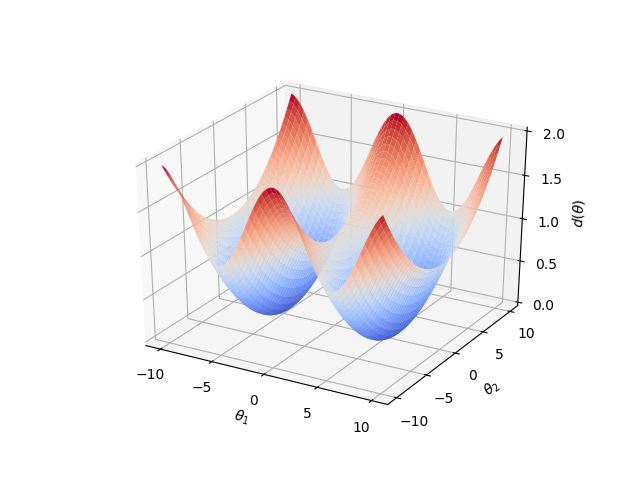}
        \caption{$d(\bm{\theta})=d_1([\theta]_1)+d_2([\theta]_2)$}
        \label{fig:d_12}
    \end{subfigure}
    \begin{subfigure}[b]{0.48\textwidth}
        \includegraphics[width=\textwidth]{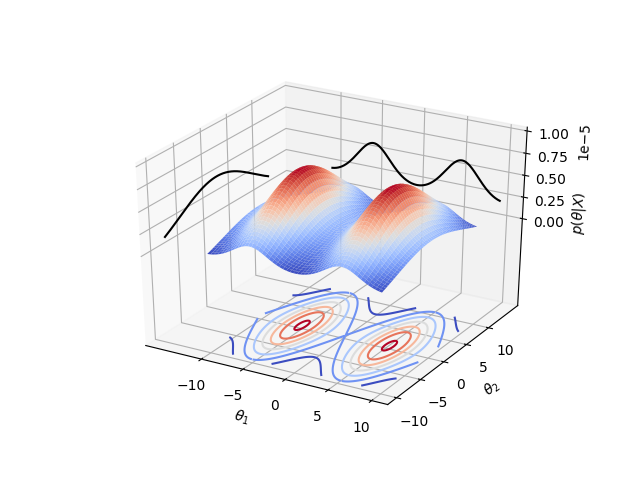}
        \caption{$p(\bm{\theta}|X)\propto\exp(-d_1([\theta]_1))\exp(-d_2([\theta]_2))$}
        \label{fig:L_12}
    \end{subfigure}
    \caption{Example additive discrepancies in two dimensions bringing about factorisable likelihood through the use of an exponential mechanism.}\label{fig:d_additive}
\end{figure}

The exponential loss likelihood mechanism has the useful property that any additive structure present in the loss becomes interpretable as multiplicative structure present in the likelihood proxy:
\begin{align}
 \tilde{\pi}(\bm{\theta}|X) &\propto \exp \left(- \sum_{j=1}^q d_j([\theta]_j)/\delta_j\right) p(\bm{\theta}) \nonumber\\
 &= \prod_{j=1}^q \exp \left(-d_j([\theta]_j)/\delta_j \right) p([\theta]_j) \nonumber\\
 &\propto  \prod_{j=1}^q \tilde{\pi}_{j}(X|[\theta]_j) p([\theta]_j),
\end{align}
using the definition of the generalised Bayesian update presented in Eq.~\eqref{eqn:gbupdate}, and assuming the additive structure in Eq.~\eqref{eqn:additiveloss}, and independent prior distributions $p([\theta]_j)$ across subsets. An example of additively combined discrepancies in two dimensions bringing out posterior structure is presented in Figure \ref{fig:d_additive}.

It is possible that the misspecification of the statistical model may vary with each parameter subsets and hence components in the sum of loss functions. As such, different parameter subsets $[\theta]_j$ may warrant different degrees of tempering, determined by setting the separate scalars $\delta_j$, depending on the goodness of specification of their associated summaries. Such reasoning has been pursued previously when combining distinct models at differing levels of misspecification \citep{jacob2017better}. Here we extend the concept to modularise subsets of parameters for a single statistical model. As such, we advocate selecting different parameters $\delta_j$ defining the degree of update for each subset, depending on the minimum value of the observed discrepancies, analogously to Section \ref{sec:misspec}. 
\section{Split-BOLFI}\label{sec:splitBOLFI}

\subsection{Gaussian Processes for Split-BOLFI}
\label{sec:implementGP}

A Gaussian Process (GP) model is a Bayesian nonparametric prior used for functions, providing flexible nonlinear regression models with analytically accessible predictive distributions. The mean function is taken to be zero without loss of generality: each GP $f_j([\theta]_j)$ has a different covariance function $K_j([\theta]_j,[\theta]'_j)$ defined over the relevant subset of parameters:
\begin{equation}
 f_j([\theta]_j)\sim\mathcal{GP}\left(0,K_j([\theta]_j,[\theta]'_j)\right)
\end{equation}

For a simulated data set $X_{\bm{\theta}}$ drawn using parameters $\bm{\theta}$, the summaries considered relevant to the inference problem are evaluated and separated according to which loss function they contribute. Each loss function is then evaluated by comparison of the simulated summary statistics and the observed summaries. Each value of the loss is then treated as an observation of the response variable, with the value of the parameters associated with the loss component used as corresponding covariates.

Each GP model trained for each discrepancy component has an analytically accessible Gaussian-distributed predictive distribution at new parameter location $[\theta^*]_j$, with mean and standard deviation $\mu_j^*([\theta^*]_j)$ and $\sigma_j^*([\theta^*]_j)$, calculated from the fitted GPs $f_j$, observed discrepancies $d^{obs}_j$, and acquired parameter values $[\theta_{acq}]_j$:

\begin{align}
p(f_j| [\theta^*]_j,& K_j, d^{obs}_j, [\theta_{acq}]_j) \sim \mathcal{N}(\mu_j^*([\theta^*]_j), \sigma_j^*([\theta^*]_j))\\
    \mu_j^*([\theta^*]_j) &= k_j([\theta^*]_j,[\theta_{acq}]_j) K_j([\theta_{acq}]_j,[\theta_{acq}]'_j)^{-1} d^{obs}_j\\
\sigma_j^*([\theta^*]_j)^2 &= k_j([\theta^*]_j,[\theta^*]_j) - k_j([\theta^*]_j,[\theta_{acq}]_j) K_j([\theta_{acq}]_j,[[\theta_{acq}]_j)^{-1}   k_j([\theta_{acq}]_j,[\theta^*]_j)  
\end{align}

The predictive mean is then used as a smooth model for the loss function for the purposes of evaluating the likelihood proxy $\tilde{\pi}_{j}(X|[\theta]_j)$. The predictive mean and variance of each loss component is used to define an acquisition function over each disjoint subset of parameters, recasting the global acquisition problem of traditional Bayesian Optimisation into multiple acquisitions over small subsets of parameters.

\subsection{Bayesian Optimization}\label{sec:bayesoptmeth}

The questions remain as to how to represent the loss functions computationally and where to evaluate simulators in parameter space. The first is addressed through the use of a GP regression model for each component in the additive loss structure: the second through the use of Bayesian Optimization to draw simulator samples informative towards the minima of the losses, in the style of BOLFI.

Each GP model trained for each discrepancy component has an analytically accessible predictive mean and standard deviation $\mu_j^*([\theta^*]_j)$ and $\sigma_j^*([\theta^*]_j)$, calculated from the model components and observed data . Through the use of the Lower Confidence Bound (LCB) acquisition function $a_{LCB}()$ \citep{srinivas2010gaussian} with an exploration-exploitation tradeoff parameter $\beta$, this gives rise to the following acquisition function minimum for the $j$th subset:
\begin{equation}
    [\theta_{acq}]_j = \underset{[\theta^*]_j}{\arg \min}~ a_{LCB}([\theta^*]_j)= \underset{[\theta^*]_j}{\arg \min}~ (\mu_j^*([\theta^*]_j)- \beta \sigma_j^*([\theta^*]_j) ).
\end{equation}

Other acquisition functions can be used in addition to the LCB, including Expected Improvement (EI) \citep{jones1998efficient}, Probability of Improvement (PI) \citep{kushner1964new}, and those specifically designed for likelihood-free inference \citep{jarvenpaa2019efficient}. LCB comes with convergence guarantees under some conditions \citep{srinivas2010gaussian}. The parameter $\beta$ in the LCB determines the compromise between exploration and exploitation in the optimization routine.Other acquisitions functions, including EI and PI, use similar hyperparameters for negotiating the tradeoff of exploration with exploitation \citep{gan2021acquisition}. For researchers looking to derive an appropriate $\beta$ parameter value for their application can follow the principle that larger parameter values emphasizing exploration do not take advantage of the current surrogate fit but evaluate more evenly in the whole optimization space, and smaller parameter values emphasizing exploitation focus the acquisition locations where surrogate fit is already optimal \citep{brochu2010tutorial}.

In the analyses in this article, we performed preliminary inference runs with varying values of $\beta$, intended to determine the appropriate value of $\beta$ for the full inference procedure. We initially tried values of several negative powers of 10, i.e. [0.001, 0.01, 0.1, 1.], and examined the posterior distributions produced to determine whether the results of inference were satisfactory. These preliminary runs were performed with simulated data drawn with known generative parameter values, allowing comparison with the underlying true values. Overly small values of $\beta$ (0.01 or smaller) could result in either convergence to a spurious local posterior mode, or overly aggressive concentration to the true posterior mode, with inadequate characterization of the rest of the posterior. Conversely, overly large values of $\beta$ resulted in a much more even selection of acquisitions within the parameter bounds, at the potential expense of thoroughly investigating, or sometimes even identifying, the true posterior mode. Given the low dimensionality of the individual subsets used in Split-BOLFI, such phenomena could be identified through diagrams such as those presented in Figure \ref{fig:daycare_30_competitive}, although quantitative metrics could also be used to evaluate the success of inference. Based on our exploratory analyses of the problems considered here we have fixed parameter $\beta=0.1$. This choice may not necessarily generalize to other models and data sets, and a wider exploration is recommended for each new application.

For users of Split-BOLFI looking to apply the methodology to new problems, we advocate a similar approach of pursuing preliminary runs with varying values of $\beta$ to establish the resulting posterior characteristics. The value of $\beta=0.1$ that we used in this article is by no means a universal solution for every simulator, but may represent a reasonable place to start an exploration of $\beta$ for preliminary analysis. The optimal value of $\beta$ will depend on the posterior geometry (conditional on choice of summary statistics), computational budget, and inferential goals. The final decision of selecting $\beta$ rests with the practising statistician and their priorities: for a given acquisition budget, smaller values of $\beta$ will result in a more thorough characterization of the posterior mode compared to the rest of the distribution (with the risk of getting caught in a local mode), whereas larger values of $\beta$ will result in a more thorough characterization of the entire posterior compared to the mode (with the risk of missing the global posterior mode entirely). The former may be preferred when a posterior mode with some degree of uncertainty is desired, whereas the latter may be preferred when full uncertainty propagation from the parameter space is the goal.

When determining the value of $\delta_j$ in practice, we use
\begin{equation}
\delta_j=\max(\underset{[\theta^*]_j}{\min}(\mu_j^*([\theta^*]_j)),\min(d^{obs}_j)),
\end{equation}
i.e. whichever is larger of the minimum GP predictive mean value $\mu_j^*([\theta^*]_j)$ and the minimum of the observed discrepancy values $d^{obs}_j$ from the simulator. Because the GP function is defined over all real numbers, the predictive mean can sometimes dip below zero, making $\underset{[\theta^*]_j}{\min}(\mu_j^*([\theta^*]_j))$ hard to interpret, so in these situations, we use the minimum observed discrepancy value $\min(d^{obs}_j)$, as this is the smallest empirically validated quantity that will lead to numerically stable solutions. It is also possible to use a non-linear transformation (e.g. logarithmic) to avoid this problem: as explored in \citep{gutmann2016bayesian} it can help in some instances of numerical instability and misspecification of the discrepancy model, although it was less influential than the introduction of noise into the acquisition process.

We consequently use the GP predictive means $\mu_j^*([\theta]_j)$ to return a posterior proxy:
\begin{align}
 \tilde{\pi}(\bm{\theta}|X)  & \propto  \prod_{j=1}^q \tilde{\pi}_{j}(X|[\theta]_j) p([\theta]_j),
\nonumber \\
 &\propto \prod_{j=1}^q \exp \left(-\mu_j^*([\theta]_j)/\delta_j \right) p([\theta]_j).
\end{align}
The inference procedure of Split-BOLFI as described in the preceding two subsections is summarised as pseudocode in Algorithm \ref{alg:Split-BOLFI}.

\begin{algorithm}
\SetAlgoLined
\KwResult{Likelihood proxies $\tilde{\pi}_{j}(X|[\theta]_j)$}
 $n_{obs}$ observations $X$, prior $p(\bm{\theta})$, simulator $p(X|\bm{\theta})$, $n_{init}$ initial parameters $\bm{\theta}_{init}$\;

\For{$i = 1\ldots n_{init}$}{
 Simulate a dataset $X_{\bm{\theta}}^{i}$ of size $n_{obs}$ from ${X_{\bm{\theta}}^{i}} \sim p(X|\bm{\theta}_{init}^{i})$  \;
 Evaluate $d_j^{i} =||[\phi (X) ]_j- [\phi ({X_{\bm{\theta}}^{i}})]_j ||_2$ \;
}
 \For{$j=1\ldots q$}{
  Initialise $D_{j} \sim \mathcal{GP}\left(\mu, K\right)$  \;
 Update GP model $D_j$ with new responses $d_j^{1:{init}}$ and covariates $[\bm{\theta}_{init}]_j$ \;}
  
 \For{$i=n_{init}+1 \ldots n_{acq}$}{
 Set $\bm{\theta}_{acq}^{i} = []$\;
  \For{$j = 1\ldots q$}{
   Minimise $ [\theta_{acq}^{i}]_j = \arg \min (\mu_j^*([\theta^*]_j)- \beta \sigma_j^*([\theta^*]_j) )$ \;
   Append $[\theta_{acq}^{i}]_j$ to $\bm{\theta}_{acq}^{i}$\;}
    Simulate a dataset $X_{\bm{\theta}}^{i}$ of size $n_{obs}$ from ${X_{\bm{\theta}}^{i}} \sim p(X|\bm{\theta}_{acq}^{i})$  \;
  \For{$j = 1 \ldots q$}{
 Evaluate $d_j^{obs, i} =||[\phi (X) ]_j- [\phi ({X_{\bm{\theta}}^{i}})]_j ||_2$ \;
   Update GP model $D_j$ with new responses $d_j^{obs, i}$ and covariates $[\theta_{acq}^{i}]_j$ \;}
}
 \For{$j = 1 \ldots q$}{
 Set $\delta_j=\max(\underset{[\theta^*]_j}{\min}(\mu_j^*([\theta^*]_j)),\min(d^\text{obs}_j))$  \;
 Set $\tilde{\pi}_{j}(X|[\theta]_j) = \exp \left(-\mu_j^*([\theta]_j)/\delta_j \right)$ \;}
 \caption{The Split-BOLFI algorithm.}\label{alg:Split-BOLFI}
\end{algorithm}

\subsection{Computational Costs}

Here we present some considerations of the computational costs of the methods used in this article, focusing on the numerical operations performed in addition to evaluations of the simulator models themselves.

For the standard BOLFI algorithm, the most expensive necessary matrix operations of Cholesky decomposition are used for evaluation of the marginal likelihood for hyperparameter optimisation and prediction for acquisition of new simulations. Cholesky decomposition is expected to scale cubically with the size of the covariance matrix and hence the number of data points, i.e. $\mathcal{O}(n^{3})$. It is more difficult to generalise about scalability with respect to the dimensionality: the likely dominant computations will be kernel construction, hyperparameter optimisation, and simulation parameter acquisition. The first would be expected to be linear for most commonly used kernels, while the latter may be harder to characterise precisely, but can be represented as a function $f(p)$, given a total scalability of $\mathcal{O}(f(p)n^{3})$.

Split-BOLFI would be expected to have the computational costs of BOLFI performed on each of the $q$ subset modules, so naively could be characterised as being at worst $\mathcal{O}(q f(p/q) n^{3}) \approx \mathcal{O}(q f(n_q) n^{3})$, where $n_q$ is the averaged number of dimensions per subset module. Consequently, if $f(n_q)$ is superlinear, then we would expect some speedup relative to a standard BOLFI algorithm performing joint inference over the complete parameter space, at least for the simulator acquisition stage of inference. However, given the independent, modular nature of the subset inference, the $q$ computational tasks can in principle be distributed across multiple cores, reducing computational times in principle to $\mathcal{O}(f(n_q) n^{3})$, plus the overhead time necessary for communication between machines.

\subsection{Summary Statistics and Partitioning}

The construction and selection of summary statistics is important for likelihood free inference. Domain-specific knowledge is often useful, as an expert who is familiar with the data or simulator often has informative intuitions concerning how the data can be condemned into its most salient lower-dimensional representations.

Pilot runs of simulations are also key to exploring the sensitivity of different choices of summary statistics and their relevance to simulator parameter values. This can be performed by investigating the sensitivity of discrepancy functions around “true” simulated parameter values, or by examining correlation structure between parameter values and simulated summaries.

Various methods also exist for the automatic construction of sets of summary statistics, either by selecting a subset of maximally informative summaries from a large initial pool of suggestions, or by algorithmically developing new statistics under various principles of informativeness. 

The first article to investigate automatically developing summary statistics for inference effectively systematized the process of investigating the behavior of various summaries under ABC pilot runs \citep{joyce2008approximately}. This was followed by articles in a similar spirit introducing stricter information-theoretic principles to filtering a large initial pool of summaries \citep{nunes2010optimal, barnes2012considerate}.

Another influential early publication \citep{fearnhead2012constructing} proposes the posterior mean as an ideal summary statistic, which can be targeted by a regression model predicting simulator parameter values given proposed summary statistics, or other transformation of the data. This work inspired much further work, generalizing the approach to model choice \citep{prangle2014semi}, expanding the regression models such to various neural net architectures \citep{jiang2017learning,wiqvist2019partially,aakesson2021convolutional,creel2017neural}, or using other higher-order posterior moments as summaries \citep{forbes2022summary}.

More recent work has looked at deriving useful summary statistics from simulations outside of an ABC framework, through various informatic principles, through the use of graph neural nets \citep{gaskell2023inferring}, variational autoencoders \citep{albert2022learning}, mutual information and neural nets \citep{chen2020neural}, and low-dimensional projections with locality constraints \citep{siren2020local}.

A further class of methods integrates learning summary statistics within the inference procedure. This has been performed within an ABC framework, selecting summaries adaptively with a nearest-neighbor distance measure between prior and posterior \citep{harrison2020automatic}, or by introducing further terms into the training loss that effectively act as a regularizer in the space of the summary statistics, of special use in cases of potential model misspecification \citep{huang2024learning}.

The partitioning of summary statistics into distinct modular groups remains an unresolved research question, explored previously in the context of likelihood-free inference \citep{li2017extending, chakraborty2023modularized, drovandi2024improving}. Like deriving summary statistics, it can be performed in an exploratory way, using a combination of automatic and manual methods. Understanding of the specific simulator itself will provide intuition and insight into appropriate partitions, especially if it is formed by combining modular components \citep{jacob2017better}. Automatic methods for generating partitions can be used through data generated during pilot runs of the simulator: the relationships between specific pairs of parameters and summary statistics can be explored by examining the behavior of proposed discrepancies under deviation from the true generative value of the parameter, or alternatively regularized network analyses such as the graphical lasso \citep{friedman2008sparse} may provide appropriately sparse solutions to the global question of associating pairs of parameters and summaries.

 We encourage the use of any of the above methods, or in some combination, to help derive sets of summary statistics appropriate for inference. Establishing a successful Bayesian workflow generally relies upon an iterative process of predictive checks, computation validation, and model evaluation, among other steps \citep{gelman2020bayesian}, with ``black box'' inference solutions often falling outside of this paradigm. \citep{brynjarsdottir2014learning} recommend further: ``We believe that formulating prior information about model discrepancy is the key to realistic calibration. In most cases, modelers have some idea about what physics or processes the simulator is missing, and we need to use the best judgments of the modelers and model users about how the simulator's deficiencies will translate into model discrepancy.'' For the examples in this article, we have used summaries derived from our understanding of the simulators, but the inference methodology is agnostic to the method used to construct them. 

\section{Experiments}
\label{sec:verify}

In this section, we present the results of the Split-BOLFI method on several simulator inference problems, both with and without model misspecification, including comparisons with the standard BOLFI method and a modularised version of Rejection ABC which is adapted from \citep{li2017extending} and described in Appendix \ref{sec:modularisedrejectionABC}. Modularised rejection ABC can be expected to scale favourably with both $n$ and $p$ at the expense of making less efficient use of simulations, and as such the cost of drawing samples from the simulator is likely to dominate. The primary algorithmic concerns are sorting the discrepancy sample to establish the quantiles, and the modular structure lends itself to distributed computation in a similar fashion to Split-BOLFI if desired. 

The implementation of Split-BOLFI used in the experiments is available in ELFI (Engine for Likelihood-Free Inference), an open-source statistical software package written in Python for likelihood-free inference \citep{lintusaari2018elfi}.

The success of the inference methods was evaluated by considering samples drawn from the posterior distributions generated by Split-BOLFI and BOLFI, and the samples returned by the rejection ABC algorithm. We considered the RMSEs (Root Mean Square Errors, averaged across random seeds i.e.~the RMSE of each posterior sample from the corresponding true generative value), AMEs (Absolute Mean Error, averaged across random seeds i.e. the absolute deviation of the sample mean from the true generative value) of the samples evaluated against the true generative values, standard deviations (SDs, averaged across posterior runs, i.e. the standard deviation of the posterior samples, representing the breadth of the posterior distribution, not considering the true generative value) of the returned samples, and 50\% posterior coverage of the true generative value compared with the first and third posterior quartiles, all averaged across the 50 random seeds used, and across the independent subsets across dimensions. Definitions for each of these are provided in \Cref{tab:metric_definition}. Further, for the Split-BOLFI method we recorded the AMAPEs (Absolute Maximum a Posterior Error, averaged across random seeds i.e. the absolute deviation of the MAP estimate from the true generative value), indicating the accuracy of the maximum a posteriori estimate of the updated belief distribution, which should be independent of the tempering level. We do not compare directly with the exact posterior of an untempered likelihood-based Bayesian update: in the presence of misspecification, then the standard Bayesian update is no longer the ideal, and we do not consider recreating it exactly to be the ultimate inferential target.

We begin with an example of overdispersion using Gaussian and Laplacian statistical and generative models and we demonstrate Split-BOLFI with a model designed for capturing competition dynamics between different species/strains of bacteria in a daycare transmission setting, featuring low- and high- dimensional versions of the model parametrisation, and using simulated and real data sets. Split-BOLFI presents interpretable and biologically sound insights into the real data set used. In the Appendix we provide an example analogous to poorly-specified causal inference, in which inference is performed on the Cholesky decomposition of a precision matrix.


\subsection{Overdispersed Gaussian}

We start with an example for estimating the mean and standard deviation vectors of a multidimensional Gaussian distribution. The statistical model is a $p$-dimensional multivariate Gaussian with unknown means and an unknown diagonal covariance matrix. The diagonal covariance lends itself to treating each of the pairs of means and variances as independent subsets: as such, this is an ideal situation to use the subset-based inference of Split-BOLFI.

Recent work \citep{drovandi2024improving} has demonstrated the need for caution when defining subsets for modular inference in this context: parameters such as the mean and variance of a Gaussian distribution that appear to contribute to the model separately, exhibit some possibly unexpected dependence when performing joint posterior inference.

Experiments were run with 500 data points as observed data sets and simulated data sets for each parameter value. We considered distributions with observation dimensions equal to 1 and 5, resulting in parameter spaces with dimensionality $p$ equal to 2 and 10. We considered both well-specified and misspecified inference cases, with the true generative distributions being multidimensional Gaussian or Laplacian distributions respectively.

The true values of the mean were drawn from $\mathcal{U}(-4,4)$, and the true values of the standard deviation (or scale parameter for the Laplacian distribution) from $\mathcal{U}(1,4)$. Each experiment was iterated over 50 random seeds to demonstrate consistency. All methods considered used $p(\mu) \sim \mathcal{U}(-5,5)$ priors for the means and $p(\sigma)\sim\mathcal{U}(0,5)$ priors for the standard deviations.

Split-BOLFI was used to perform inference in this situation, with comparison to the standard BOLFI algorithm and modular rejection ABC algorithm. Both Split-BOLFI and BOLFI were evaluated after acquiring $n_s=$ 50, 100, and 250 simulated data sets.

For the rejection ABC algorithm, a large pool of 100,000 simulations from the prior were used to draw samples from the subset distributions defined by the same discrepancies used by Split-BOLFI using a quantile $q=0.01$, resulting in 1000 accepted samples for every run.

The summaries used to construct the discrepancies were the means, standard deviations and kurtoses of each dimension of the simulated data. The data means and standard deviations represent sufficient statistics to perform inference over the mean and standard deviation parameters, while the kurtosis statistic gave an indication of the misspecification level of heaviness of the tails of the data. BOLFI was run without the kurtosis statistics, and Split-BOLFI and rejection ABC were performed separately both with and without the kurtosis included to demonstrate the influence of the kurtosis summary statistic on  the inference.

Split-BOLFI with the kurtosis statistics would consequently be expected to provide tempered belief updates when the data generative process is the Laplacian distribution and hence the tails of the Gaussian statistical model are too light. Split-BOLFI without the kurtosis summary statistics would not be expected to detect the heavier tails in the presence of misspecification.

Mat\'{e}rn kernels were used for the kernel components of the discrepancy models in Split-BOLFI and BOLFI, with Gamma(2, 2) hyperpriors for the lengthscales and exp(1) hyperpriors for the kernel variances.

The inference result statistics of the $\mu$ and $\sigma$ parameters for maximum number of acquisitions used, with observed data sets of 500 data points. are presented in \Cref{GAUSSIAN500AME,GAUSSIAN500RMSE,GAUSSIAN500SD,GAUSSIAN500COVERAGE}. The inference result statistics of the $\mu$ and $\sigma$ parameters for maximum number of acquisitions used, with observed data sets of 5000 data points, are presented in  \Cref{GAUSSIAN5000AME,GAUSSIAN5000RMSE,GAUSSIAN5000SD,GAUSSIAN5000COVERAGE}. The sizes of these datasets were chosen to demonstrate the behavior of the inference with the summary statistics exhibiting varying levels of stability, and do not necessarily reflect real-world investigative conditions.

\begin{table}[!h]
\centering
\begin{tabular}{c|ccccc}
 & $\text{ABC}_{m,s}$ & $\text{ABC}_{m,s,k}$ & $\text{SB}_{m,s}$ & $\text{SB}_{m,s,k}$ & $\text{BOLFI}_{m,s}$ \\\hline
$\mu_1$ & 0.13 (0.72) & 0.12 (0.64) & 0.03 (0.03) & 0.04 (0.03) & 0.03 (0.03) \\
$\sigma_1$ & 0.05 (0.29) & 0.06 (0.31) & 0.02 (0.02) & 0.02 (0.02) & 0.02 (0.02) \\
$\mu_5$ & 0.13 (0.71) & 0.14 (0.75) & 0.03 (0.02) & 0.03 (0.03) & 1.79 (1.53) \\
$\sigma_5$ & 0.05 (0.28) & 0.05 (0.25) & 0.02 (0.02) & 0.02 (0.02) & 0.73 (0.48) \\
\hline
$\mu_1$ & 0.13 (0.71) & 0.001 (0.006) & 0.03 (0.03) & 0.02 (0.02) & 0.03 (0.02) \\
$\sigma_1$ & 0.05 (0.26) & 0.002 (0.009) & 0.03 (0.03) & 0.03 (0.02) & 0.03 (0.03) \\
$\mu_5$ & 0.14 (0.74) & 0.001 (0.009) & 0.03 (0.03) & 0.03 (0.03) & 1.59 (1.41) \\
$\sigma_5$ & 0.05 (0.28) & 0.002 (0.009) & 0.03 (0.03) & 0.03 (0.03) & 0.79 (0.46) \\
\end{tabular}
\caption{Results for a Gaussian statistical model with parameters $\mu_p$ and $\sigma_p$ in the well-specified case (above line) and misspecified case (below line) for $p=1$ and $p=5$. Algorithms are ABC, Split-BOLFI, and BOLFI using mean and standard deviation summaries $m,s$ and additionally kurtosis summary $m,s,k$. Results of the mean(sd) of AME or AMAPE with a data set of 5000 observations. Error is based on the point estimate which was sample mean for ABC and BOLFI and MAP for Split-BOLFI.}
\label{GAUSSIAN5000AME}
\end{table}

\begin{table}[!h]
\centering
\begin{tabular}{c|ccccc}
 & $\text{ABC}_{m,s}$ & $\text{ABC}_{m,s,k}$ & $\text{SB}_{m,s}$ & $\text{SB}_{m,s,k}$ & $\text{BOLFI}_{m,s}$ \\\hline
$\mu_1$ & 0.14 (0.72) & 0.12 (0.64) & 0.28 (0.03) & 0.41 (0.04) & 0.1 (0.03) \\
$\sigma_1$ & 0.05 (0.29) & 0.06 (0.31) & 0.27 (0.03) & 0.4 (0.04) & 0.1 (0.02) \\
$\mu_5$ & 0.13 (0.71) & 0.14 (0.75) & 0.27 (0.04) & 0.39 (0.05) & 2.93 (1.44) \\
$\sigma_5$ & 0.05 (0.28) & 0.05 (0.25) & 0.27 (0.03) & 0.39 (0.05) & 1.39 (0.34) \\
\hline
$\mu_1$ & 0.13 (0.71) & 0.01 (0.05) & 0.28 (0.02) & 2.54 (0.25) & 0.1 (0.02) \\
$\sigma_1$ & 0.05 (0.27) & 0.01 (0.05) & 0.27 (0.02) & 1.43 (0.11) & 0.1 (0.03) \\
$\mu_5$ & 0.14 (0.74) & 0.01 (0.05) & 0.27 (0.03) & 2.52 (0.24) & 2.75 (1.33) \\
$\sigma_5$ & 0.05 (0.28) & 0.01 (0.05) & 0.27 (0.03) & 1.44 (0.12) & 1.43 (0.33) \\
\end{tabular}
\caption{Results for a Gaussian statistical model with parameters $\mu_p$ and $\sigma_p$ in the well-specified case (above line) and misspecified case (below line) for $p=1$ and $p=5$. Algorithms are ABC, Split-BOLFI, and BOLFI using mean and standard deviation summaries $m,s$ and additionally kurtosis summary $m,s,k$. Results of the mean(sd) of RMSE with a data set of 5000 observations. Error is based on the point estimate which was sample mean for ABC and BOLFI and MAP for Split-BOLFI.}
\label{GAUSSIAN5000RMSE}
\end{table}

\begin{table}[!h]
\centering
\begin{tabular}{c|ccccc}
 & $\text{ABC}_{m,s}$ & $\text{ABC}_{m,s,k}$ & $\text{SB}_{m,s}$ & $\text{SB}_{m,s,k}$ & $\text{BOLFI}_{m,s}$ \\\hline
$\mu_1$ & 0.01 (0.05) & 0.01 (0.05) & 0.27 (0.04) & 0.41 (0.04) & 0.17 (0.02) \\
$\sigma_1$ & 0.01 (0.05) & 0.01 (0.05) & 0.27 (0.04) & 0.4 (0.04) & 0.17 (0.02) \\
$\mu_5$ & 0.01 (0.05) & 0.01 (0.05) & 0.27 (0.05) & 0.39 (0.05) & 2.02 (0.83) \\
$\sigma_5$ & 0.01 (0.05) & 0.01 (0.05) & 0.27 (0.05) & 0.38 (0.05) & 1.13 (0.29) \\
\hline
$\mu_1$ & 0.01 (0.05) & 0.01 (0.05) & 0.27 (0.1) & 2.27 (0.1) & 0.17 (0.02) \\
$\sigma_1$ & 0.01 (0.05) & 0.01 (0.05) & 0.27 (0.02) & 1.31 (0.02) & 0.18 (0.02) \\
$\mu_5$ & 0.01 (0.05) & 0.01 (0.05) & 0.27 (0.1) & 2.3 (0.1) & 2.05 (0.82) \\
$\sigma_5$ & 0.01 (0.05) & 0.01 (0.05) & 0.27 (0.02) & 1.31 (0.02) & 1.13 (0.3) \\
\end{tabular}
\caption{Results for a Gaussian statistical model with parameters $\mu_p$ and $\sigma_p$ in the well-specified case (above line) and misspecified case (below line) for $p=1$ and $p=5$. Algorithms are ABC, Split-BOLFI, and BOLFI using mean and standard deviation summaries $m,s$ and additionally kurtosis summary $m,s,k$. Results of the mean(sd) of SD with a data set of 5000 observations. Error is based on the point estimate which was sample mean for ABC and BOLFI and MAP for Split-BOLFI.}
\label{GAUSSIAN5000SD}
\end{table}

In \Cref{GAUSSIAN5000AME,GAUSSIAN5000RMSE,GAUSSIAN5000SD}, we see that the SB algorithm performs well in terms of the Average Mean Error on the MAP in both the low- and high-dimensional cases, the well- and misspecified cases and with and without the kurtosis included as a summary statistic. The method further introduces a large amount of posterior tempering in the misspecified case when the kurtosis is included as a summary statistics, correctly identifying the misspecification present in the model. The RMSE for SB with kurtosis in the misspecified case sees a corresponding increase due to the posterior tempering. SB without kurtosis sees no increase in tempering in the misspecified case, effectively ignoring the misspecification present. We see that the BOLFI algorithm demonstrates accurate inference in terms of the AME in the lower-dimensional case, while substantially losing accuracy in the 10-dimensional case.

The BOLFI results show no substantial difference between the well- and mis-specified cases, which may be problematic if tempering is considered desirable. The ABC shows some intriguing characteristics on this problen: the accepted samples are all show very low variation per posterior, but lower accuracy in terms of the AME compared to the active acquisition methods, with the exception of misspecified inference with the kurtosis statistic included. In this instance, the simple rejection algorithm appears to achieve high accuracy and precision: this may be due to the kurtosis statistic in the misspecified case actively selecting simulated data sets that explore the tails of the Gaussian distribution more thoroughly, thereby generating a more stable estimate of the mean and standard deviation, while the other summary statistics ensure that a bias is not introduced that might be expected when using data sets with extra outliers. We do not know how this phenomenon might generalise outside of this example, but note it to be of interest.

\begin{figure}[h!]
    \centering
    \centering
    \begin{subfigure}[b]{0.44\textwidth}
        \includegraphics[width=\textwidth]{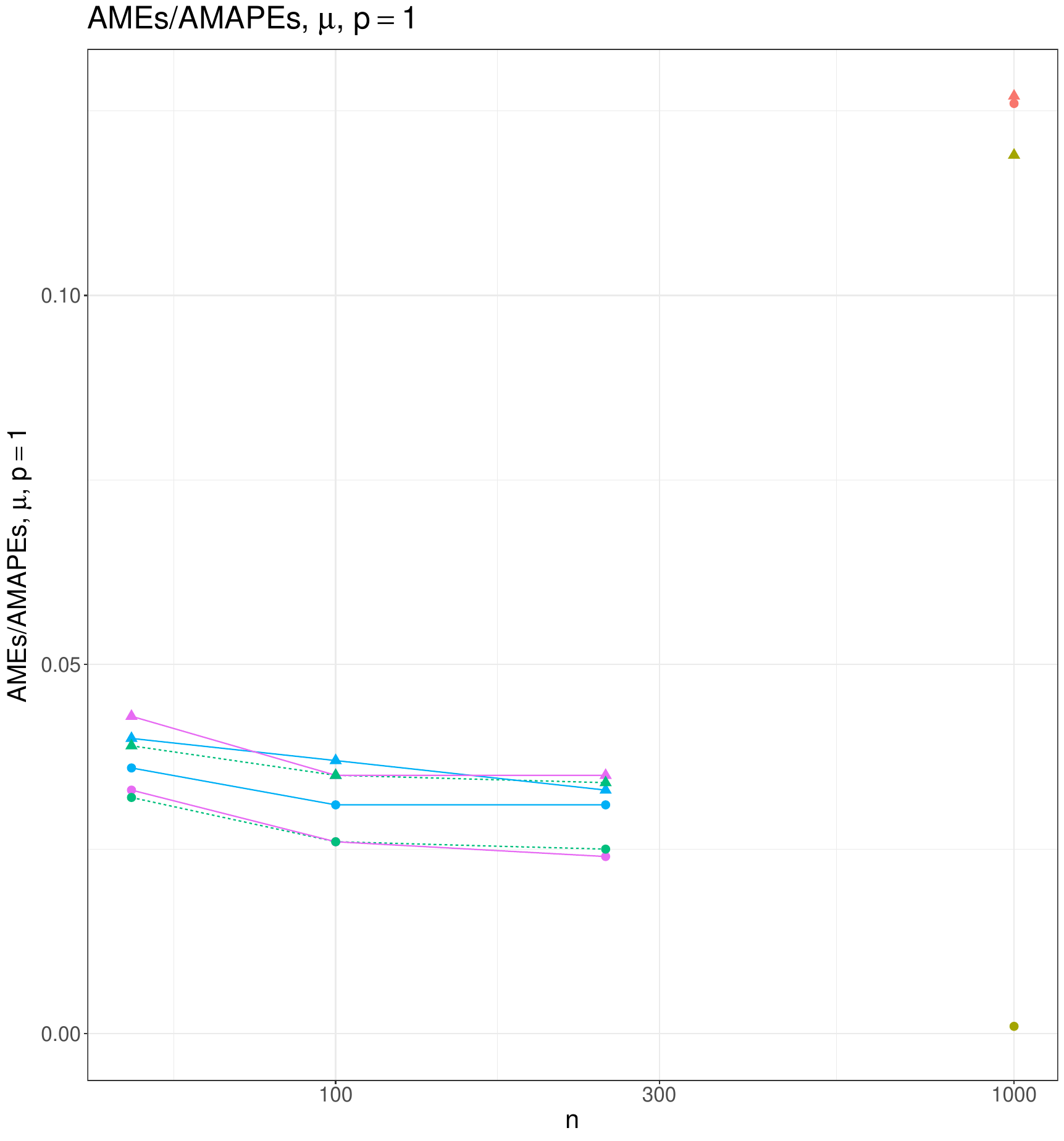}
    \end{subfigure}
    ~ 
    \begin{subfigure}[b]{0.44\textwidth}
        \includegraphics[width=\textwidth]{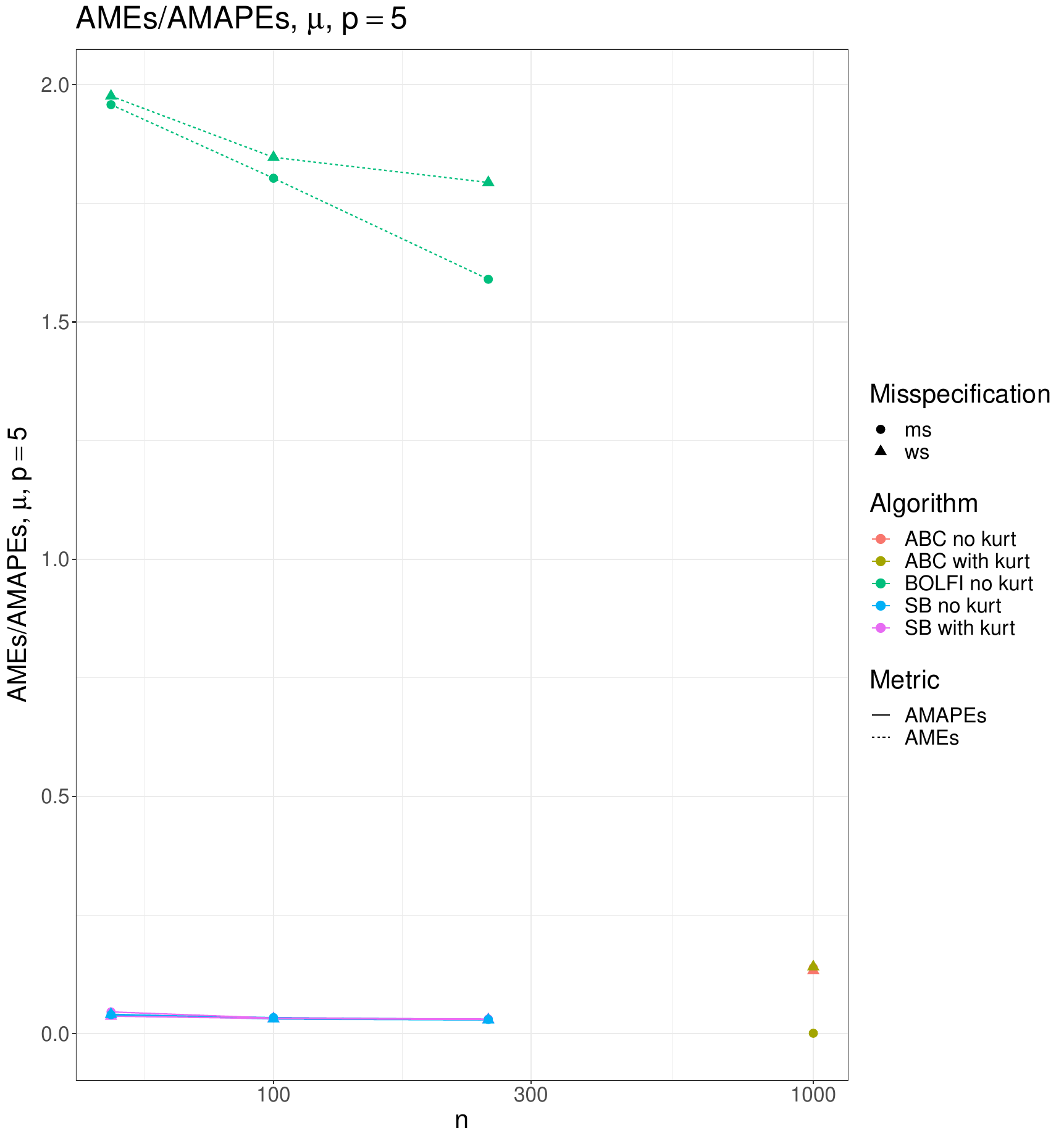}
    \end{subfigure}
    
    \begin{subfigure}[b]{0.44\textwidth}
        \includegraphics[width=\textwidth]{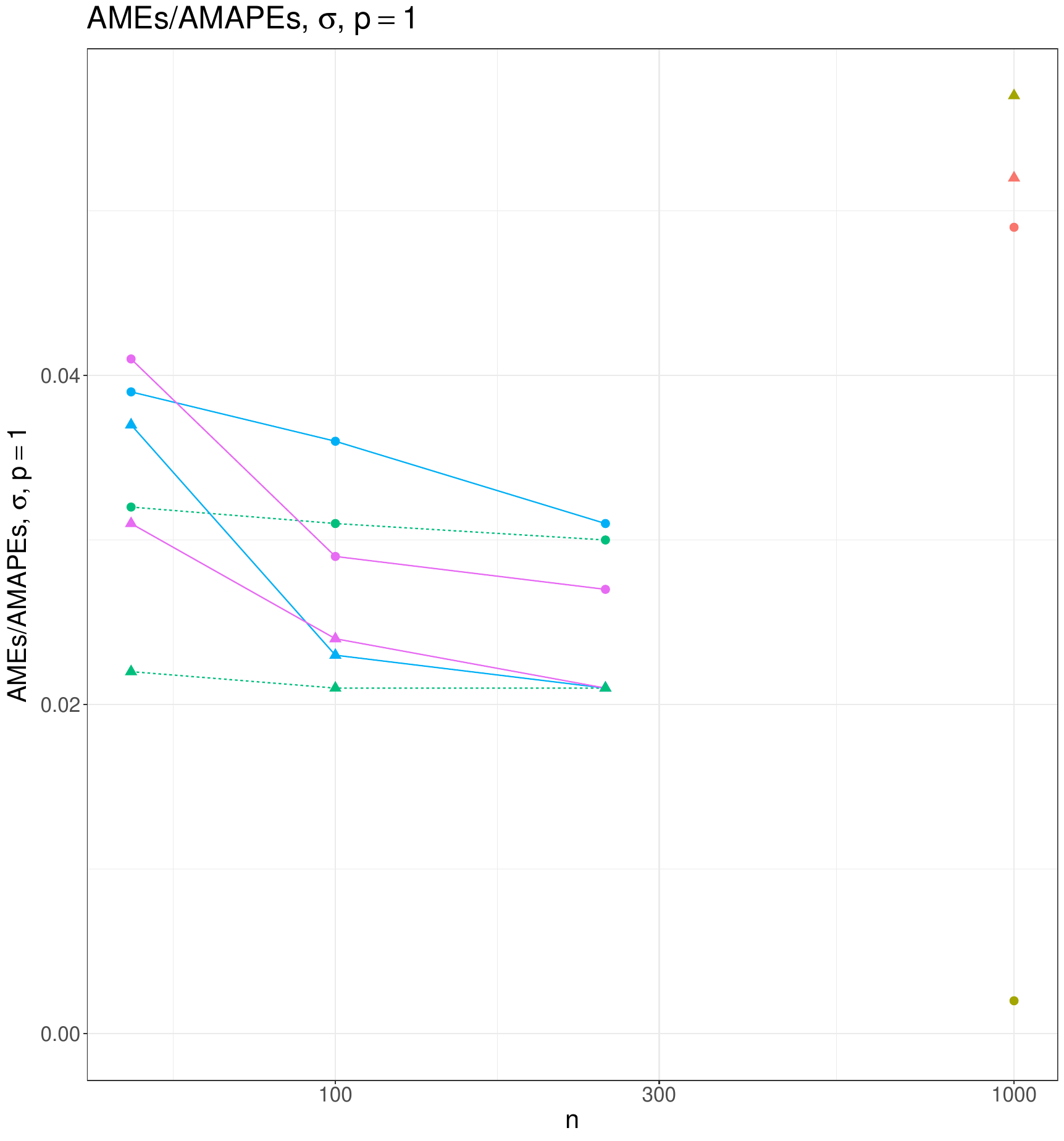}
    \end{subfigure}
    ~ 
    \begin{subfigure}[b]{0.44\textwidth}
        \includegraphics[width=\textwidth]{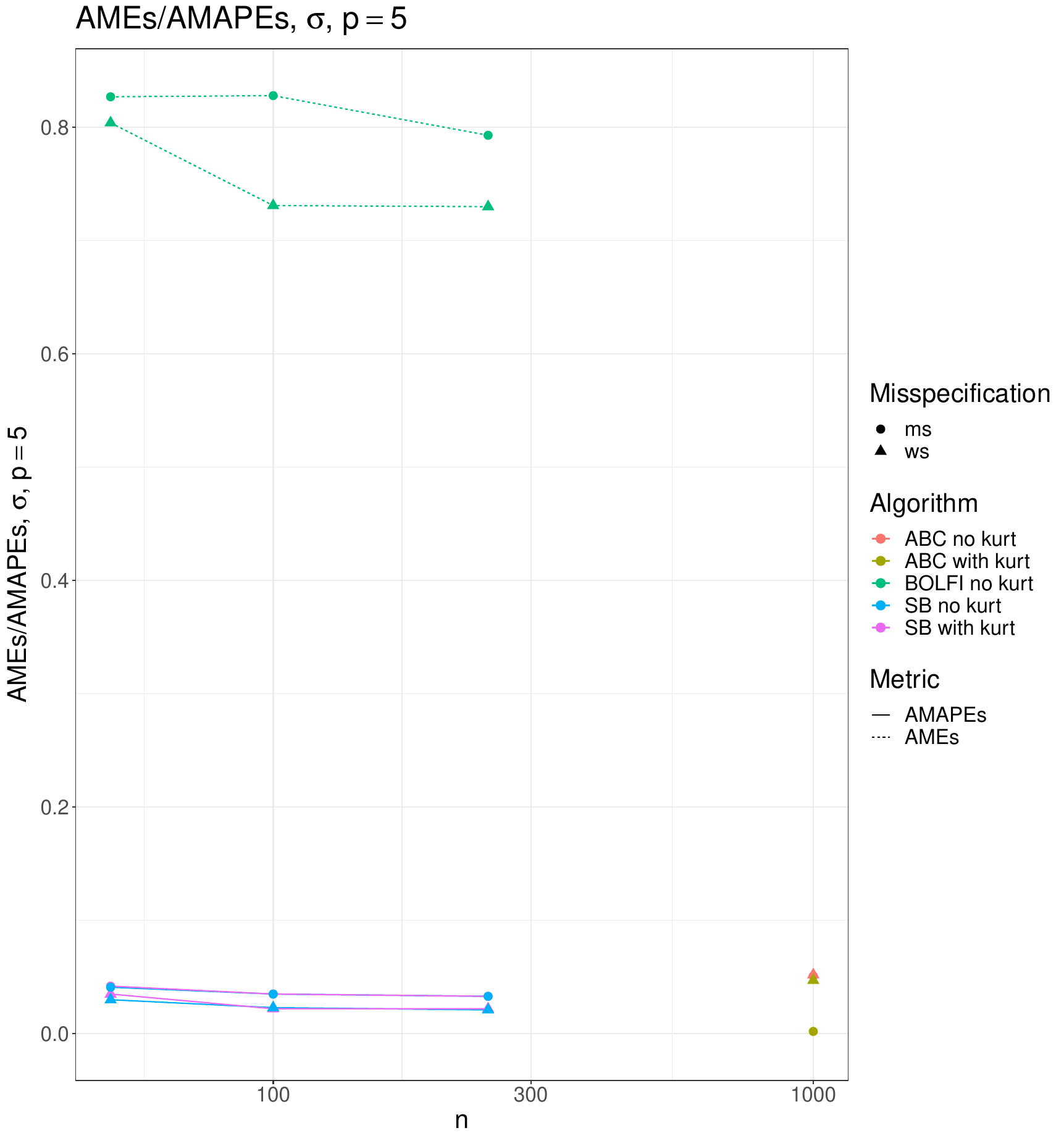}
    \end{subfigure}
    \caption{Results for a Gaussian model with $n_{obs}=5000$, as a function of number of simulations $n_{sim}$. Diagrams vary by dimensionality $p$, and parameter $\mu$ or $\sigma$. Plotted are AME for and AMAPE, varying model specification, with algorithms ABC, Split-BOLFI, and BOLFI using mean and standard deviation summaries and with or without the kurtosis summary. \label{fig:gaussian_5000_results_appendix_new}}
\end{figure}

In Figure \ref{fig:gaussian_5000_results_appendix_new} we see that the point estimates associated with SB and BOLFI make effective use of increasing number of simulations, appearing to reach a plateau by $n_{sim} = 250$, except for BOLFI in the higher-dimensional case which clearly has not converged.

In \Cref{GAUSSIAN500AME,GAUSSIAN500RMSE,GAUSSIAN500SD}, we see that in the well-specified case the point estimate provided by SB performs comparably well with BOLFI and the ABC estimate performed used much more simulations. The BOLFI algorithm loses accuracy substantially in the 10-dimensional parameter space compared with ABC and SB. We see that the SB method assigns more breadth to the posterior than the other methods when successful, especially when the kurtosis statistics is included, resulting in larger SD and RMSE measures. This is not necessarily appropriate, as the model is well-specified in this case, but may simply indicate that $n_{obs}=500$ is insufficiently large to reject the possibility of misspecification, and consequently some uncertainty is added. Comparing with the misspecified case, the results are substantially similar to the well-specified case except that SB with kurtosis adds substantially more posterior breadth having clearly identified the misspecification present. None of the other results change substantially, which is not necessarily desirable if we would like to identify misspecification and add posterior uncertainty when it is present. We see corresponding results as a function of the number of simulators acquisitions in Figure  \ref{fig:gaussian_500_results_appendix_new}.

We also investigated coverages and they indicated that the ABC methods generally report less than 50\% coverage, whereas Split-BOLFI and BOLFI show generally larger than 50\% coverage. We observe that the ABC methods contribute slightly larger coverage when the kurtosis statistic is introduced in the misspecified case, whereas the other methods' coverages do not vary substantially with changes in experimental setup. Details can be found in Appendix \Cref{GAUSSIAN500COVERAGE,GAUSSIAN5000COVERAGE}.

\subsection{Daycare transmission dynamics competition model}

To illustrate the Split-BOLFI method with a challenging real-world simulator model lacking a tractable likelihood, we developed an extension of the pneumococcal competition model introduced in \citep{numminen2013estimating}.  Each child can be colonized by a number of strains of bacteria, and each of these strains may compete within each daycare centre to colonize the children who are assumed to interact with one another, and to transmit the bacteria further when they have established a stable colonization. The observations at any given time are encoded as a binary matrix $\mathbf{I}$, indicating which children are colonized by which pathogens. 

The model is defined by a set of transition equations determining the probabilities of a colonization event taking place for a particular strain given the absences and presences of the other strains. The observed data are assumed to have been sampled from an equilibrium state distribution defined by the model parameters. The original model in \citep{numminen2013estimating} is parameterised by scalars $\beta, \Lambda, \gamma$ and $\mathbf{\theta}$, while the vector $\mathbf{P}$ indicates the background prevalence of each strain in the population, with the following transition equations:
\begin{align}
 P(I_{is}(t+\delta t) = 1 | I_{is}(t)=0) &= \beta \mathit{E}_s[I(t)] + \Lambda P^{s}  + o(\delta t),  \quad \text{if }  \sum_j I_{ij}(t) = 0 \nonumber \\
 P(I_{is}(t+\delta t) = 1 | I_{is}(t)=0) &= \theta \left(\beta \mathit{E}_s[I(t)] + \Lambda P^{s}\right) + o(\delta t) , \quad \text{otherwise} \nonumber \\
  P(I_{is}(t+\delta t) = 0 | I_{is}(t)=1) &= \gamma+ o(\delta t),
\end{align}
where $\beta$ determines the rate of transmission within the daycare population, $\Lambda$ determines the rate of transmission from the general population outside the daycare center, $\gamma$ determines the natural clearance rate of a strain within in a host, and $\theta$ is a single parameter describing the level of competition between every pair of different observed strains. The parameters are estimated relative to the parameter $\gamma$, without loss of generality.

We propose an extension to the model in \citep{numminen2013estimating}, designed to capture the strain-specific competition dynamics of bacterial colonization. We replace the scalar competition parameter $\theta$ used previously with a competition matrix $\bm{\theta}$  that contains the strain-specific pairwise competition parameters $\theta_{ij}=\theta_{ji}$, which suppresses the transmission of a strain $i$ if strain $j$ is already present in each host. A non-zero value of $\theta_{ij}$ consequently implies competition between the two particular strains. We propose the following updated transmission equations:
\begin{align}
 P(I_{is}(t+\delta t) = 1 | I_{is}(t)=0) &= \beta \mathit{E}_s[I(t)] + \Lambda P^{s}  + o(\delta t),  \quad \text{if }  \sum_j I_{ij}(t) = 0 \nonumber \\
 P(I_{is}(t+\delta t) = 1 | I_{is}(t)=0) &= 2 \Phi\left( - \sum_j^{p} I_{ij} \theta_{js} \right)\left(\beta \mathit{E}_s[I(t)] + \Lambda P^{s}\right) + o(\delta t) , \quad \text{otherwise} \nonumber \\
  P(I_{is}(t+\delta t) = 0 | I_{is}(t)=1) &= \gamma+ o(\delta t).
\end{align}

With the new competition matrix $\bm{\theta}$, the parameter space now scales with the number of observed strains: for a number of observed strains $n_s$, the parameter space of the entire model is $n_s^2 -n_s +2$ dimensional.

We followed \cite{numminen2013estimating} and their carefully constructed summary statistics were used for this model. The first used was the Shannon index of diversity of the distribution of observed strains calculated as:
\begin{equation}
    \text{SID}(X)= - \sum_k p_k \log p_k,\quad p_k=\sum_{ij} X_{ijk}/\sum_{ijk} X_{ijk}.
\end{equation}

Further summaries were the number of different strains observed, the prevalence of each strain and the prevalence of individuals with multiple colonizations. In addition, to characterise the pairwise competitions, three summary statistics were derived for each pair of strains to determine the  co-prevalence and co-anti-prevalence of each pair, with a normalisation included to account for each strain's general prevalence or scarcity, i.e.
\begin{align}
    \phi^{(0)}_{ab}(X)&=\frac{\sum_{ij}X_{ija}X_{ijb}}{\sqrt{n(0.01+\sum_{ij}X_{ija})(0.01+\sum_{ij}X_{ijb})}}\label{eq:daycaress_1}\\
    \phi^{(1)}_{ab}(X)&=\frac{\sum_{ij}(1-X_{ija})X_{ijb}}{\sqrt{n(0.01+\sum_{ij}(1-X_{ija}))(0.01+\sum_{ij}X_{ijb})}}\label{eq:daycaress_2}\\
    \phi^{(2)}_{ab}(X)&=\frac{\sum_{ij}X_{ija}(1-X_{ijb})}{\sqrt{n(0.01+\sum_{ij}X_{ija})(0.01+\sum_{ij}(1-X_{ijb}))}}\label{eq:daycaress_3}.
\end{align}

The sums are calculated over the individual daycare centres and children, and the addition of $0.01$ ensures stability when zero prevalence or zero absence of a strain is observed. The normalisation ensures that joint sparsity is not confused for competitiveness. The three pair-specific summary statistics were used to construct discrepancies for each competition parameter, and the sum of the other summary statistics was used to construct a single discrepancy to perform joint inference for the parameters $\beta$ and $\Lambda$.

\subsubsection{Simulator Study}

We consider a simulator study to evaluate the performance of various algorithms on the daycare simulator in the well- and misspecified cases. Here we use the high-dimensional pairwise interaction model as the data generative process in each case, and vary the statistical model between the low-dimensional model presented originally in \citep{numminen2013estimating} and the higher dimensional variation used to generate the data. We would not expect the lower-dimensional model to be able to accurately capture the variation in competition between difference pairs and hence exhibit misspecification, identifiable from the individual summary statistics from Equations \eqref{eq:daycaress_1}-\eqref{eq:daycaress_3}. By contrast the higher-dimensional statistical model should be able to capture the extra variation present in the model and hence not exhibit misspecification.

Experiments were performed averaged over 50 seeds with 53 individuals and 29 simulated daycare centers. The true parameters were fixed at $\beta=3.6$, $\Lambda=0.6$ and alternately 0 or 3 for the vectorised pairwise competition parameters $\theta_{ij}$. Modular rejection ABC was not performed in this instance as the simulator evaluations were too expensive to generate enough samples for stable posterior characterisation. BOLFI and Split-BOLFI were evaluated at the number of acquisitions $n_{acq}=$ 50, 100 and 250. The parameter $\beta$ was given a prior $\mathcal{U}(0,5)$, the parameter $\Lambda$ a prior $\mathcal{U}(0,1)$, and each pairwise competition parameter $\theta_{ij}$ was given a prior $\mathcal{U}(0,3)$, while the lower-dimensional averaged competition parameter was given a prior $\mathcal{U}(0,1)$.

The inference was assessed with the RMSE, AME and SD scores of the resulting samples for Split-BOLFI and BOLFI. In the misspecified case, the ``true value'' of the competition parameter was taken to be the average of $2 \Phi(\theta_{ij})$, to represent the averaged competition between all strains in the true model. For the higher-dimensional model the inference results were averaged over all of the competition parameters.

For Split-BOLFI in the lower-dimensional case, all three parameters were combined as one subset, while in the higher-dimensional case then $\beta$ and $\Lambda$ were combine as a two-dimensional subset, while each competition parameter formed its own univariate subset.

\begin{figure}[h!]
    \centering
        \includegraphics[width=.3\textwidth]{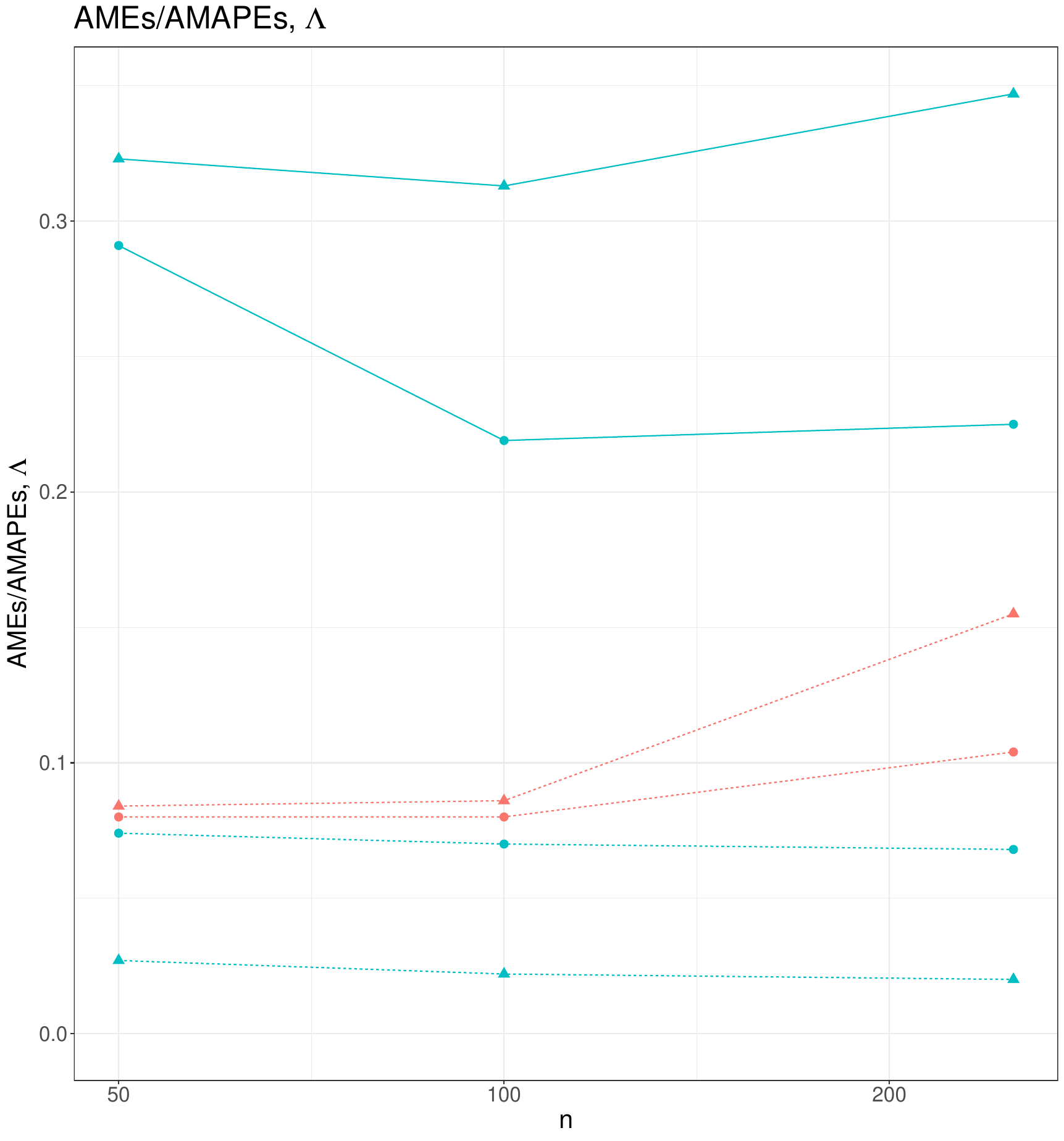}
        \includegraphics[width=.3\textwidth]{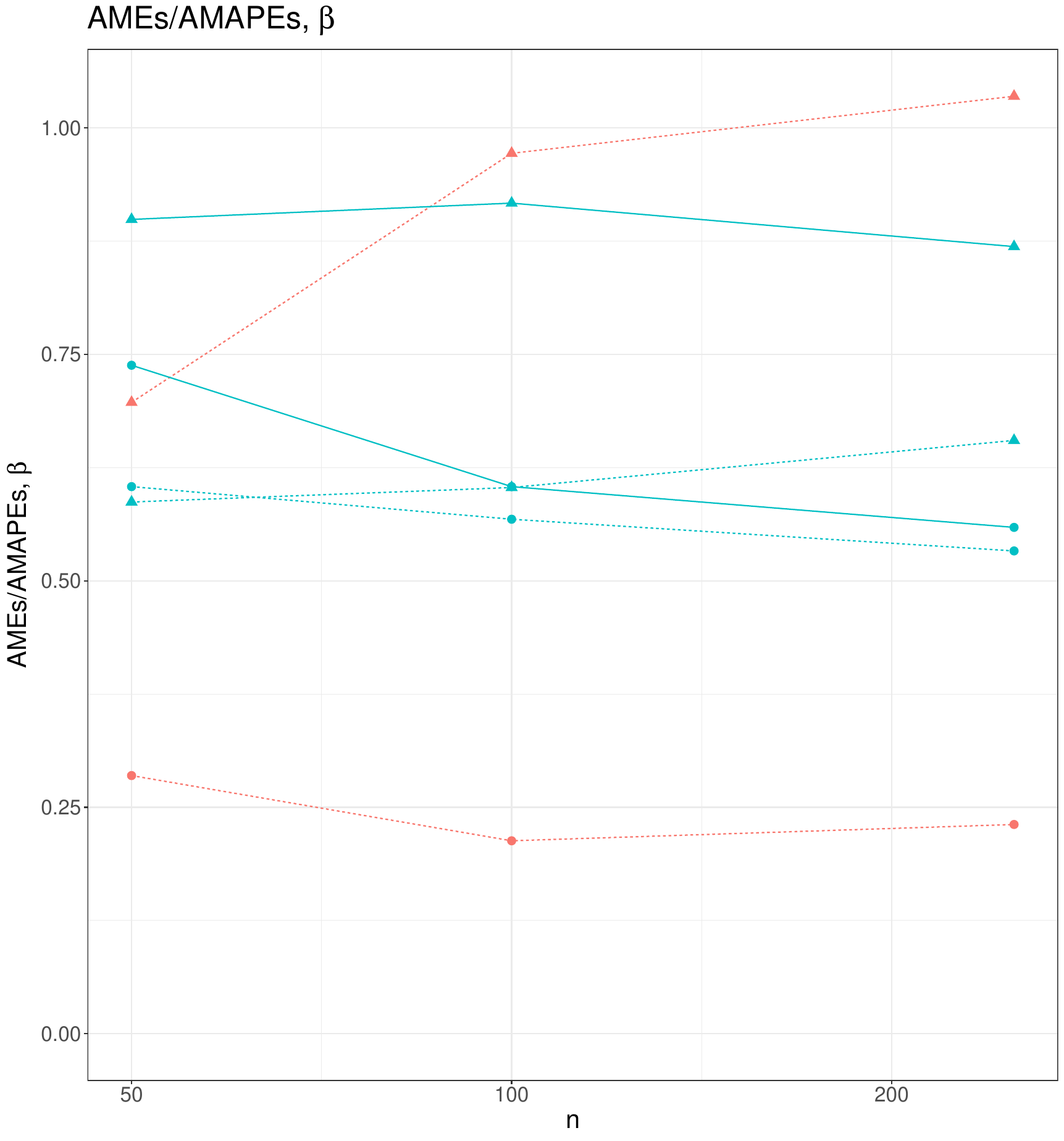}
        \includegraphics[width=.3\textwidth]{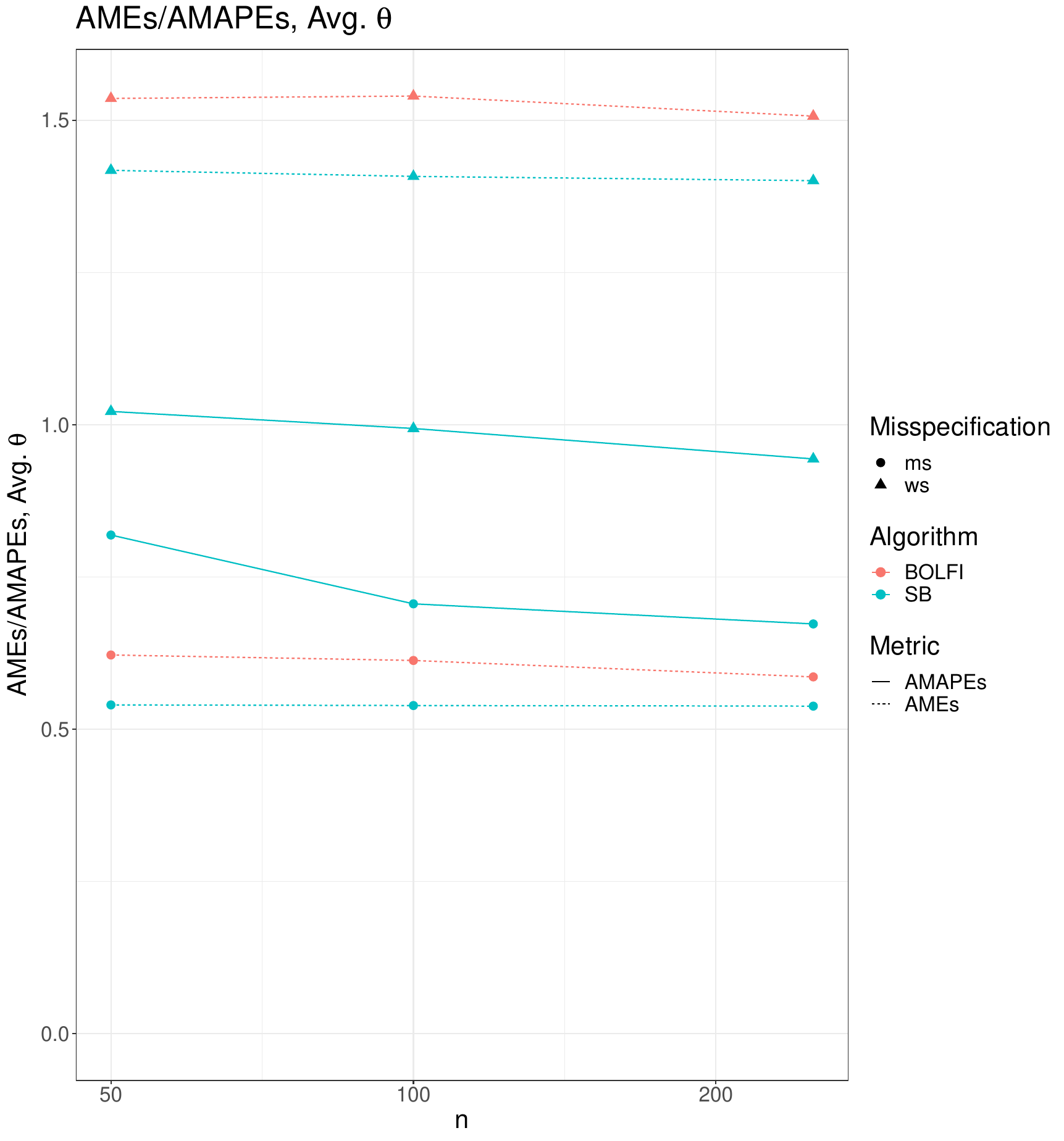}
\caption{Results for the Daycare model, as a function of number of simulations $n_{sim}$. Diagrams vary by parameters ($\beta$, $\Lambda$ and $\theta)$. Plotted are AME and AMAPE, varying model specification, with algorithms Split-BOLFI and BOLFI.} \label{fig:daycare_results_appendix_new}
\end{figure}

Results for the simulator study are presented in Table \ref{DAYCARETABLE} and Figure \ref{fig:daycare_results_appendix_new}. We see the slightly counterintuitive result that some of the AME for BOLFI and Split-BOLFI are smaller in the misspecified case, but this is because the misspecified model is a lower-dimensional inference task. We observe that the AMEs for all of the parameters in the well-specified case are smaller for Split-BOLFI than for BOLFI, suggesting that the modularised approach has helped negotiate the higher-dimensional space. We do not observe a larger posterior breadth for Split-BOLFI when misspecification is introduced, but since the misspecification is brought about by a substantial change in the statistical model rather than the true generative process, then these posterior statistics are not necessarily comparable. Figure \ref{fig:daycare_results_appendix_new} suggests that Split-BOLFI makes constructive use of increasing number of acquisitions. BOLFI appears to become unstable with increasing number of simulations in the higher-dimensional case, which is not evident for the Split-BOLFI algorithm.  The coverage analysis shows consistently greater than 50\% coverage for Split-BOLFI, and conflicting results for BOLFI in the well and misspecified cases: details can be found in Appendix \Cref{DAYCARECOVERAGE}.

\begin{table}[!h]
\centering
\begin{tabular}{c|cc|cc|cc|c}
& \multicolumn{2}{c}{AME} & \multicolumn{2}{c}{RMSE} & \multicolumn{2}{c}{SD} & AMAPE \\
& BOLFI & SB & BOLFI & SB & BOLFI & SB & SB \\ \hline
$\beta$ & 1.04 (.15) & .68 (.2) & 1.08 (.12) & 1.5 (.13) & .26 (.11) & 1.33 (.07) & .87 (.49) \\
$\Lambda$ & .16 (.09) & .02 (.02) & .27 (.03) & .27 (.01) & .2 (.06) & .26 (.01) & .35 (.07) \\
Avg. $\theta$ & 1.51 (.53) & 1.41 (.18) & 1.77 (.49) & 1.64 (.16) & .87 (.26) & .83 (.04) & .94 (1.08) \\ \hline \hline
$\beta$ & .23 (.17) & .53 (.23) & .72 (.1) & 1.33 (.18) & .66 (.11) & 1.2 (.12) & .56 (.43) \\
$\Lambda$ & .1 (.07) & .07 (.03) & .25 (.05) & .29 (.01) & .21 (.03) & .28 (.01) & .23 (.14) \\
Avg. $\theta$ & .59 (.09) & .54 (.03) & .61 (.08) & .6 (.03) & .18 (.02) & .27 (.01) & .67 (.2) \\
\end{tabular}
\caption{Results for a daycare simulator model in the well-specified (above double line) and misspecified case (below double line)  i.e~with a high-dimensional and low-dimensional statistical model and generative models, respectively. Results are presented for estimators Average Posterior Mean Error(AME)/RMSE/posterior SD/Average MAP Error (AMAPE) for BOLFI and Split-BOLFI for the parameters $\beta$, $\Lambda$ and averaged over $\theta$. Results are presented as mean(sd) of the estimators.}
\label{DAYCARETABLE}
\end{table}



\subsubsection{Real Data}

Here we use the higher-dimensional extension of the daycare simulator model from \citep{numminen2013estimating} on a real data set with data measured recording the presence of different strains of bacteria, lending itself to the pairwise competition parametrisation that we have pursued.

The samples from each time point are considered independent draws from the underlying dynamic process and are compared to independently generated equilibrium states from the simulator implementing the transition equations. The observed data contained 20\% missing values resulting from specific children being absent during a given week of a nurse visiting the daycare center. The missing data problem was addressed by making a missing at random assumption, such that the summary statistics of the observed data were used for the whole population, which should be a reasonable approximation to the missingness generating mechanism in the observation process.

The study design for the data collection was previously described in \citep{sa2008high}. Forty-seven children attending a daycare center in Lisbon, Portugal were sampled monthly during one year. The presence of \textit{H. influenzae}, \textit{S. pneumoniae}, \textit{M. catarrhalis}, \textit{S. aureus} and \textit{S. pyogenes} were recorded, with 14 serotypes of \textit{S. pneumoniae} also identified and recorded when present. Further details are found in Appendix. Some of the serotypes were observed very infrequently and were discarded from our analysis, leading to the four most common being retained. These are the serotypes 19F, 23F, 6B and 14, which are common in pre-pneumococcal conjugate vaccine infant populations. These four most common serotypes of \textit{S. pneumoniae} were considered as distinct strains in the simulations alongside \textit{H. influenzae}, \textit{M. catarrhalis}, \textit{S. aureus} and \textit{S. pyogenes}. The strains will be referred to using the letters in Table \ref{tab:strain_labels} for brevity.

\begin{table}[h!]
\centering
\begin{tabular}{c c |c c}
Label&Strain name&Label&Strain name\\
A&\textit{H. influenzae}&E& \textit{S. pneumoniae}, serotype 23F\\
B&\textit{M. catarrhalis}&F& \textit{S. pyogenes}\\
C& \textit{S. pneumoniae}, serotype 19F & G &\textit{S. pneumoniae}, serotype 6B\\
D& \textit{S. aureus}&H&\textit{S. pneumoniae}, serotype 14\\
\end{tabular}
\caption{The strains used in the high-dimensional data analysis, with labels used for reference.}
\label{tab:strain_labels}
\end{table}

As a result, we performed inference on a model with 28 competition parameters and the two parameters $\beta$ and $\Lambda$. Joint inference was performed over parameters $\beta$ and $\Lambda$, while each competition parameters was treated as a univariate distribution.

We performed inference with Split-BOLFI as the algorithm, as it showed the best inferential ability of the pairwise competition parameters $\theta_{ij}$.  Split-BOLFI used 250 simulator acquisitions, to fit models for the discrepancies associated with each parameter and to generate posterior proxies for each of the subset distributions.

Competition parameters were judged to show evidence of competition if either the mean or the mode of the posterior proxy distribution were larger than 1.5, the centre of the uniform prior distribution. Under this criterion, all of \textit{M. catarrhalis}, \textit{S. pneumoniae} (serotype 19F), \textit{S. pneumoniae} (serotype 23F), \textit{S. pyogenes}, \textit{S. pneumoniae} (serotype 6B) and \textit{S. pneumoniae} (serotype 14) were found to be mutually competitive with each other.

\begin{figure}[h!]
    \centering
     \begin{subfigure}[b]{0.45\textwidth}
        \includegraphics[width=\textwidth]{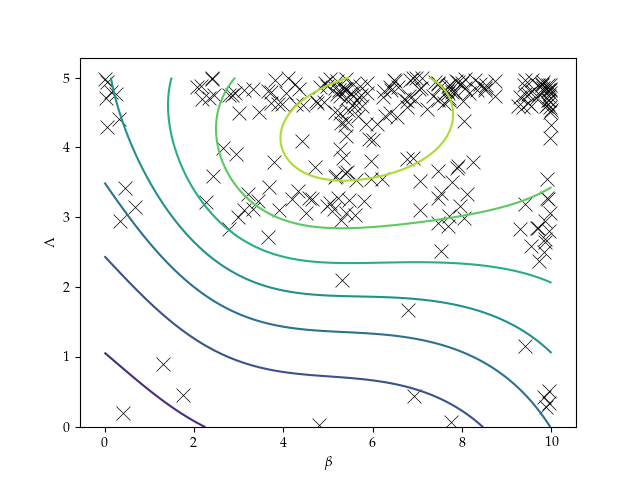}
    \end{subfigure}
    
     \begin{subfigure}[b]{0.22\textwidth}
        \includegraphics[width=\textwidth]{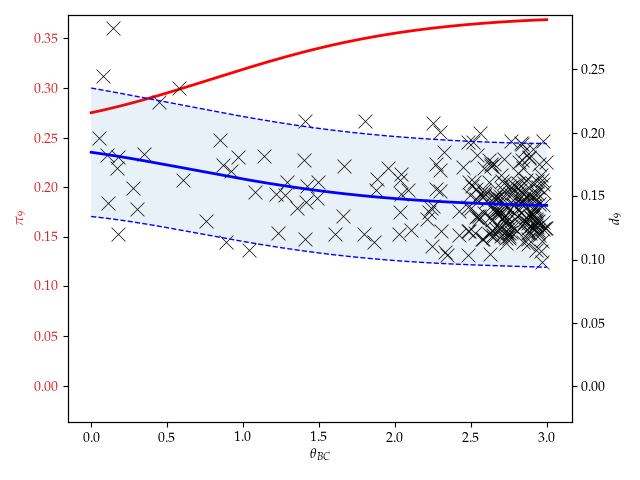}
    \end{subfigure}
    ~ 
     \begin{subfigure}[b]{0.22\textwidth}
        \includegraphics[width=\textwidth]{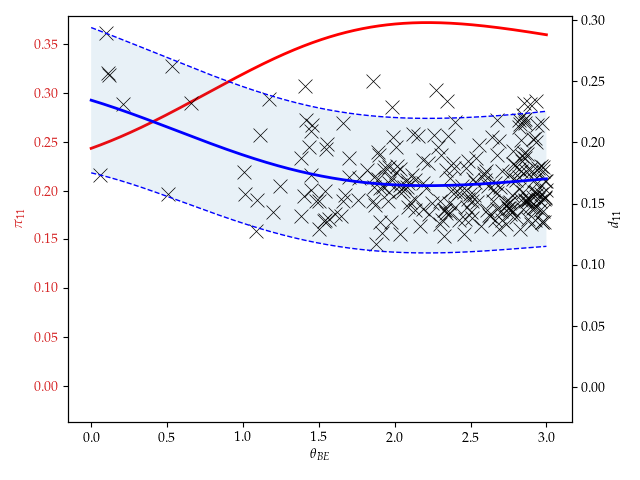}
    \end{subfigure}
    ~ 
     \begin{subfigure}[b]{0.22\textwidth}
        \includegraphics[width=\textwidth]{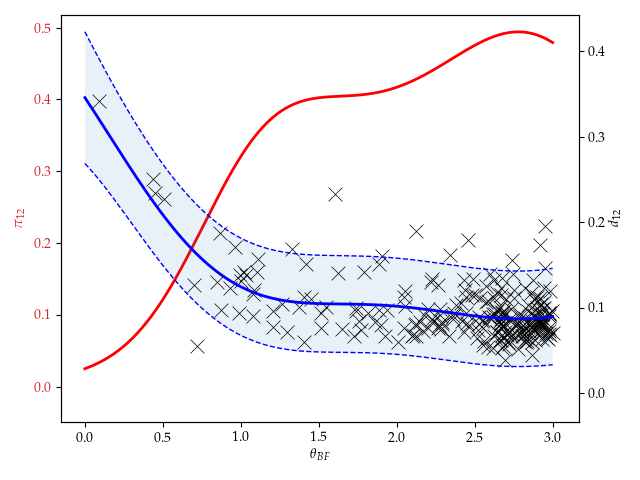}
    \end{subfigure}
    ~ 
     \begin{subfigure}[b]{0.22\textwidth}
        \includegraphics[width=\textwidth]{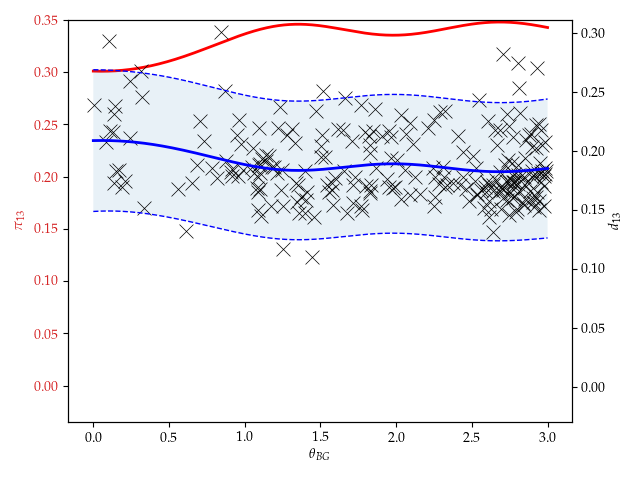}
    \end{subfigure}
    ~ 
     \begin{subfigure}[b]{0.22\textwidth}
        \includegraphics[width=\textwidth]{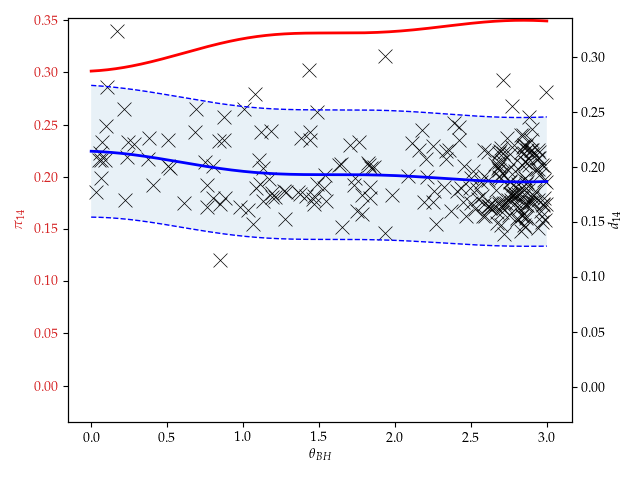}
    \end{subfigure}
    ~ 
     \begin{subfigure}[b]{0.22\textwidth}
        \includegraphics[width=\textwidth]{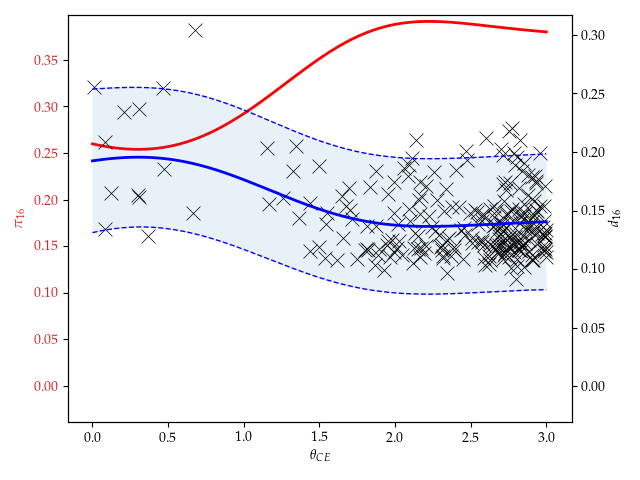}
    \end{subfigure}
    ~ 
     \begin{subfigure}[b]{0.22\textwidth}
        \includegraphics[width=\textwidth]{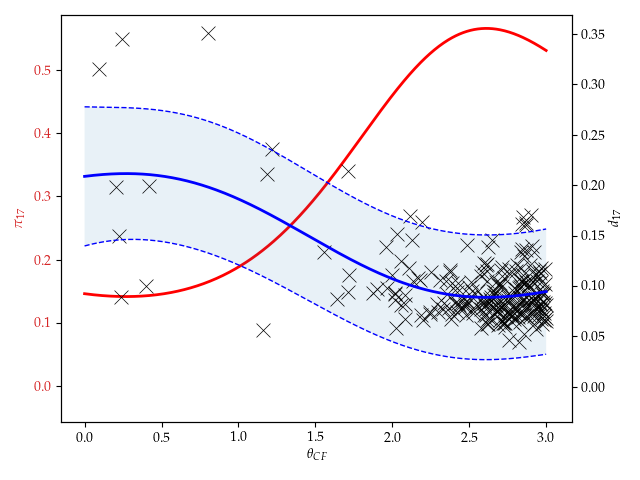}
    \end{subfigure}
    ~ 
     \begin{subfigure}[b]{0.22\textwidth}
        \includegraphics[width=\textwidth]{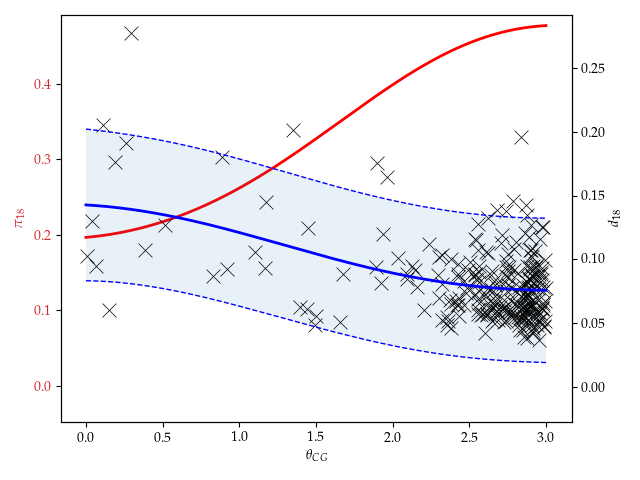}
    \end{subfigure}
    ~ 
     \begin{subfigure}[b]{0.22\textwidth}
        \includegraphics[width=\textwidth]{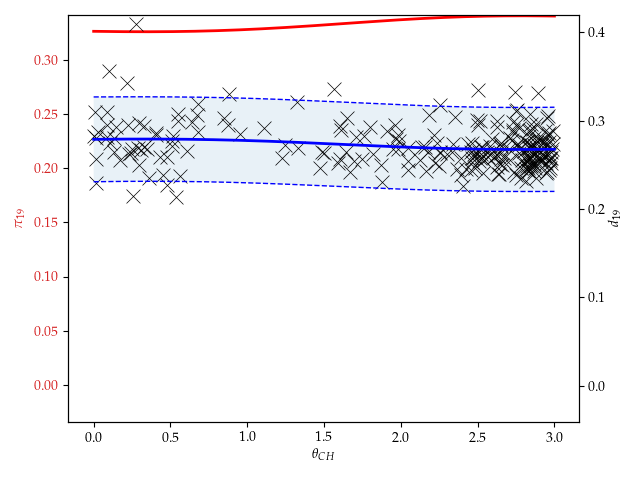}
    \end{subfigure}
    ~ 
     \begin{subfigure}[b]{0.22\textwidth}
        \includegraphics[width=\textwidth]{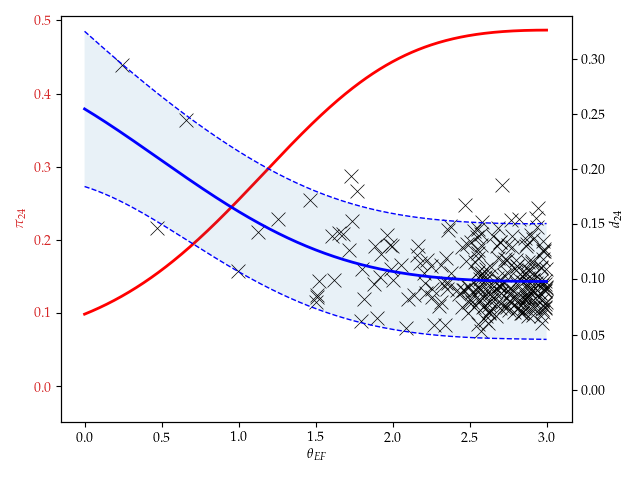}
    \end{subfigure}
    ~ 
     \begin{subfigure}[b]{0.22\textwidth}
        \includegraphics[width=\textwidth]{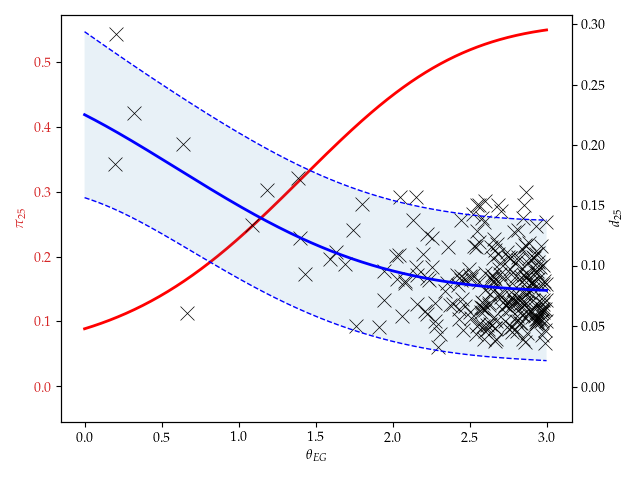}
    \end{subfigure}
    ~ 
     \begin{subfigure}[b]{0.22\textwidth}
        \includegraphics[width=\textwidth]{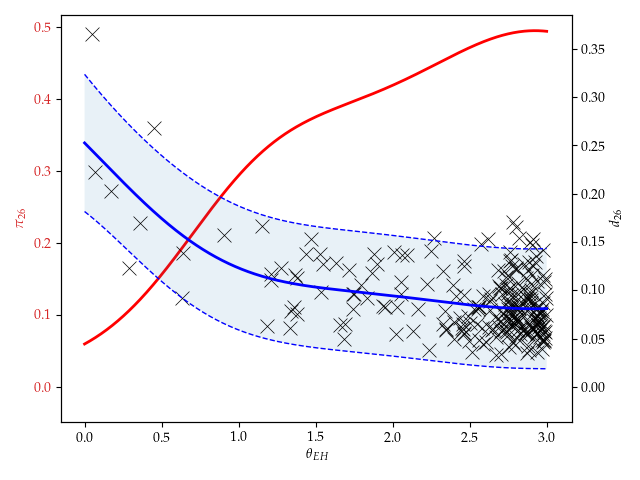}
    \end{subfigure}
    ~ 
     \begin{subfigure}[b]{0.22\textwidth}
        \includegraphics[width=\textwidth]{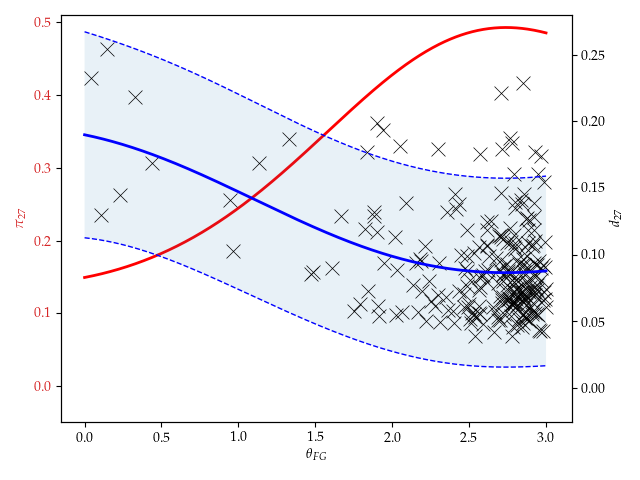}
    \end{subfigure}
    ~ 
     \begin{subfigure}[b]{0.22\textwidth}
        \includegraphics[width=\textwidth]{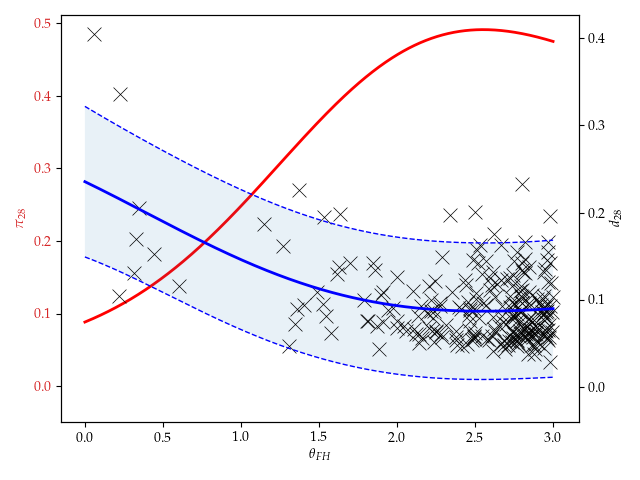}
    \end{subfigure}
    ~ 
     \begin{subfigure}[b]{0.22\textwidth}
        \includegraphics[width=\textwidth]{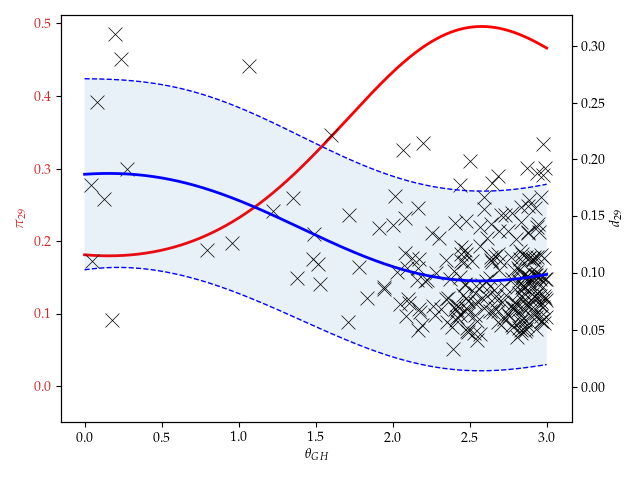}
    \end{subfigure}
    \caption{Joint proxy posterior contours and acquisition locations for $\beta$ and $\Lambda$, and proxy posteriors, and discrepancy models and acquisitions for the competition parameters identified as competitive from a 30-dimensional daycare model trained on real data.}\label{fig:daycare_30_competitive}   
\end{figure}

\begin{figure}[h!]
    \centering
     \begin{subfigure}[b]{0.22\textwidth}
        \includegraphics[width=\textwidth]{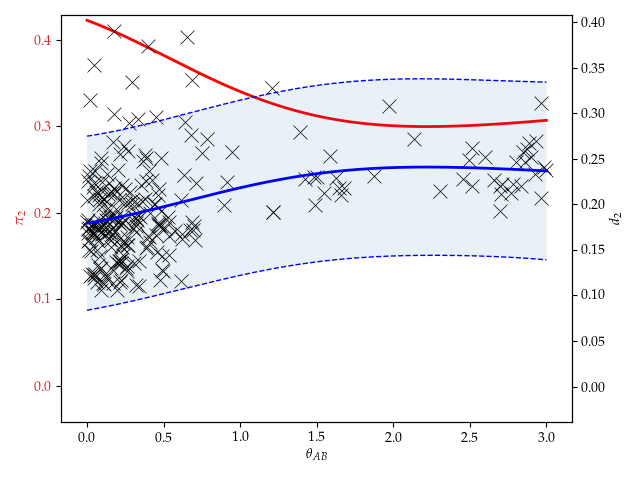}
    \end{subfigure}
    ~ 
     \begin{subfigure}[b]{0.22\textwidth}
        \includegraphics[width=\textwidth]{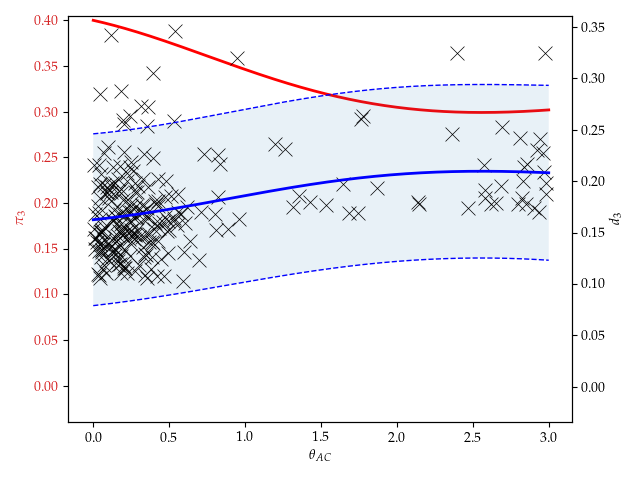}
    \end{subfigure}
    ~ 
     \begin{subfigure}[b]{0.22\textwidth}
        \includegraphics[width=\textwidth]{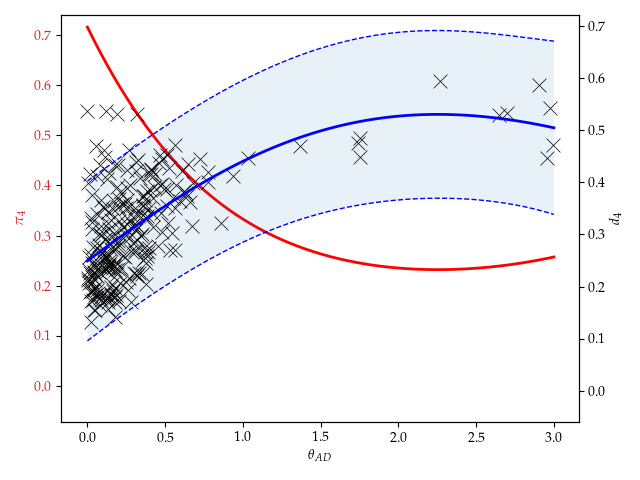}
    \end{subfigure}
    ~ 
     \begin{subfigure}[b]{0.22\textwidth}
        \includegraphics[width=\textwidth]{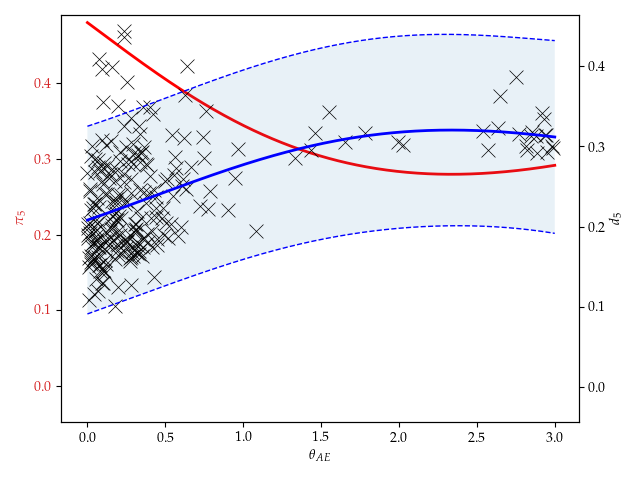}
    \end{subfigure}
    ~ 
     \begin{subfigure}[b]{0.22\textwidth}
        \includegraphics[width=\textwidth]{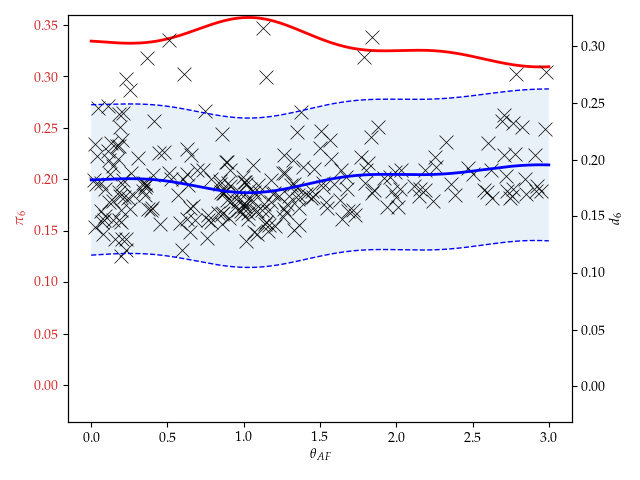}
    \end{subfigure}
    ~ 
     \begin{subfigure}[b]{0.22\textwidth}
        \includegraphics[width=\textwidth]{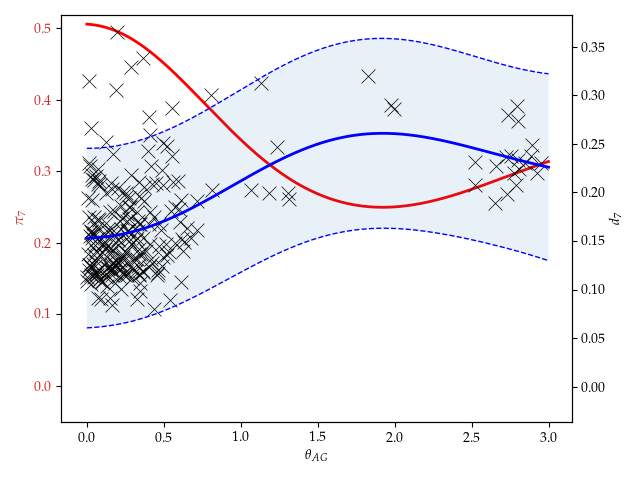}
    \end{subfigure}
    ~ 
     \begin{subfigure}[b]{0.22\textwidth}
        \includegraphics[width=\textwidth]{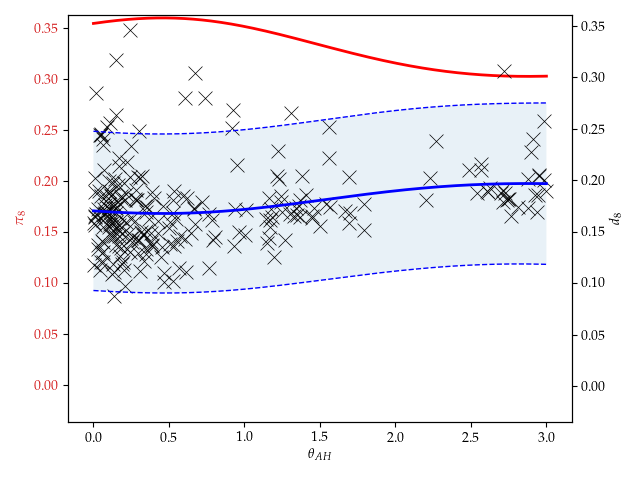}
    \end{subfigure}
    ~ 
     \begin{subfigure}[b]{0.22\textwidth}
        \includegraphics[width=\textwidth]{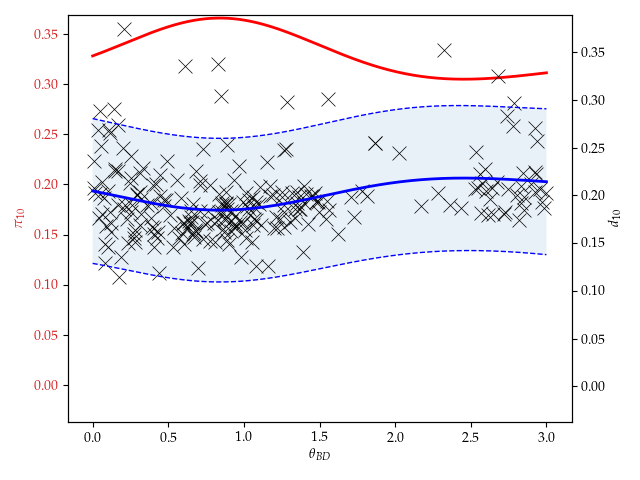}
    \end{subfigure}
    ~ 
     \begin{subfigure}[b]{0.22\textwidth}
        \includegraphics[width=\textwidth]{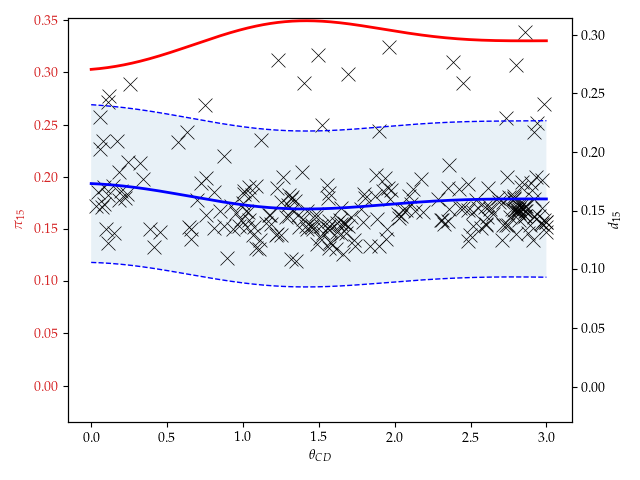}
    \end{subfigure}
    ~ 
     \begin{subfigure}[b]{0.22\textwidth}
        \includegraphics[width=\textwidth]{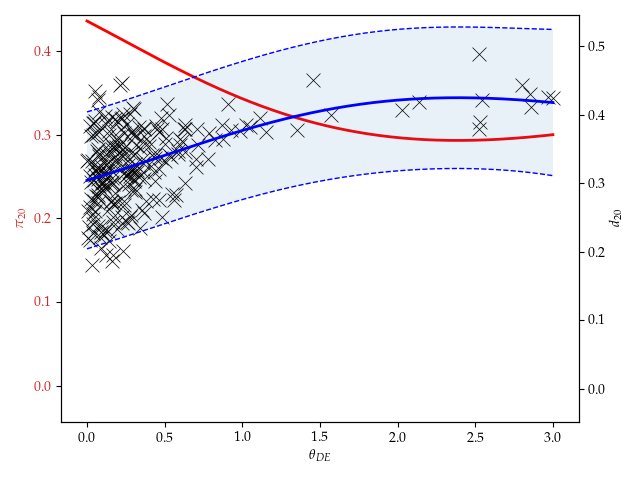}
    \end{subfigure}
    ~ 
     \begin{subfigure}[b]{0.22\textwidth}
        \includegraphics[width=\textwidth]{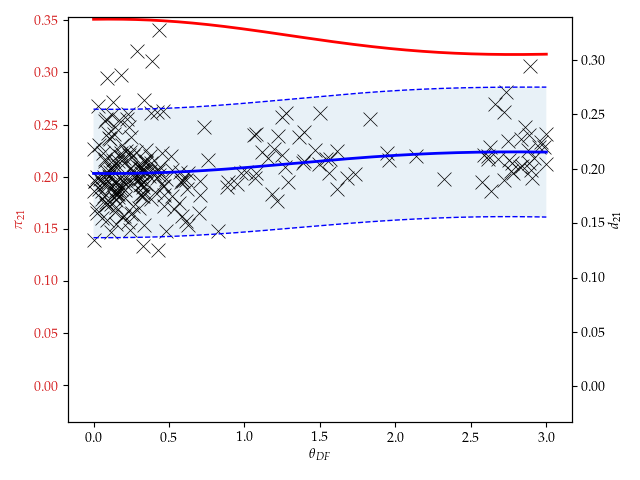}
    \end{subfigure}
    ~ 
     \begin{subfigure}[b]{0.22\textwidth}
        \includegraphics[width=\textwidth]{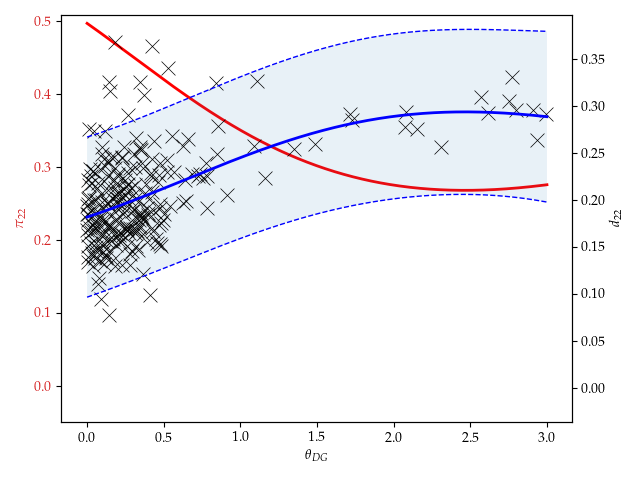}
    \end{subfigure}
    ~ 
     \begin{subfigure}[b]{0.22\textwidth}
        \includegraphics[width=\textwidth]{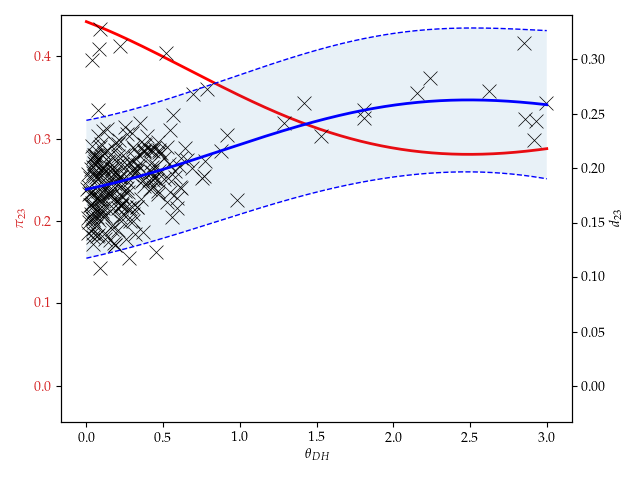}
    \end{subfigure}
    ~ 
    \caption{Proxy posteriors, and discrepancy models and acquisitions for the competition parameters identified as non-competitive from a 30-dimensional daycare model trained on real data.}\label{fig:daycare_30_noncompetitive}   
\end{figure}

The means, modes and standard deviations of $\beta$, $\Lambda$ and the $\theta_{ij}$ parameters showing evidence of competition are presented in Table \ref{tab:daycare_rewrite_competitive} and of the non-competitive $\theta_{ij}$ parameters in Table \ref{tab:daycare_rewrite_noncompetitive}. The posterior proxies and acquisitions for $\beta$, $\Lambda$ and the $\theta_{ij}$ parameters showing evidence of competition are plotted in Figure \ref{fig:daycare_30_competitive}, and those of the $\theta_{ij}$ parameters not showing evidence of competition are plotted in Figure \ref{fig:daycare_30_noncompetitive}.

\begin{table}[h!]
\centering
\begin{tabular}{c c c c c}
Mean&Mode&SD&Parameter\\ \hline
5.492 & 5.65 & 2.75 & $\beta$ \\
3.07 & 4.24 & 1.30 & $\Lambda$\\ \hline
1.58 & 3. & 0.86 & B, C \\
1.60 & 2.22 & 0.85 & B, E \\
1.87 & 2.79 & 0.73 & B, F \\
1.53 & 2.67 & 0.87 & B, G \\
1.54 & 2.85 & 0.87 & B, H \\
1.64 & 2.22 & 0.85 & C, E \\
1.92 & 2.61 & 0.79 & C, F \\
1.76 & 3. & 0.84 & C, G \\
1.51 & 2.79 & 0.87 & C, H \\
1.85 & 3. & 0.77 & E, F \\
1.91 & 3. & 0.76 & E, G \\
1.85 & 2.91 & 0.76 & E, H \\
1.82 & 2.73 & 0.81 & F, G \\
1.86 & 2.55 & 0.76 & F, H \\
1.81 & 2.58 & 0.82 & G, H 
\end{tabular}
\caption{Means, modes, standard deviations of $\log \beta$, $\log \Lambda$, and competition parameters of pairs of strains that put significant mass away from zero (both of the posterior mean and mode are greater than 1.5). The posterior covariance between $\beta$ and $\Lambda$ was -0.25.}
\label{tab:daycare_rewrite_competitive}
\end{table}

\begin{table}[h!]
\centering
\begin{tabular}{c c c c}
Mean&Mode& SD & Parameter\\ \hline
1.41 & 0. & 0.89 & A, B \\
1.42 & 0. & 0.88 & A, C \\
1.21 & 0. & 0.90 & A, D \\
1.36 & 0. & 0.89 & A, E \\
1.47 & 1.02 & 0.87 & A, F \\
1.33 & 0. & 0.92 & A, G \\
1.44 & 0.45 & 0.87 & A, H \\
1.45 & 0.84 & 0.87 & B, D \\
1.52 & 1.41 & 0.86 & C, D \\
1.39 & 0. & 0.89 & D, E \\
1.47 & 0.12 & 0.88 & D, F \\
1.33 & 0. & 0.89 & D, G \\
1.37 & 0. & 0.88 & D, H \\
\end{tabular}
\caption{Means, modes, standard deviations of competition parameters of pairs of strains that did not put significant posterior mass away from zero, i.e. either the mean or mode is less than 1.5.}
\label{tab:daycare_rewrite_noncompetitive}
\end{table}

The proxy likelihoods for $\beta$, $\Lambda$, and competitive $\theta_{ij}$ parameters in Figure \ref{fig:daycare_30_competitive} demonstrate concentration at a discrepancy minimum away from zero, with the Bayesian optimisation algorithm drawing from near the posterior mode. Some boundary effects are possibly evident, but are difficult to distinguish from the true value of the parameter lying at the edge of the prior. The competition parameters posteriors assigning most of the mass close to zero are presented in Figure \ref{fig:daycare_30_noncompetitive}, demonstrating concentration to a competition parameter value close to zero.

The results make sound biological sense in the light of existing literature on colonization competition between successful pneumococcal strain and between the species \textit{M. catarrhalis}, \textit{S. pyogenes} \citep{dahlblom2012bacterial,xu2012nasopharyngeal,jourdain2011differences,man2017microbiota,gjini2016direct,valente2016impact}. It is also reassuring that the inference has identified a complete subset of strains as being mutually competitive, as such a structure was not \textit{a priori} encoded into the model parameters. The results presented here result goes beyond previous analyses that were restricted to lower-dimensional spaces and hence considered all of the bacterial strains to be equally competitive with one another \citep{numminen2013estimating}. The extended results here illustrate the potential of the Split-BOLFI approach to infer parameters for models representing complex processes observed in reality.

\section{Conclusion}
\label{sec:conc}

In this work, we have extended existing LFI methodology to make new links with model misspecification, scalability to high-dimensional parameter spaces, and modularised inference: this is achieved through the use of additively defined generalised losses and exponentiated loss proxy likelihoods. The methodology exhibits efficient use of simulation resources in high-dimensional spaces, targeting parameter acquisitions local to the marginal modes through the use of Bayesian optimisation, and also contains robustness to misspecification through the use of exponentiated loss likelihoods and tempering in the case of large minimum discrepancies. The robustness properties of the inference procedure integrate naturally within a subjective Bayesian framework, in which the practising statistician is able to include summary statistics representing their suspicions regarding the misspecification of their statistical model, presenting an interpretable method for detecting and accounting for model shortcomings.

Split-BOLFI also offers a flexible framework to explore various structures in the joint posterior of the high-dimensional parameter space. The approach sidesteps the curse of dimensionality with a divide-and-conquer strategy, still allowing for \textit{a priori} specification of posterior correlation between specific parameters.

The same kernel does not need to be used for the purposes of simulation acquisition and posterior characterization, depending on the structure anticipated. It is possible to use separate GP priors to make efficient parameter acquisitions for the subset distributions, then use the resulting parameter values and discrepancy evaluations to build a likelihood proxy with an overlapping correlation structure. The acquisitions will not have been performed completely optimally for modelling the joint distribution, but we consider drawing simulations optimally for subset characterisation to be a generally reasonable acquisition strategy.

We also demonstrated the ability of the Split-BOLFI framework to work with real data on a generative model relevant to questions in contemporary epidemiology. We successfully perform inference in a 30-dimensional space (also a large parameter space for likelihood-free methods), to identify novel and biologically consistent competition dynamics between pathogens in a real world problem.

\section*{Acknowledgements}

OT, HP and JC are supported by  European Research Council grant [742158] (SCARABEE, Scalable inference algorithms for Bayesian evolutionary epidemiology).
RS-L and HdL were supported by Projects LISBOA-01-0145-FEDER-007660 (Microbiologia Molecular, Estrutural e Celular, funded by FEDER funds through COMPETE2020 - Programa Operacional Competitividade e Internacionalização (POCI) and LISBOA-01-0145-FEDER-016417 (ONEIDA project, co-funded by FEEI - "Fundos Europeus Estruturais e de Investimento" from "Programa Operacional Regional Lisboa 2020"). SK was supported by the Academy of Finland Flagship programme: Finnish Center for Artificial Intelligence, FCAI, by grant [292334].

\bibliographystyle{plainnat}
\bibliography{main_manuscript}

\clearpage

\appendix

\section*{Appendix}

\subsection*{Modularised rejection ABC}\label{sec:modularisedrejectionABC}

\begin{algorithm}[h!]
\SetAlgoLined
\KwResult{Posterior samples $\bm{\theta}$}
 observations $X$, prior $p(\bm{\theta})$, quantile $\kappa$, simulator $X_{\theta} \sim p(X|\bm{\theta})$ \;
 $i=0, \bm{\theta}=[\ ]$\;
\For{$i = 1\ldots n_{sim}$}{
   Sample $\bm{\theta}_{sim}^{i} \sim p(\bm{\theta})$ \;
   Sample ${X_{\bm{\theta}}^{i}} \sim p(X|\bm{\theta}_{sim}^{i})$ \;
}
 \For{$j = 1\ldots q$}{
   $\bm{d}_j = [\ ]$\;
\For{$i = 1\ldots n_{sim}$}{
Set $d^{i}_{j} = ||[\phi (X) ]_j- [\phi ({X_{\bm{\theta}}^{i}})]_j ||_2$  \% Discrepancy for subset j\;
  Append $d^{i}_{j}$ to $\bm{d}_j$ 
}
Find ascending order $s_j = \text{arg\ sort}(\bm{d}_j)$ \;
Select $a_j = s_j[1:\kappa n_{sim}]$ \;
Append $[\bm{\theta}]_{j} = [\bm{\theta}_{sim}]_{j}[,a_j]$  \;
}
 \caption{Rejection modular ABC algorithm as used in this article. Samples are selected by finding quantiles of discrepancies associated with distinct subsets of the parameters.
}\label{alg:modrejABC}
\end{algorithm}

\subsection*{Microbiological Study Design}
\label{sec:implementmicro}

The children occupied three rooms of the same daycare centre, and are here treated as a single population. Nasopharyngeal samples were taken eleven times between February 1998 to February 1999. Nasopharyngeal swabs were obtained and inoculated into agar media in order to select and identify \textit{Streptococcus pneumoniae}, \textit{Streptococcus pyogenes}, \textit{Staphylococcus aureus}, \textit{Moraxella catarrhalis} and \textit{Haemophilus influenzae}. \textit{S. pneumoniae} and \textit{H. influenzae} were selectively cultured in blood agar with gentamicin and chocolate blood agar containing iso-vitalex and bacitracin, respectively. \textit{S. aureus} was selectively cultured in mannitol-salt agar and \textit{S. pyogenes} and \textit{M. catarrhalis} were isolated from trypticase-soy blood agar plates. Pneumococcal identification was based on $\alpha$-hemolysis and optochin susceptibility; \textit{H. influenzae} were identified based on the requirement of X and V factors for growth; \textit{S. aureus} was identified on the basis of mannitol fermentation and positive coagulase test; \textit{S. pyogenes} was identified based on $\beta$-hemolysis and susceptibility to bacitracin and \textit{M. catarrhalis} was identified based on colony morphology and positive tributyrin test. Routinely, a single colony of each species was isolated, cultured and frozen. \textit{S. pneumococci} strains were serotyped by the Quellung reaction using commercially available antisera (Statens Serum Institut, Copenhagen, Denmark).

\subsection*{Example: Cholesky Precision Inference}
\label{sec:cholesky}

In this Example we pursue inference over a representation of the precision matrix $\Omega$ of a multivariate Gaussian, i.e. $X \sim \mathcal{N}(0, \Omega^{-1})$. We can consider this to be a simple form of causal inference using a Bayesian Network \citep{scutari2021bayesian}: assuming the data is joint Gaussian distributed and there are no unobserved confounders, the elements of the precision matrix can be interpreted as the partial correlations, indicating the causal associations between the variables in an implicit Directed Acyclic Graph (DAG), with potential confounding by other observed variables conditionally removed \cite{loh2014high}. Unlike the covariance elements, the individual off-diagonal elements of the precision matrix depend globally on all of the observed data variables.

In order to constrain the dimensionality of the problem, we consider pairwise subsets of the data dimensions. Without loss of generality, we order the dimensions and form subsets of consecutive pairs, i.e. $i$ and $i+1$ for even $i$. In this case, the estimation of the associated off-diagonal precision matrix element $\Omega_{i,i+1}$ is potentially biased, as every data dimension outside of $i, i+1$ becomes a potential confounder. Failing to account for the potential non-zero off-diagonal elements between subsets can be considered an example of misspecification, which we expect Split-BOLFI to identify and ameliorate with tempering.

In practice, we perform inference over the Cholesky decomposition $L$ of the precision matrix $\Omega$, defined as an lower triangular matrix solving $L L^{T} = \Omega$.
The decomposition allows for independent uniform priors over the elements of the matrix $L$, and no need to maintain positive-definiteness when exploring parameter values. In contrast to $\Omega$, $L$ is a triangular matrix whose non-zero elements can vary freely, except for the diagonal entries which must be non-negative.

When considering pairwise subsets of the data variables, we can expect the estimation of $L$ to be biased in a similar way to $\Omega$ by neglected cross-precisions between subsets, except that the definition of $L$ means that estimation of elements of $L$ at index $i$ are only biased by non-zero off-diagonal precision with index greater than $i$, i.e. potentially ignored cross-precisions will only influence Cholesky subsets of lower index. The downstream influence of cross-precisions has the effect of introducing varying levels of misspecification between subsets in order of the indexing, presenting an interesting challenge for inference procedures concerned with identifying or adjusting for misspecification.

When selecting summary statistics for inference for each subset of index $i$, $i+1$, then the variances $\sigma^2_i$, and $\sigma^2_{i+1}$ and correlations $r_{i,i+1}$ within the subset are clear choices.

In addition, the statistician may suspect that the model is misspecified by not representing the causal effect between the subset and dimension $j$, with $j>i+1$. In this case, it is possible to include correlations $r_{i,j}$ or $r_{i+1,j}$. These additional correlations should contain the information necessary to identify model misspecification, and then the Split-BOLFI should introduce tempering to avoid an overly confident update.

In this instance, it would be possible to perform inference using the likelihood without using subsets. However, this example is illustrative for demonstrating the relevant features of conditional independence, scalability, and differing levels of misspecification in modular inference. This treatment of conditional dependence between modules offers an alternative to the Cut-Bayesian or similar frameworks that target the conditional distribution directly.

\begin{figure}[h]
    \centering
    \begin{subfigure}[b]{0.44\textwidth}
        \includegraphics[width=\textwidth]{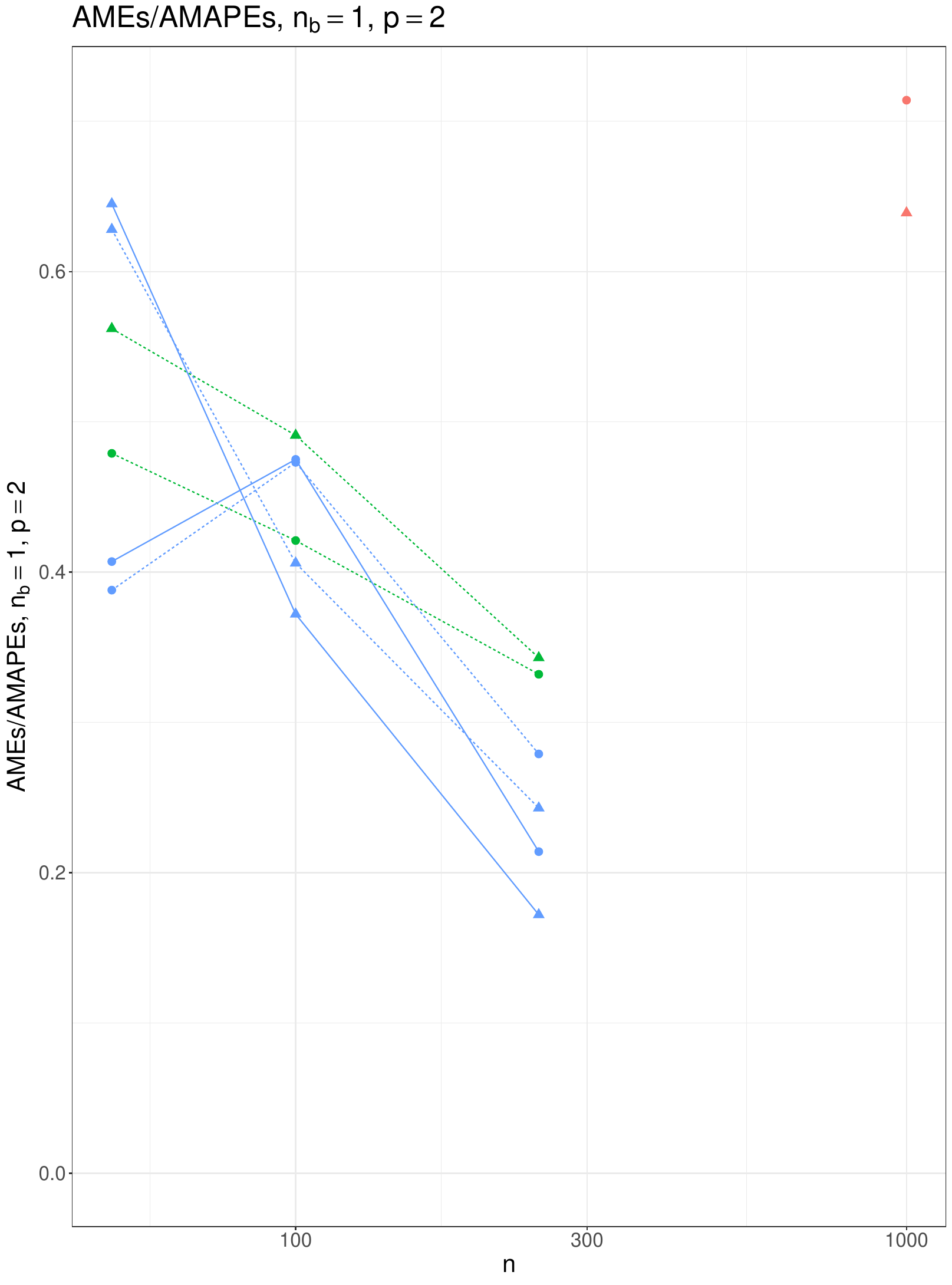}
    \end{subfigure}
    ~ 
    \begin{subfigure}[b]{0.44\textwidth}
        \includegraphics[width=\textwidth]{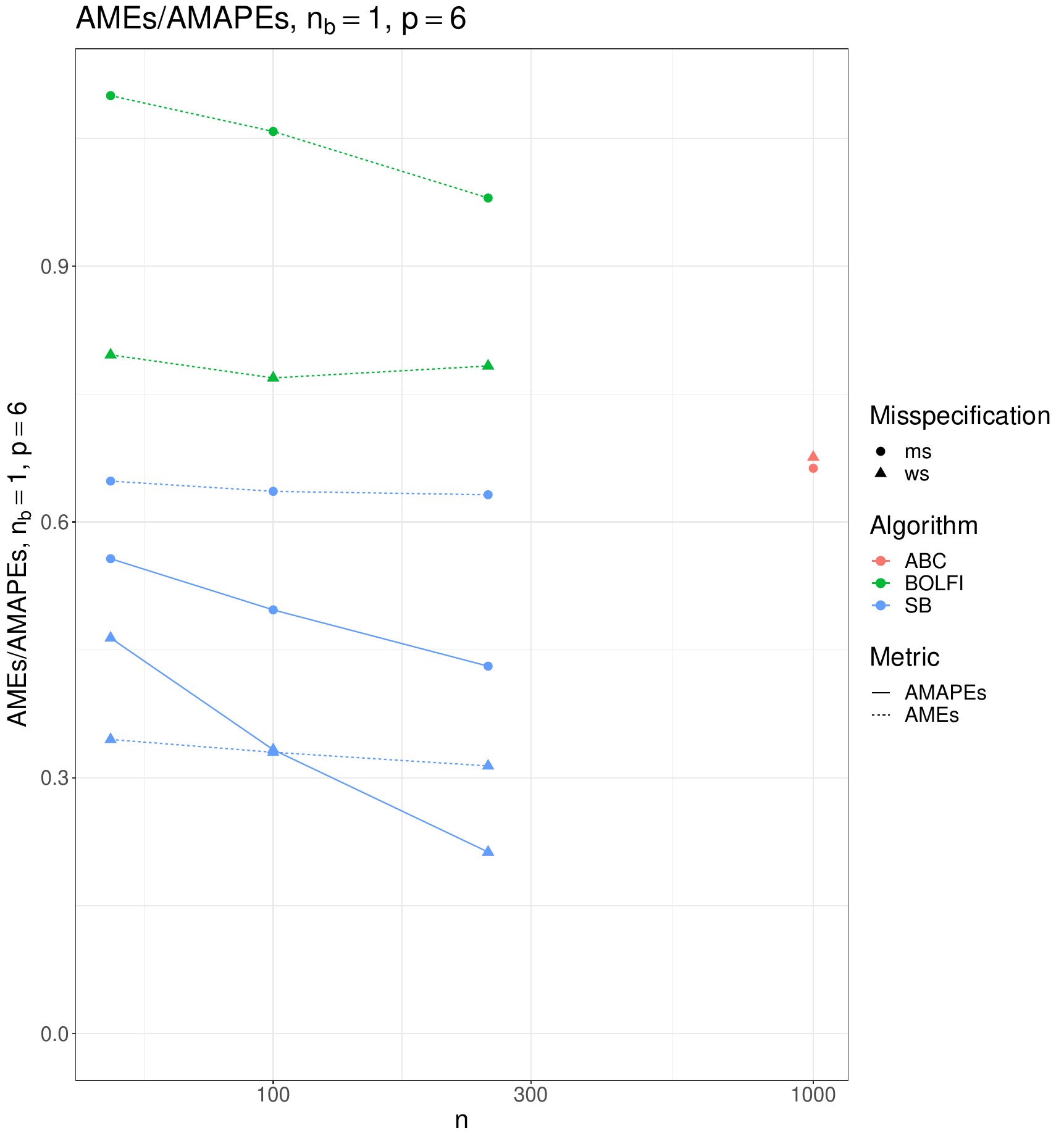}
    \end{subfigure}
    
    \begin{subfigure}[b]{0.44\textwidth}
        \includegraphics[width=\textwidth]{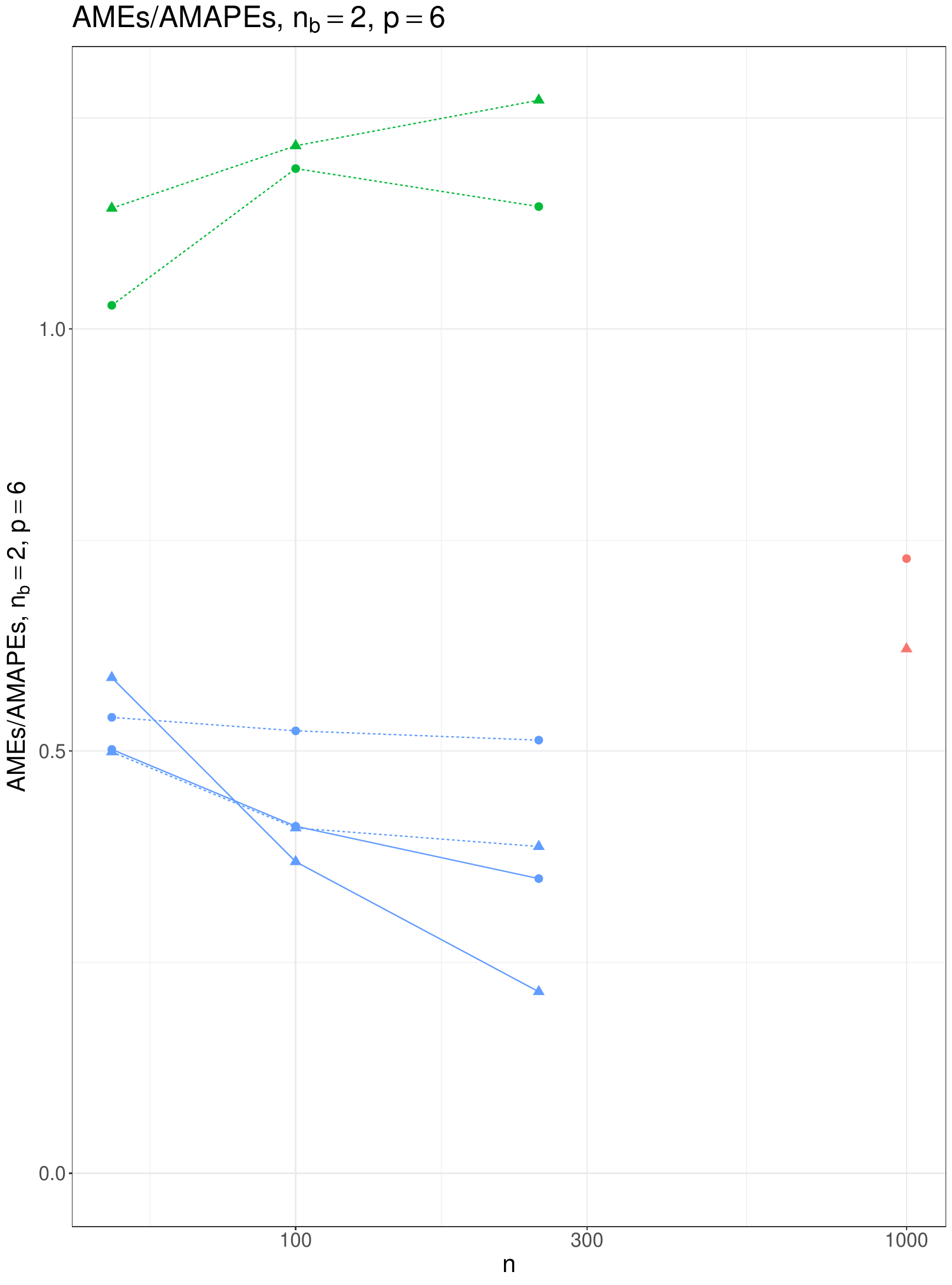}
    \end{subfigure}
    ~ 
    \begin{subfigure}[b]{0.44\textwidth}
        \includegraphics[width=\textwidth]{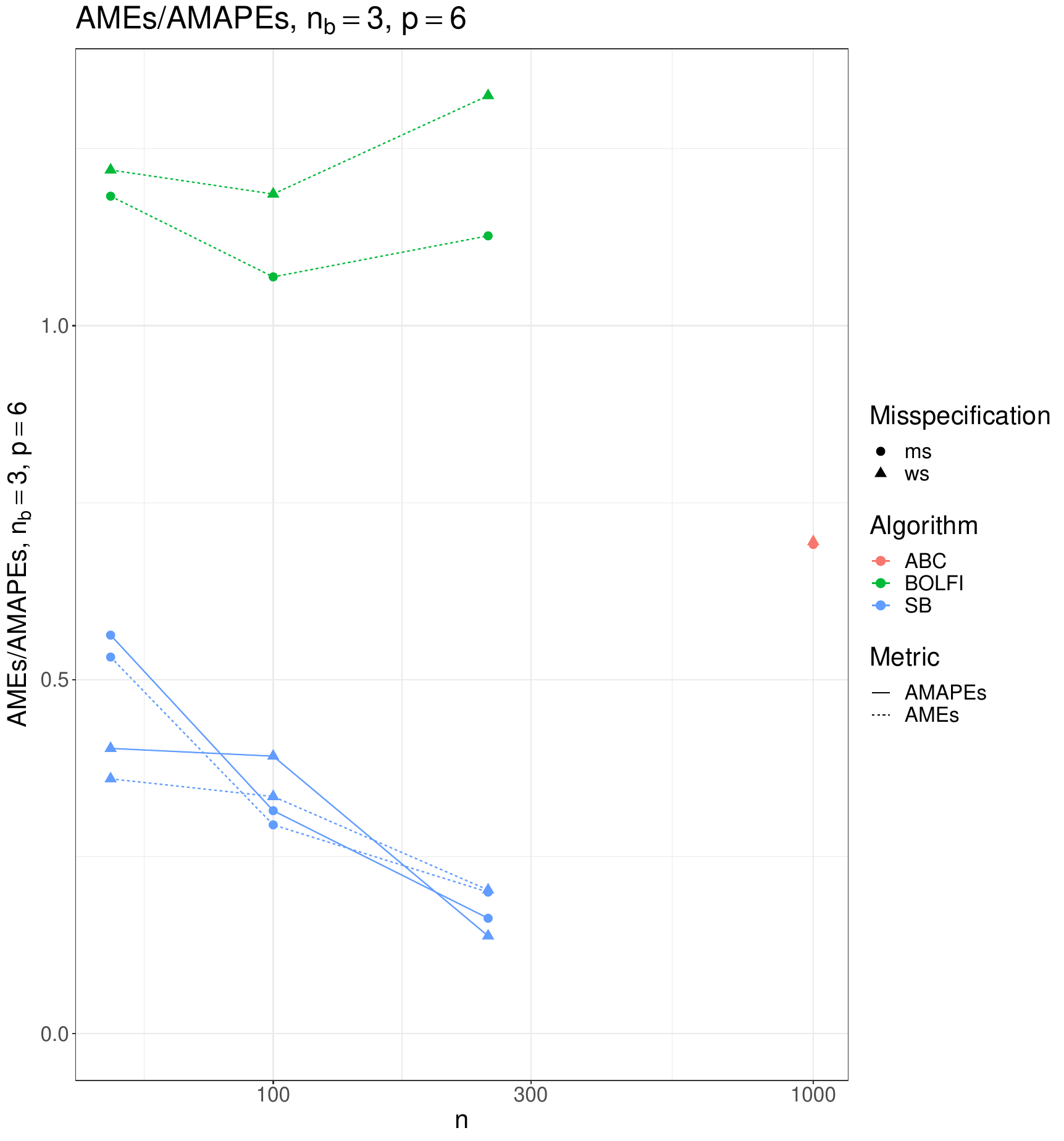}
    \end{subfigure}
    \caption{Results for a pairwise blocky Cholesky precision model with $n_{obs}=500$, as a function of number of simulations $n_{sim}$. Diagrams vary by dimensionality $p$ equal to 2 or 6, producing  1 or 3 blocks in each case, indexed by $n_b$.  Plotted are AME and AMAPE, varying model specification, with algorithms ABC, Split-BOLFI, and BOLFI. .}\label{fig:cholesky_results_appendix_new}
\end{figure}

The results are presented in \Cref{CHOLESKYAME,CHOLESKYRMSE,CHOLESKYSD}, and visually in Figure \ref{fig:cholesky_results_appendix_new}. The well-specified and misspecified cases in the $p=2$ case are essentially equivalent, and as such present similar behaviour.

\begin{table}[!h]
\centering
\begin{tabular}{cc|ccc}
& & $\text{ABC}$ & $\text{SB}$ & $\text{BOLFI}$ \\ \hline
$p=2$ & $n_{b}=1$ & 0.64 (0.49) & 0.17 (0.39) & 0.15 (0.14) \\ \hline
\multirow{3}{*}{ $p=6$ } & $n_{b}=1$ & 0.68 (0.45) & 0.21 (0.2) & 0.80 (0.42) \\
& $n_{b}=2$ & 0.62 (0.46) & 0.22 (0.21) & 0.90 (0.77) \\
& $n_{b}=3$ & 0.70 (0.46) & 0.14 (0.12) & 0.89 (0.81) \\ \hline \hline
$p=2$ & $n_{b}=1$ & 0.71 (0.56) & 0.21 (0.43) & 0.17 (0.16) \\ \hline
\multirow{3}{*}{ $p=6$ } & $n_{b}=1$ & 0.66 (0.49) & 0.43 (0.40) & 0.81 (0.53) \\
& $n_{b}=2$ & 0.73 (0.52) & 0.35 (0.31) & 0.85 (0.66) \\
& $n_{b}=3$ & 0.69 (0.56) & 0.16 (0.16) & 0.85 (0.72) \\
\end{tabular}
\caption{Results for a pairwise blocky Cholesky precision model, in the well-specified case (above double line) and misspecified case (below double line). We consider data dimensionalities $p$ equal to 2 and 6, producing  1 or 3 blocks in each case, indexed by $n_b$. Algorithms are ABC, Split-BOLFI, and BOLFI.  Results of the mean(sd) of AME and AMAPE with a data set of 500 observations. Error is based on the point estimate which was sample mean for ABC and BOLFI and MAP for Split-BOLFI.}
\label{CHOLESKYAME}
\end{table}

\begin{table}[!h]
\centering
\begin{tabular}{cc|ccc}
& & $\text{ABC}$ & $\text{SB}$ & $\text{BOLFI}$ \\ \hline
$p=2$ & $n_{b}=1$ & 0.8 (0.41) & 0.55 (0.32) & 0.36 (0.17)  \\ \hline
\multirow{3}{*}{ $p=6$ } & $n_{b}=1$ & 0.82 (0.38) & 0.82 (0.14) & 0.96 (0.36)\\
& $n_{b}=2$ & 0.77 (0.40) & 0.81 (0.16) & 1.37 (0.72) \\
& $n_{b}=3$ & 0.83 (0.39) & 0.52 (0.15) & 1.39 (0.79)\\ \hline \hline
$p=2$ & $n_{b}=1$ &0.89 (0.47) & 0.58 (0.39) & 0.4 (0.2)  \\ \hline
\multirow{3}{*}{ $p=6$ } & $n_{b}=1$ & 0.84 (0.40) & 1.27 (0.46) & 1.01 (0.47) \\
& $n_{b}=2$ & 0.90 (0.43) & 1.17 (0.36) & 1.33 (0.67) \\
& $n_{b}=3$ & 0.87 (0.47) & 0.52 (0.16) & 1.34 (0.67) \\
\end{tabular}
\caption{Results for a pairwise blocky Cholesky precision model, in the well-specified case (above double line) and misspecified case (below double line). We consider data dimensionalities $p$ equal to 2 and 6, producing  1 or 3 blocks in each case, indexed by $n_b$. Algorithms are ABC, Split-BOLFI, and BOLFI.  Results of the mean(sd) of RMSE with a data set of 500 observations. Error is based on the point estimate which was sample mean for ABC and BOLFI and MAP for Split-BOLFI.}
\label{CHOLESKYRMSE}
\end{table}

\begin{table}[!h]
\centering
\begin{tabular}{cc|ccc}
& & $\text{ABC}$ & $\text{SB}$ & $\text{BOLFI}$ \\ \hline
$p=2$ & $n_{b}=1$ & 0.38 (0.1) & 0.50 (0.12) & 0.31 (0.14)  \\ \hline
\multirow{3}{*}{ $p=6$ } & $n_{b}=1$ & 0.38 (0.1) & 0.77 (0.09) & 0.46 (0.16)\\
& $n_{b}=2$ & 0.38 (0.1) & 0.71 (0.07) & 0.93 (0.36) \\
& $n_{b}=3$ & 0.38 (0.1) & 0.47 (0.10) & 0.96 (0.42)\\ \hline \hline
$p=2$ & $n_{b}=1$ & 0.42 (0.14) & 0.53 (0.23) & 0.34 (0.15)   \\ \hline
\multirow{3}{*}{ $p=6$ } & $n_{b}=1$ & 0.41 (0.13) & 1.27 (0.33) & 0.52 (0.18) \\
& $n_{b}=2$ & 0.41 (0.13) & 1.20 (0.307) & 0.94 (0.41) \\
& $n_{b}=3$ & 0.41 (0.13) & 0.47 (0.09) & 0.94 (0.35)  \\
\end{tabular}
\caption{Results for a pairwise blocky Cholesky precision model, in the well-specified case (above double line) and misspecified case (below double line). We consider data dimensionalities $p$ equal to 2 and 6, producing  1 or 3 blocks in each case, indexed by $n_b$. Algorithms are ABC, Split-BOLFI, and BOLFI.  Results of the mean(sd) of SD with a data set of 500 observations. Error is based on the point estimate which was sample mean for ABC and BOLFI and MAP for Split-BOLFI.}
\label{CHOLESKYSD}
\end{table}

\begin{table}[!h]
\centering
\begin{tabular}{cc|ccc}
& & $\text{ABC}$ & $\text{SB}$ & $\text{BOLFI}$ \\ \hline
$p=2$ &$n_{b}=1$ &0.24 (0.43) &0.82 (0.48) &0.64 (0.48) \\ \hline
\multirow{3}{*}{$p=6$} &$n_{b}=1$ &0.15 (0.35) &0.85 (0.35) &0.15 (0.35) \\
&$n_{b}=2$ &0.14 (0.35) &0.69 (0.48) &0.35 (0.48) \\
&$n_{b}=3$ &0.12 (0.33) &0.83 (0.48) &0.37 (0.48) \\ \hline \hline
$p=2$ &$n_{b}=1$ &0.25 (0.44) &0.83 (0.46) &0.7 (0.46) \\ \hline
\multirow{3}{*}{$p=6$} &$n_{b}=1$ &0.19 (0.39) &0.72 (0.4) &0.19 (0.4) \\
&$n_{b}=2$ &0.18 (0.38) &0.84 (0.48) &0.37 (0.48) \\
&$n_{b}=3$ &0.19 (0.4) &0.85 (0.49) &0.42 (0.49) \\
\end{tabular}
\caption{Results for a pairwise blocky Cholesky precision model, in the well-specified case (above double line) and misspecified case (below double line). We consider data dimensionalities $p$ equal to 2 and 6, producing  1 or 3 blocks in each case, indexed by $n_b$. Algorithms are ABC, Split-BOLFI, and BOLFI.  Results of the mean(sd) of 50\% coverage with a data set of 500 observations. Error is based on the point estimate which was sample mean for ABC and BOLFI and MAP for Split-BOLFI.}
\label{CHOLESKYCOVERAGE}
\end{table}

In the well-specified case in \Cref{CHOLESKYAME,CHOLESKYRMSE,CHOLESKYSD}, there are no cross-precisions between groups in the generative model: we consequently expect and mostly observe consistent statistics between dimensionalities and blocks. We observe that the Split-BOLFI approach achieves the most accurate point estimate, while adding more posterior breadth than rejection ABC, resulting in a larger RMSE. BOLFI performs reasonably well in the $p=2$ case but fails in the higher-dimensional case. There is some curious variation between subsets in the $p=6$ case, with BOLFI seemingly performing best in the $n_{b} = 1$ case, and Split-BOLFI in the $n_{b}=3$ case: there is no theoretical reason for this to be the case, and may be a sign that the number of observations is small enough for finite-sample effects to become evident.

In the misspecified case in \Cref{CHOLESKYAME,CHOLESKYRMSE,CHOLESKYSD} we again observe that Split-BOLFI provides the most accurate point estimations, with posterior variance proportionally inflated in the $n{b} = $ 1 and 2 blocks that are most influenced by the between-block precision elements. We see no such structure in the ABC of BOLFI results, which respectively show approximately constant or even opposite trends with misspecification in accuracy and posterior uncertainty.

In the visually represented results in Figure \ref{fig:cholesky_results_appendix_new}, we see that Split-BOLFI makes efficient use of increasing number of acquisitions in every case, whereas BOLFI only appears to make effective use of extra acquisitions in the $p=2$ situation.

The 50\% coverage metric in \Cref{CHOLESKYCOVERAGE} suggests that ABC shows consistently lower than 50\% coverage, Split-BOFLI exhibits somewhat higher than 50\% coverage, and BOLFI exhibits lower than 50\% coverage on the high dimensional problem, and higher than 50\% coverage on the lower-dimensional problem. We do not observe substantial variation in coverage in any of the methods between the well and misspecified conditions.

\begin{figure}[h]
    \centering
    \begin{subfigure}[b]{0.2\textwidth}
        \includegraphics[width=\textwidth]{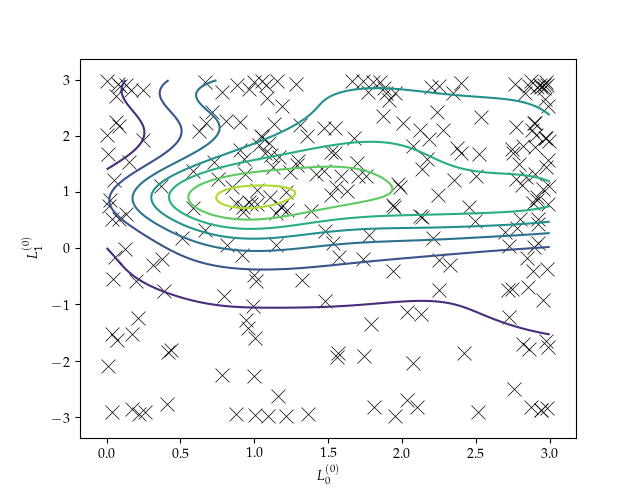}
    \end{subfigure}
    ~ 
    \begin{subfigure}[b]{0.2\textwidth}
        \includegraphics[width=\textwidth]{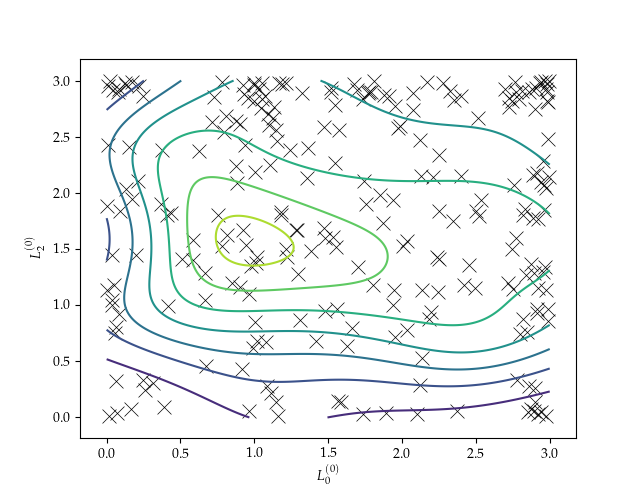}
    \end{subfigure}
    ~ 
    \begin{subfigure}[b]{0.2\textwidth}
        \includegraphics[width=\textwidth]{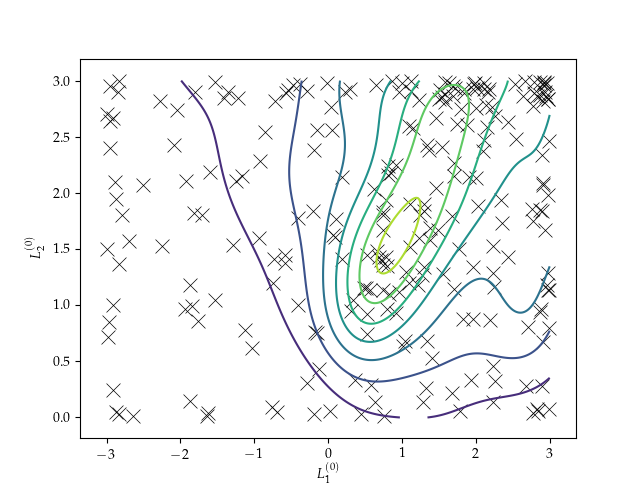}
    \end{subfigure}

    \begin{subfigure}[b]{0.2\textwidth}
        \includegraphics[width=\textwidth]{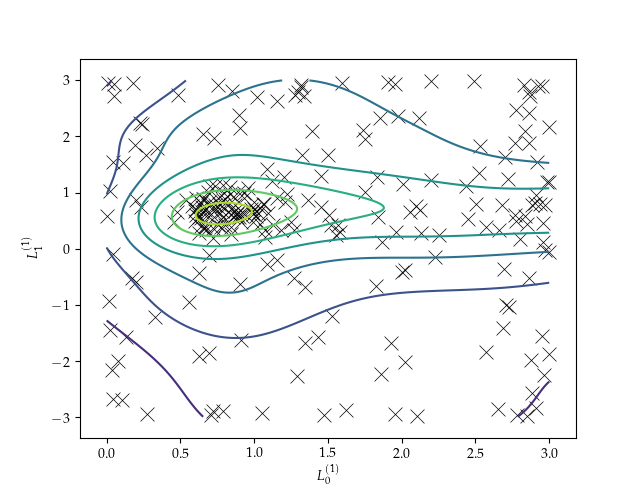}
    \end{subfigure}
    ~ 
    \begin{subfigure}[b]{0.2\textwidth}
        \includegraphics[width=\textwidth]{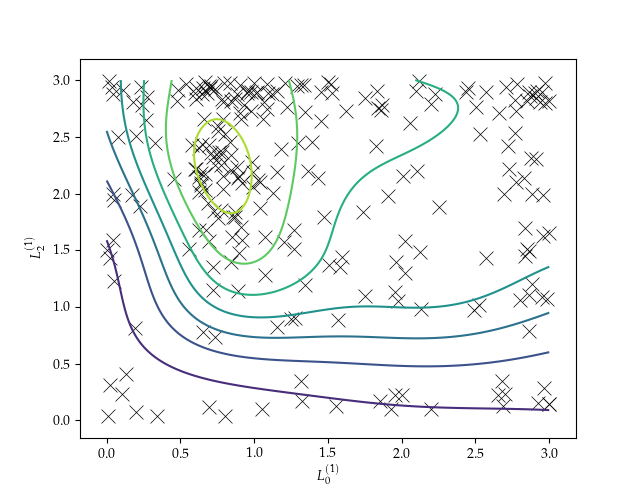}
    \end{subfigure}
    ~ 
    \begin{subfigure}[b]{0.2\textwidth}
        \includegraphics[width=\textwidth]{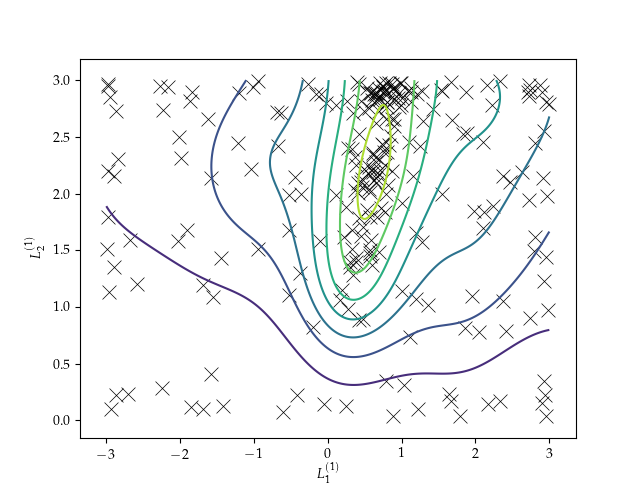}
    \end{subfigure}

    \begin{subfigure}[b]{0.2\textwidth}
        \includegraphics[width=\textwidth]{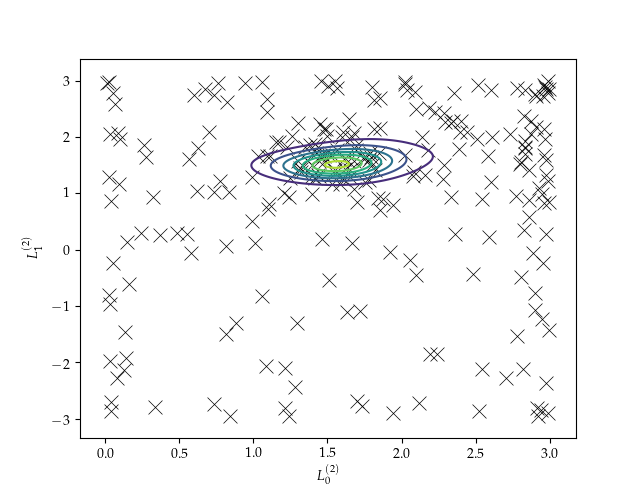}
    \end{subfigure}
    ~ 
    \begin{subfigure}[b]{0.2\textwidth}
        \includegraphics[width=\textwidth]{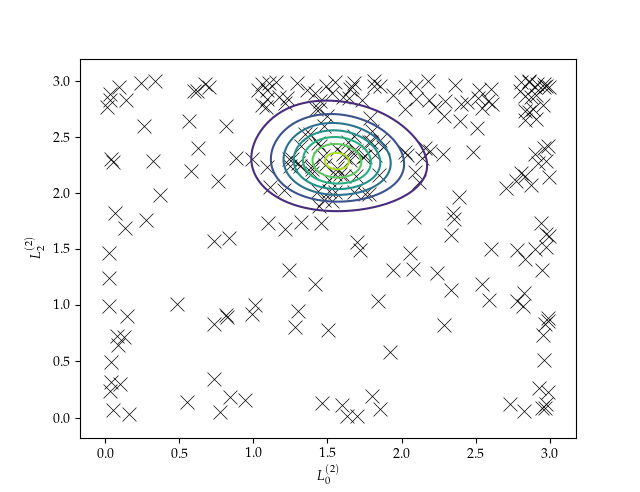}
    \end{subfigure}
    ~ 
    \begin{subfigure}[b]{0.2\textwidth}
        \includegraphics[width=\textwidth]{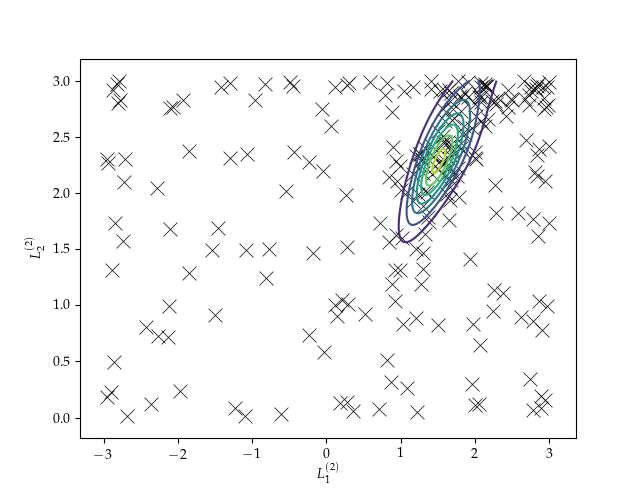}
    \end{subfigure}

    \begin{subfigure}[b]{0.2\textwidth}
        \includegraphics[width=\textwidth]{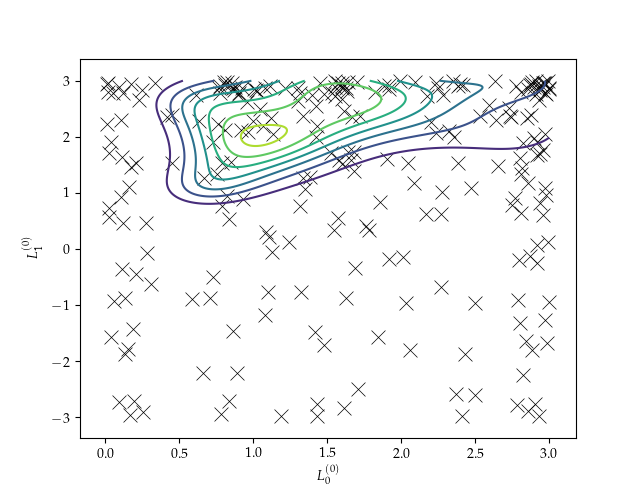}
    \end{subfigure}
    ~ 
    \begin{subfigure}[b]{0.2\textwidth}
        \includegraphics[width=\textwidth]{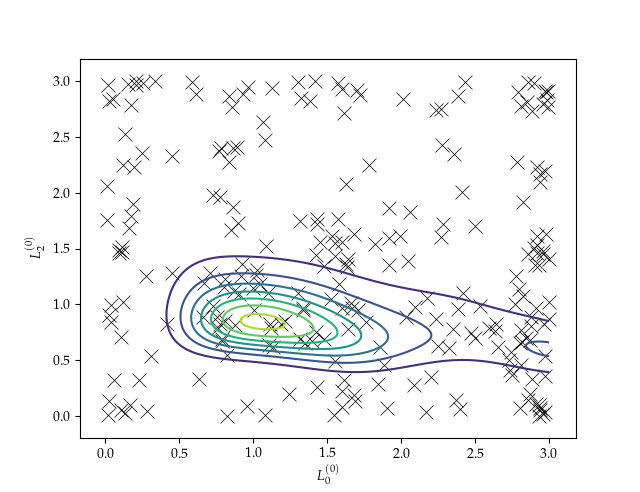}
    \end{subfigure}
    ~ 
    \begin{subfigure}[b]{0.2\textwidth}
        \includegraphics[width=\textwidth]{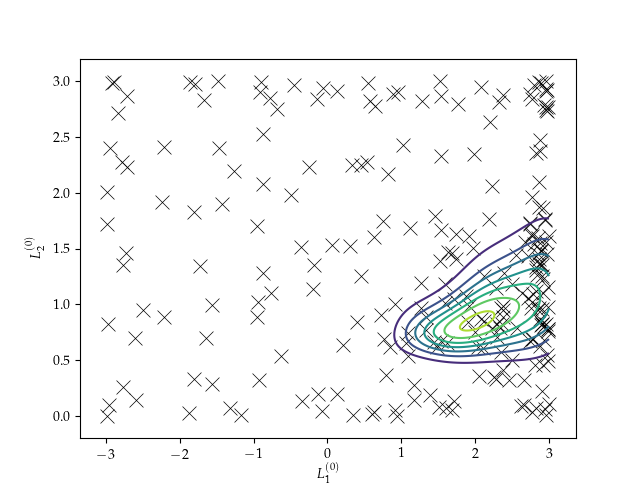}
    \end{subfigure}

    \begin{subfigure}[b]{0.2\textwidth}
        \includegraphics[width=\textwidth]{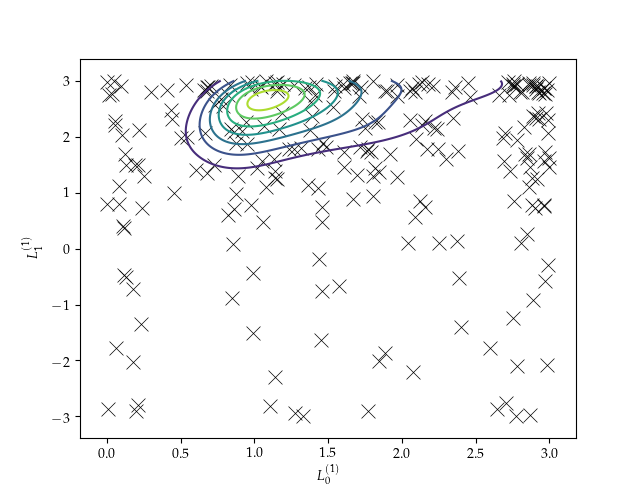}
    \end{subfigure}
    ~ 
    \begin{subfigure}[b]{0.2\textwidth}
        \includegraphics[width=\textwidth]{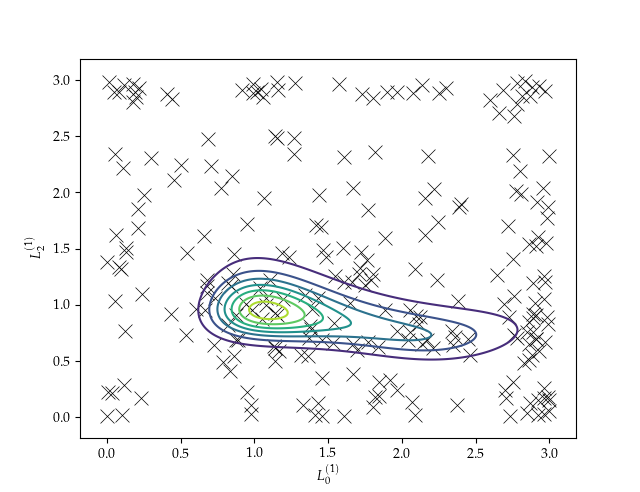}
    \end{subfigure}
    ~ 
    \begin{subfigure}[b]{0.2\textwidth}
        \includegraphics[width=\textwidth]{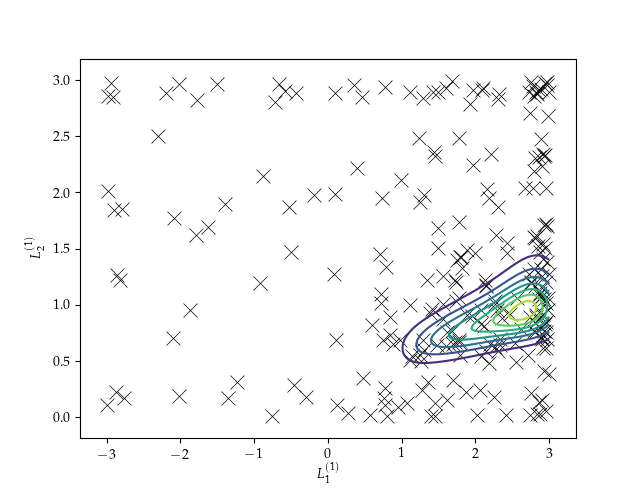}
    \end{subfigure}

    \begin{subfigure}[b]{0.2\textwidth}
        \includegraphics[width=\textwidth]{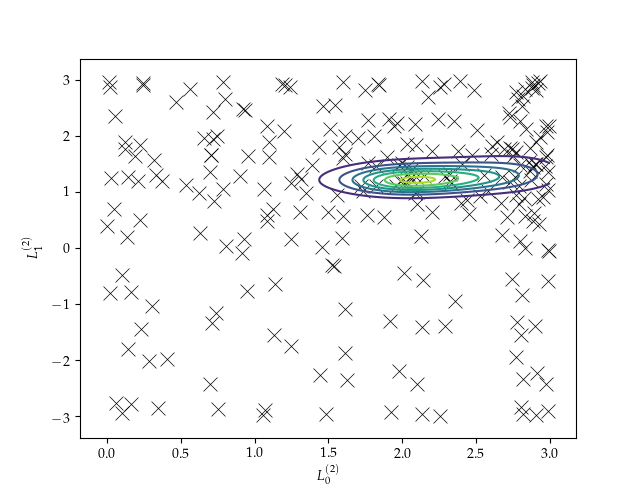}
    \end{subfigure}
    ~ 
    \begin{subfigure}[b]{0.2\textwidth}
        \includegraphics[width=\textwidth]{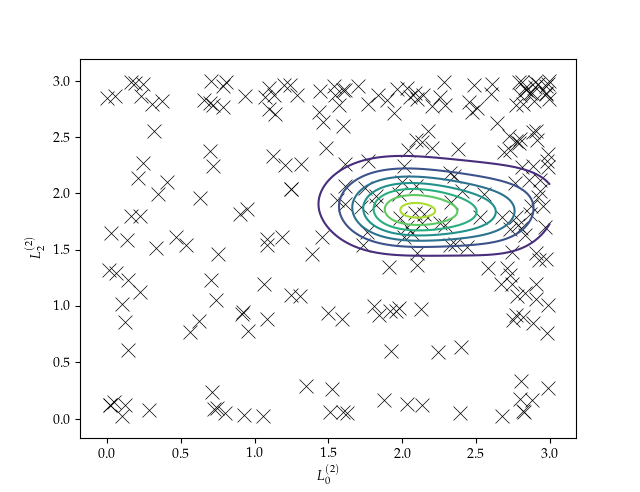}
    \end{subfigure}
    ~ 
    \begin{subfigure}[b]{0.2\textwidth}
        \includegraphics[width=\textwidth]{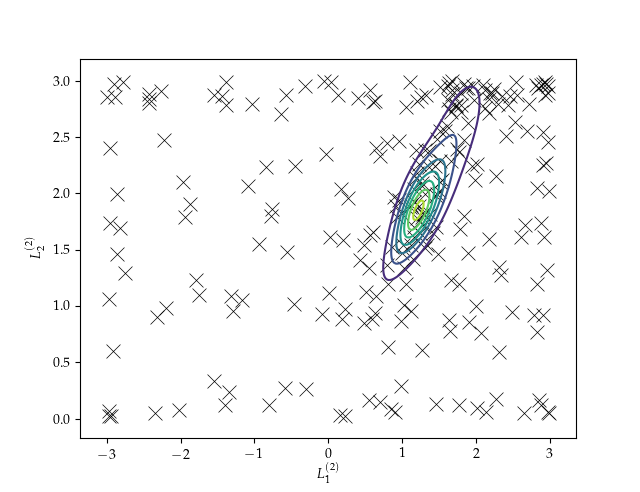}
    \end{subfigure}
    \caption{Proxy posteriors contours with $n_\text{acq}=250$ black acqusition points for each pair of precision Cholesky elements for each block from the example described in Section \ref{sec:cholesky}, with $n_\text{obs}=5000$. The top three rows  represent a misspecified model, and those below are performed on a well-specified model. Each row corresponds to different blocks in the blockwise covariance model.}\label{fig:cholesky}
\end{figure}

\subsection*{Supplementary Information}

\begin{table}[]
    \centering
    \begin{tabular}{c|c}
      metric   &  formula \\ \hline
      $\text{RMSE}^i$ &  $\sqrt{ \frac{1}{n_{j}} \sum_j ( \theta_j^i -  \theta_T^i )^2 }$ \\
      $\text{AME}^i$ & $\lvert \theta_T^i - \sum_j \theta_j^i \rvert$ \\
      $\text{AMAPE}^i$ & $\lvert \theta_T^i - \theta_{MAP}^i\rvert$ \\
      $\text{SD}^i$ & $\sqrt{ \frac{1}{n_{j}} \sum_j ( \theta_j^i -  ( \frac{1}{n_j} \sum_k \theta_k^i ) )^2 }$ \\ 
      $\text{50\% coverage}^i$ & $( \theta_T^i > Q1[\theta_j^i] ) ( \theta_T^i < Q3[\theta_j^i] )$  \\ \hline
     mean(metric)   &  $\frac{1}{n_i} \sum_i \text{metric}^i$ \\
     sd(metric)  & $\sqrt{\frac{1}{n_i} \sum_i (\text{metric}^i -  \text{mean(metric)} )^2}$\\
    \end{tabular}
    \caption{Definitions of the various performance metrics used. Performance metrics were computed for each random seed, then means and sds were computed across random seeds. AMAPE was only computed for Split-BOLFI, as this was the only method for which $\theta_{MAP}^i$ were recorded. Posterior samples $\theta_j^i$ are indexed by $n_i$ random seeds $i$ and $n_j$ within-posterior sample indices $j$. True generative values $\theta_T^i$ are indexed by random seed $i$. $Q1[\theta_j^i]$ and $Q3[\theta_j^i]$ are respectively the first and third quartiles of the $i$th posterior sample $\theta_j^i$.}
    \label{tab:metric_definition}
\end{table}

\begin{table}[!h]
\centering
\begin{tabular}{c|ccccc}
 & $\text{ABC}_{m,s}$ & $\text{ABC}_{m,s,k}$ & $\text{SB}_{m,s}$ & $\text{SB}_{m,s,k}$ & $\text{BOLFI}_{m,s}$ \\\hline
$\mu_1$ & 0.12 (0.65) & 0.16 (0.81) & 0.1 (0.08) & 0.09 (0.08) & 0.09 (0.07) \\
$\sigma_1$ & 0.05 (0.26) & 0.05 (0.26) & 0.06 (0.04) & 0.06 (0.04) & 0.05 (0.04) \\
$\mu_5$ & 0.14 (0.74) & 0.12 (0.67) & 0.09 (0.08) & 0.1 (0.08) & 1.82 (1.48) \\
$\sigma_5$ & 0.05 (0.27) & 0.05 (0.28) & 0.06 (0.06) & 0.07 (0.06) & 0.68 (0.47) \\ \hline
$\mu_1$ & 0.13 (0.7) & 0.12 (0.64) & 0.1 (0.09) & 0.1 (0.09) & 0.1 (0.08) \\
$\sigma_1$ & 0.06 (0.29) & 0.04 (0.24) & 0.09 (0.08) & 0.09 (0.08) & 0.09 (0.08) \\
$\mu_5$ & 0.13 (0.7) & 0.13 (0.7) & 0.09 (0.07) & 0.09 (0.07) & 1.88 (1.44) \\
$\sigma_5$ & 0.05 (0.27) & 0.05 (0.28) & 0.1 (0.09) & 0.1 (0.09) & 0.71 (0.44) \\
\end{tabular}
\caption{Results for a Gaussian statistical model with parameters $\mu_p$ and $\sigma_p$ in the well-specified case (above line) and misspecified case (below line) for $p=1$ and $p=5$. Algorithms are ABC, Split-BOLFI, and BOLFI using mean and standard deviation summaries $m,s$ and additionally kurtosis summary $m,s,k$. Results of the mean(sd) of AME or AMAPE with a data set of 500 observations. Error is based on the point estimate which was sample mean for ABC and BOLFI and MAP for Split-BOLFI.}
\label{GAUSSIAN500AME}
\end{table}

\begin{table}[!h]
\centering
\begin{tabular}{c|ccccc}
 & $\text{ABC}_{m,s}$ & $\text{ABC}_{m,s,k}$ & $\text{SB}_{m,s}$ & $\text{SB}_{m,s,k}$ & $\text{BOLFI}_{m,s}$ \\\hline
$\mu_1$ & 0.12 (0.66) & 0.16 (0.82) & 0.4 (0.07) & 0.71 (0.15) & 0.2 (0.06) \\
$\sigma_1$ & 0.05 (0.26) & 0.05 (0.27) & 0.38 (0.06) & 0.64 (0.1) & 0.18 (0.05) \\
$\mu_5$ & 0.14 (0.74) & 0.13 (0.67) & 0.38 (0.38) & 0.72 (0.15) & 2.88 (1.38) \\
$\sigma_5$ & 0.05 (0.28) & 0.06 (0.29) & 0.36 (0.36) & 0.65 (0.11) & 1.39 (0.31) \\
\hline
$\mu_1$ & 0.13 (0.7) & 0.12 (0.64) & 0.39 (0.07) & 2.39 (0.32) & 0.21 (0.07) \\
$\sigma_1$ & 0.06 (0.3) & 0.05 (0.25) & 0.38 (0.05) & 1.4 (0.14) & 0.21 (0.07) \\
$\mu_5$ & 0.13 (0.7) & 0.13 (0.71) & 0.38 (0.08) & 2.4 (0.4) & 2.92 (1.38) \\
$\sigma_5$ & 0.05 (0.27) & 0.06 (0.29) & 0.38 (0.08) & 1.39 (0.14) & 1.41 (0.29) \\
\end{tabular}
\caption{Results for a Gaussian statistical model with parameters $\mu_p$ and $\sigma_p$ in the well-specified case (above line) and misspecified case (below line) for $p=1$ and $p=5$. Algorithms are ABC, Split-BOLFI, and BOLFI using mean and standard deviation summaries $m,s$ and additionally kurtosis summary $m,s,k$. Results of the mean(sd) of RMSE with a data set of 500 observations. Error is based on the point estimate which was sample mean for ABC and BOLFI and MAP for Split-BOLFI.}
\label{GAUSSIAN500RMSE}
\end{table}

\begin{table}[!h]
\centering
\begin{tabular}{c|ccccc}
 & $\text{ABC}_{m,s}$ & $\text{ABC}_{m,s,k}$ & $\text{SB}_{m,s}$ & $\text{SB}_{m,s,k}$ & $\text{BOLFI}_{m,s}$ \\\hline
$\mu_1$ & 0.01 (0.05) & 0.01 (0.06) & 0.38 (0.06) & 0.7 (0.15) & 0.17 (0.04) \\
$\sigma_1$ & 0.01 (0.05) & 0.01 (0.06) & 0.38 (0.06) & 0.64 (0.1) & 0.17 (0.04) \\
$\mu_5$ & 0.01 (0.05) & 0.01 (0.06) & 0.36 (0.06) & 0.71 (0.15) & 2.02 (0.81) \\
$\sigma_5$ & 0.01 (0.05) & 0.01 (0.05) & 0.36 (0.06) & 0.65 (0.1) & 1.13 (0.27) \\
\hline
$\mu_1$ & 0.01 (0.05) & 0.02 (0.07) & 0.37 (0.06) & 2.2 (0.23) & 0.17 (0.05) \\
$\sigma_1$ & 0.01 (0.05) & 0.02 (0.06) & 0.37 (0.05) & 1.29 (0.06) & 0.18 (0.05) \\
$\mu_5$ & 0.01 (0.05) & 0.02 (0.07) & 0.37 (0.07) & 2.21 (0.26) & 2.05 (0.81) \\
$\sigma_5$ & 0.01 (0.05) & 0.02 (0.06) & 0.36 (0.06) & 1.29 (0.07) & 1.13 (0.29) \\
\end{tabular}
\caption{Results for a Gaussian statistical model with parameters $\mu_p$ and $\sigma_p$ in the well-specified case (above line) and misspecified case (below line) for $p=1$ and $p=5$. Algorithms are ABC, Split-BOLFI, and BOLFI using mean and standard deviation summaries $m,s$ and additionally kurtosis summary $m,s,k$. Results of the mean(sd) of SD with a data set of 500 observations. Error is based on the point estimate which was sample mean for ABC and BOLFI and MAP for Split-BOLFI.}
\label{GAUSSIAN500SD}
\end{table}

\begin{table}[!h]
\centering
\begin{tabular}{c|ccccc}
 & $\text{ABC}_{m,s}$ & $\text{ABC}_{m,s,k}$ & $\text{SB}_{m,s}$ & $\text{SB}_{m,s,k}$ & $\text{BOLFI}_{m,s}$ \\\hline
$\mu_1$ &0 (0.05) &0 (0) &0.94 (0.24) &0.98 (0.14) &0.74 (0.44) \\
$\sigma_1$ &0.01 (0.08) &0.01 (0.07) &1 (0) &1 (0) &0.92 (0.27) \\
$\mu_5$ &0 (0.05) &0 (0.05) &0.94 (0.25) &0.99 (0.09) &0.5 (0.5) \\
$\sigma_5$ &0.01 (0.07) &0.01 (0.07) &0.99 (0.09) &1 (0) &0.64 (0.48) \\
\hline
$\mu_1$ &0 (0.03) &0 (0.05) &0.94 (0.24) &0.96 (0.2) &0.68 (0.47) \\
$\sigma_1$ &0 (0.05) &0.01 (0.1) &0.94 (0.24) &0.98 (0.14) &0.76 (0.43) \\
$\mu_5$ &0 (0.05) &0 (0.04) &0.94 (0.23) &0.95 (0.22) &0.49 (0.5) \\
$\sigma_5$ &0.01 (0.08) &0.01 (0.08) &0.89 (0.31) &1 (0.06) &0.65 (0.48) \\
\end{tabular}
\caption{Results for a Gaussian statistical model with parameters $\mu_p$ and $\sigma_p$ in the well-specified case (above line) and misspecified case (below line) for $p=1$ and $p=5$. Algorithms are ABC, Split-BOLFI, and BOLFI using mean and standard deviation summaries $m,s$ and additionally kurtosis summary $m,s,k$. Results of the mean(sd) of 50\% coverage with a data set of 500 observations. Error is based on the point estimate which was sample mean for ABC and BOLFI and MAP for Split-BOLFI.}
\label{GAUSSIAN500COVERAGE}
\end{table}

\begin{table}[!h]
\centering
\begin{tabular}{c|ccccc}
 & $\text{ABC}_{m,s}$ & $\text{ABC}_{m,s,k}$ & $\text{SB}_{m,s}$ & $\text{SB}_{m,s,k}$ & $\text{BOLFI}_{m,s}$ \\\hline
$\mu_1$ &0 (0) &0 (0.05) &1 (0) &1 (0) &0.92 (0.27) \\
$\sigma_1$ &0.01 (0.09) &0.01 (0.07) &1 (0) &1 (0) &1 (0) \\
$\mu_5$ &0 (0.02) &0 (0.03) &1 (0.06) &1 (0) &0.53 (0.5) \\
$\sigma_5$ &0 (0.06) &0.01 (0.07) &1 (0) &1 (0) &0.63 (0.48) \\
\hline
$\mu_1$ &0 (0.06) &0.05 (0.22) &1 (0) &0.96 (0.2) &0.96 (0.2) \\
$\sigma_1$ &0.01 (0.07) &0.05 (0.22) &1 (0) &1 (0) &0.9 (0.3) \\
$\mu_5$ &0 (0.04) &0.05 (0.22) &0.99 (0.09) &0.94 (0.23) &0.6 (0.49) \\
$\sigma_5$ &0.01 (0.08) &0.05 (0.22) &1 (0) &0.99 (0.09) &0.57 (0.5) \\
\end{tabular}
\caption{Results for a Gaussian statistical model with parameters $\mu_p$ and $\sigma_p$ in the well-specified case (above line) and misspecified case (below line) for $p=1$ and $p=5$. Algorithms are ABC, Split-BOLFI, and BOLFI using mean and standard deviation summaries $m,s$ and additionally kurtosis summary $m,s,k$. Results of the mean(sd) of 50\% coverage with a data set of 5000 observations. Error is based on the point estimate which was sample mean for ABC and BOLFI and MAP for Split-BOLFI.}
\label{GAUSSIAN5000COVERAGE}
\end{table}

\begin{table}[!h]
\centering
\begin{tabular}{c|cc}
& BOLFI & SB  \\ \hline
$\beta$ &0 (0) &1 (0) \\
$\Lambda$ &0.44 (0.5) &1 (0) \\
Avg. $\theta$ &0 (0) &0 (0) \\  \hline \hline
$\beta$ &0.8 (0.4) &0.96 (0.2) \\
$\Lambda$ &0.74 (0.44) &1 (0) \\
Avg. $\theta$ &0 (0) &0 (0) \\
\end{tabular}
\caption{Results for a daycare simulator model in the well-specified (above double line) and misspecified case (below double line)  i.e~with a high-dimensional and low-dimensional statistical model and generative models, respectively. Results are presented as mean(sd) of the 50\% coverage for BOLFI and Split-BOLFI for the parameters $\beta$, $\Lambda$ and averaged over $\theta$. The coverage for the $\theta$ parameter is not meaningful since it lies on the boundary of the parameter space}
\label{DAYCARECOVERAGE}
\end{table}

\begin{figure}[h!]
    \centering
    \centering
    \begin{subfigure}[b]{0.44\textwidth}
        \includegraphics[width=\textwidth]{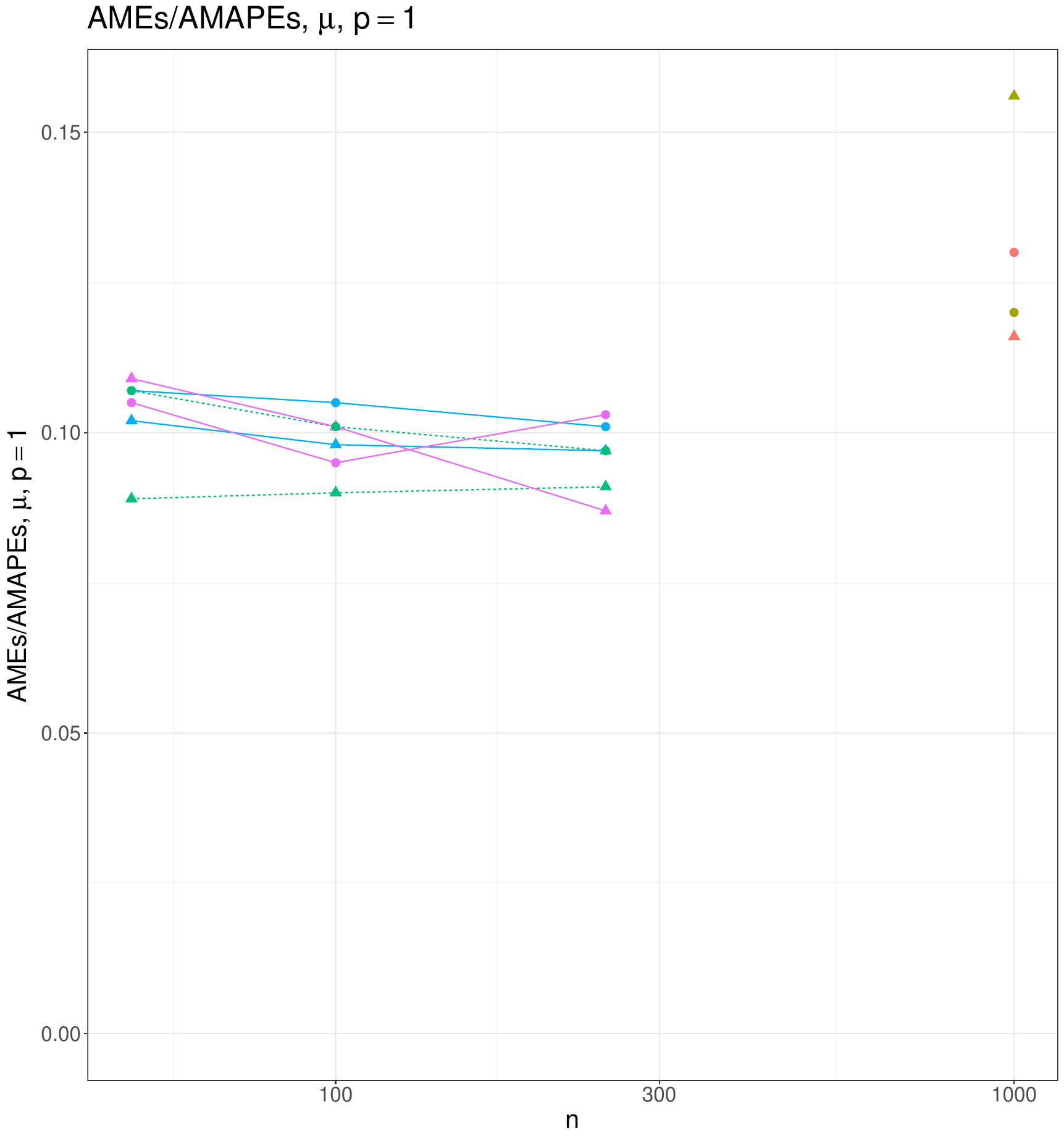}
    \end{subfigure}
    ~ 
    \begin{subfigure}[b]{0.44\textwidth}
        \includegraphics[width=\textwidth]{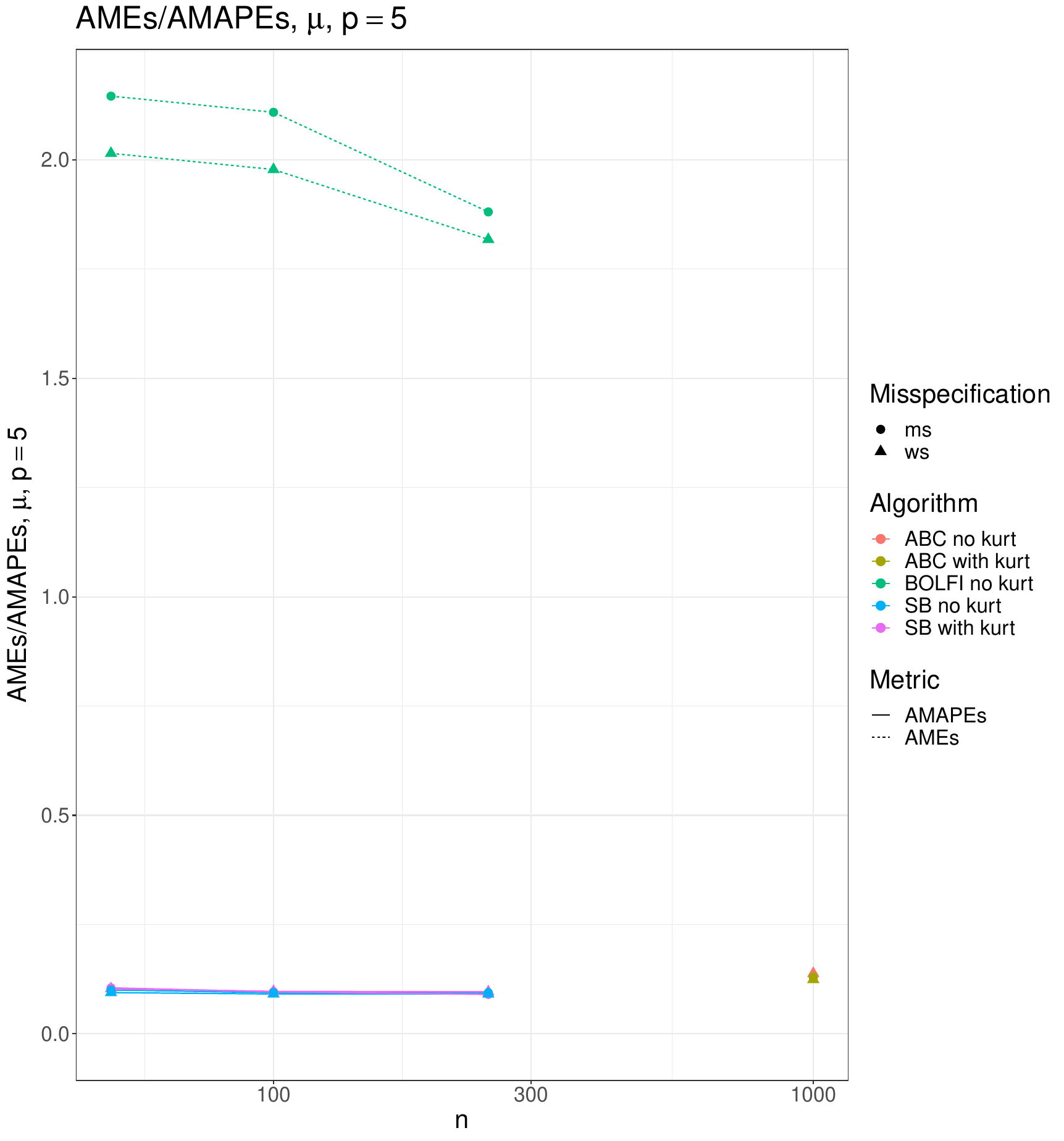}
    \end{subfigure}
    
    \begin{subfigure}[b]{0.44\textwidth}
        \includegraphics[width=\textwidth]{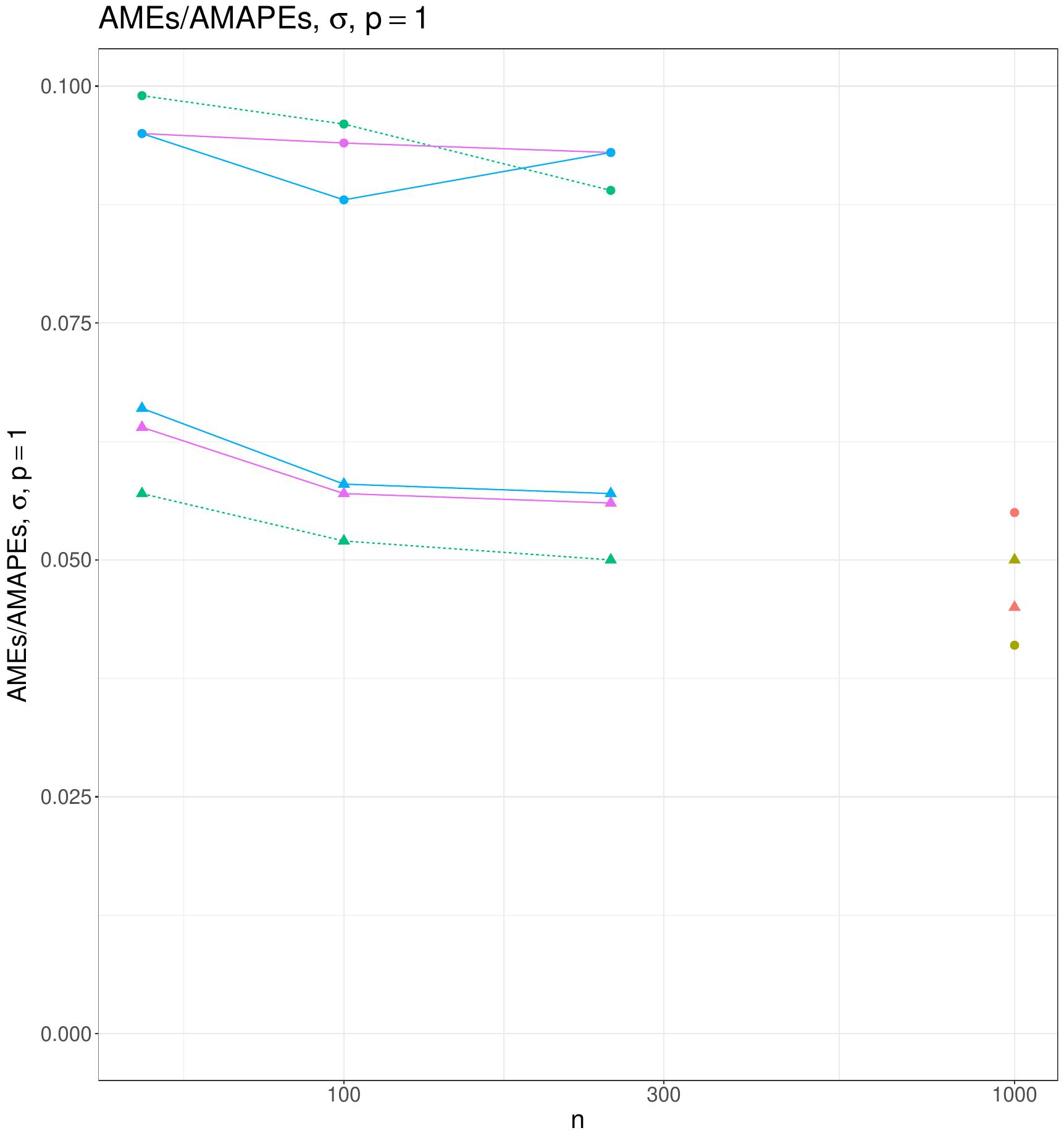}
    \end{subfigure}
    ~ 
    \begin{subfigure}[b]{0.44\textwidth}
        \includegraphics[width=\textwidth]{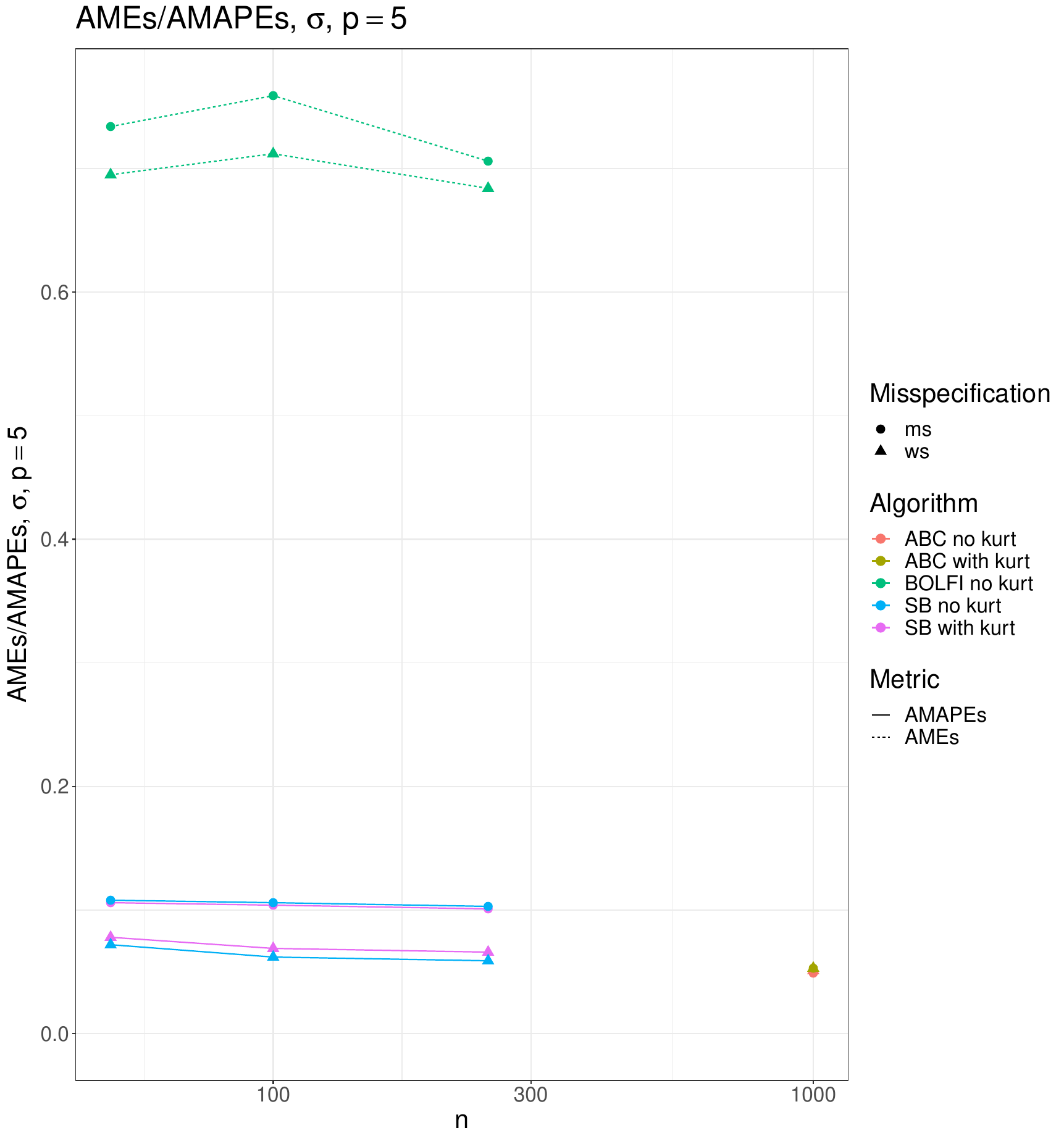}
    \end{subfigure}
    \caption{Results for a Gaussian model with $n_{obs}=500$, as a function of number of simulations $n_{sim}$. Diagrams vary by dimensionality $p$, and parameter $\mu$ or $\sigma$. Plotted are AME for and AMAPE, varying model specification, with algorithms ABC, Split-BOLFI, and BOLFI using mean and standard deviation summaries and with or without the kurtosis summary. \label{fig:gaussian_500_results_appendix_new}}
\end{figure}

\end{document}